\documentclass[preprint]{imsart}
\pdfoutput=1
\usepackage[top=2.8cm, bottom=2.8cm, left=2.8cm, right=2.8cm]{geometry}
\usepackage[utf8]{inputenc}
\usepackage{xr}

\usepackage{tikz}

\usepackage{amsthm,amsmath,amssymb, booktabs}
\RequirePackage{threeparttable}
\usepackage{parskip}
\usepackage[authoryear]{natbib}
\usepackage{hypernat}
\usepackage{marvosym}
\usepackage[hang,small,bf]{caption}
\usepackage{mathtools}
\usepackage{courier}
\usepackage{rotating}
\usepackage[ruled,vlined]{algorithm2e}
\usepackage{epsfig}
\usepackage{hyperref}
\hypersetup{
  colorlinks,%
  citecolor=blue,%
  filecolor=black,%
  linkcolor=blue,%
  urlcolor=blue
}
\usepackage{hypernat}
\usepackage[none]{hyphenat}
\usepackage{enumerate}
\usepackage{color}
\usepackage{bbm}
\usepackage{thmtools}
\usepackage{xcolor}
\usepackage{subcaption}
\usepackage{textcomp}
\usepackage{graphicx}
\usepackage{float}
\usepackage{algpseudocode}

\RequirePackage{adjustbox}
\RequirePackage{blindtext}
\newcommand{\indep}{\rotatebox[origin=c]{90}{$\models$}}
\RequirePackage{color}  
\numberwithin{equation}{section}

\newtheoremstyle{exampstyle}
{8pt} 
{8pt} 
{\it} 
{} 
{\bfseries} 
{.} 
{.5em} 
{} 

\theoremstyle{exampstyle}

\newtheorem{theorem}{Theorem}[section]

\newtheorem{lemma}{Lemma}
\newtheorem{corollary}[theorem]{Corollary}

\startlocaldefs

\newcommand{\eat}[1]{}

\renewcommand{\bar}[1]{\overline{#1}}
\renewcommand{\hat}[1]{\widehat{#1}}

\theoremstyle{plain}

\def\beq{\begin{equation}}
\def\eeq{\end{equation}}
\def\ba{\begin{enumerate}[(a)]}
\def\bei{\begin{enumerate}[(i)]}
\def\be{\begin{enumerate}[(1)]}
\def\ee{\end{enumerate}}
\def\bi{\begin{itemize}}
\def\ei{\end{itemize}}
\def\beg{\begin{eg}}
\def\eeg{\end{eg}}
\def\bd{\begin{defn}}
\def\ed{\end{defn}}
\def\bt{\begin{thm}}
\def\et{\end{thm}}
\def\bl{\begin{lemma}}
\def\el{\end{lemma}}
\def\bfac{\begin{fact}}
\def\efac{\end{fact}}

\def\bc{\begin{corollary}}
\def\ec{\end{corollary}}
\def\bp{\begin{prop}}
\def\ep{\end{prop}}
\def\bo{\begin{observe}}
\def\eo{\end{observe}}
\def\bas{\begin{assumption}}
\def\eas{\end{assumption}}

\endlocaldefs

\begin{document}

\begin{frontmatter}
\title{Robust Model-based Inference for Non-Probability Samples}


\runtitle{Robust Model-based Inference for Non-Probability Samples}

\begin{aug}
\author[A]{\fnms{Ali} \snm{Rafei}},  
\author[A]{\fnms{Michael R.} \snm{Elliott}\thanks{\textit{Corresponding author; address: 426 Thompson St. Ann Arbor, MI 48109. Rm 4068 ISR, email: \href{mailto:mrelliot@umich.edu}{mrelliot@umich.edu}.}}},
\and
\author[B]{\fnms{Carol A. C.} \snm{Flannagan}} 

\address[A]{Survey Methodology Program, University of Michigan}
\vspace{-10pt}
\address[B]{University of Michigan Transportation Research Institute}
\vspace{10pt}
\end{aug}

\begin{abstract}
With the ubiquitous availability of unstructured data, growing attention is paid as how to adjust for selection bias in such non-probability samples. The majority of the robust estimators proposed by prior literature are either fully or partially design-based, which may lead to inefficient estimates if outlying (pseudo-)weights are present. In addition, correctly reflecting the uncertainty of the adjusted estimator remains a challenge when the available reference survey is complex in the sample design. This article proposes a fully model-based method for inference using non-probability samples where the goal is to predict the outcome variable for the entire population units. We employ a Bayesian bootstrap method with Rubin's combing rules to derive the adjusted point and interval estimates. Using Gaussian process regression, our method allows for kernel matching between the non-probability sample units and population units based on the estimated selection propensities when the outcome model is misspecified. The repeated sampling properties of our method are evaluated through two Monte Carlo simulation studies. Finally, we examine it on a real-world non-probability sample with the aim to estimate crash-attributed injury rates in different body regions in the United States.
\end{abstract}


\end{frontmatter}


\section{Introduction}\label{S:1}
With rapid advances in automated measurement technologies, massive amount of unstructured data become ubiquitously available in various fields. Political views shared on social media, Google searches for particular terms, payment transactions recorded by online stores, electronic health records of patients admitted to a chain of hospitals, videos captured by traffic cameras, and GPS trajectory data by satellite are among the common examples of this revolution. This phenomenon coincides with downward trends in the response rate of traditional probability surveys over recent decades, which not only escalate the survey implementation costs, but imperil the validity of their findings in the absence of relevant auxiliary information for non-response adjustment \citep{groves2011three, miller2017there}. As a result, there is growing demand to leverage these novel data sources, as a cheaper and faster alternative to probability surveys, to inform policy \citep{johnson2017big, senthilkumar2018big}. This has led to a surge of research into methods of inference for non-probability samples to mitigate the inherent selection bias in such data \citep{meng2018statistical, kim2021data, rao2021making}.\par

Aside from this shift, there are situations where data collection is expensive or must be administered onsite, and therefore conducting a probability survey may not be practical. For instance, the national automotive sampling system (NASS) is responsible for collecting detailed medical and biomechanical information about the occupants of traffic accidents in the U.S., which demands for well-equipped labs. Although the Crash Injury Research Engineering Network (CIREN) fills this gap in NASS, where more succinct data are gathered regularly through larger probability surveys, the additional data items in CIREN are achieved at the expense of a lack of representativity \citep{elliott2010appropriate}. While it is unknown how well the pool of the traumatic patients referring to CIREN's level 1 trauma centers represents the population of all occupants involved in traffic accidents across the U.S., the collaborating centers tend to be located in the urban areas. More importantly, CIREN tends to recruit severely injured occupants, which may result in overestimated risk of injuries \citep{elliot2013combining, flannagan2009comparison}.\par

The prior literature proposed various robust approaches for finite population inference based on a non-probability sample under situations where a parallel ``reference survey'' is available as the external benchmark \citep{chen2019doubly, rafei2021robust, rafei2022robust}. One common assumption among these approaches is that units of the reference survey have been selected independently with unequal probabilities of selection. This, however, may not hold in practice as probability samples often come with additional complexity in their design. For instance, U.S. federal statistical agencies widely use a stratified multi-stage cluster design in the conduct of their large-scale surveys, such as the American Community Survey (ACS), Current Population Survey (CPS), and National Health Interview Survey (NHIS). These sampling designs are expected to improve sampling efficiency not only statistically but with respect to cost and logistics \citep{kish1965survey, cochran1977sampling}. To maximize statistical efficiency while assuring broad representativeness, a common practice involves partitioning the population into a large number of strata and then selecting a few clusters, ideally two, randomly within each stratum \citep{zhou2014accounting}.\par

To mitigate the potential selection bias in CIREN, which shapes the goal of the application of this article, the literature suggests combining the data with the CDS as a reference survey, because the two samples potentially represent the same population and have a wide range of auxiliary variables in common \citep{elliott2010appropriate}. However, one notable challenge is that units of CDS are selected through a stratified three-stage cluster sampling design \citep{chen2015nhtsa}. \cite{elliott2010appropriate} recognize the importance of adjusting for the cluster-level variance of CDS, and suggest using a jackknife repeated replication method to incorporate the uncertainty of the estimated pseudo-weights into the variance estimator. Under a pseudo-likelihood setting, a modified linearization technique is proposed by \cite{wang2020efficient} to estimate the variance under inverse propensity score weighting (IPSW). When prediction modeling (PM) is incorporated, it is also essential to account for the fixed and random effects of stratification and clustering, respectively, in addition to the sampling weights \citep{little2007bayesian, little2004model}. Failure to do so may cause some degrees of model misspecification, which appear as bias in point and variance estimates.\par

Furthermore, to avoid imposing a pseudo-likelihood structure under a Bayesian framework while adjusting for unequal sampling weights of the reference survey in estimating the PS, \cite{rafei2022robust} employed a two-step pseudo-weighting approach proposed by \cite{Elliott2017Inference}. A noteworthy complication arose from the first step where the sampling weights of the reference survey had to be modeled conditional on the observed auxiliary information explicating the selection mechanism in the non-probability sample. However, the correct structure of such a model may look complicated in the finite population as the weights are deemed to be a deterministic function of the design features of the reference survey. The association between the sampling weights and auxiliary information thus depends on how the latter correlates with those design features. Unless the strength of this association is trivial, misspecifying the working model for sampling weights is likely to induce some degree of residual bias in the final estimates \citep{rafei2020big}.\par

More importantly, the ultimate form of the adjusted estimators in previous literature mostly involve design-based terms, and therefore, are subject to the general drawbacks of design-based methods. For instance, both the QR and PM terms in the proposed estimator of \cite{rafei2021robust} appeared as (pseudo-)weighted sums, and that proposed in \cite{rafei2022robust} still contained a similar structure in the PM term. Despite being design-consistent, it is well-understood that any presence of influential (pseudo-)weights may cause inflated variance in such estimators, especially when the sample size is small \citep{zhang2011comparative, zangeneh2012model, chen2016weights}. From a model-based perspective, \cite{zheng2003penalized} argue that the \textit{HT}-estimator is equivalent to a rigid no-intercept regression model with heteroscedasticity. Although the \textit{HT}-estimator is unbiased, if the true data model is far from the implied model, the resulting estimator will be highly inefficient \citep{Basu1971}.\par 

Alternatively, one can use a fully model-based approach where the goal is to impute the outcome variable(s) for the entire non-sampled units of the population \citep{smith1983validity, little2007bayesian}. Given the observed outcome for the entire population, one can directly quantify the unknown population parameter of interest. This idea clearly eliminates the need for the remaining design-based term in \cite{rafei2022robust}'s adjusted estimator, and therefore not only fills the above-mentioned gaps in bias adjustment when using more flexible models but fully satisfies the likelihood principle \citep{berger1988likelihood, little2004model}. As a direct advantage, one can easily expand such a method to a Bayesian setting. Alternative pseudo-likelihood approaches are limited to distributions with an exponential family and have known problems in uncertainty propagation under a Bayesian setting \citep{savitsky2016bayesian, gunawan2020bayesian, williams2021uncertainty}. \par

The present article proposes a fully model-based approach for robust inference based on non-probability samples when the available reference survey is based on a complex probability sample. To this end, a partially linear Gaussian process regression is employed to non-parametrically link the estimated propensity scores (PS) to the response surface \citep{kammann2003geoadditive}. This approach was termed the Gaussian Process of Propensity Prediction (GPPP) by \cite{rafei2022robust}. However, because the auxiliary variables are unmeasured for the non-sampled units, we generate synthetic populations beforehand by undoing the sampling mechanism of the reference survey through a P\'olya urn scheme \citep{dong2014nonparametric}. The use of a synthetic population eliminates the need for a pseudo-likelihood approach and for modeling the sampling weights when estimating the PS. To be able to correctly account for the cluster-level variance of the reference survey, we choose to use the two-step bootstrap-P\'olya posterior technique proposed by \cite{zhou2016two}. \cite{mercer2018selection} uses a similar idea for inference under a non-probability setting, though with limited theoretical details expressed. Nevertheless, his method utilizes Bayesian Additive Regression Trees (BART) and penalized spline regression for modeling the outcome while our method is based on a GP regression model.\par

The rest of the article is organized as follows: In Section~\ref{S:5.2}, we start by defining the necessary notation and assumptions required for valid inference under a non-probability sample setting and then describe the proposed method under a rigorous mathematical framework. Section~\ref{S:5.3} evaluates the repeated sampling properties of the proposed method and contrasts its performance with alternative approaches through a simulation study. Section~\ref{S:5.4} describes the datasets and auxiliary variables we utilize for the empirical study and reports results of bias adjustment on CIREN/CDS data. Finally, Section~\ref{S:5.5} reviews the strengths and weaknesses of the study in more detail and suggests some future research directions. Supplemental information about the simulation and empirical studies are provided in Appendix~\ref{S:5.6}.\par


\section{Methods}\label{S:5.2}
\subsection{Notation and Assumptions}\label{S:5.2.1}
\noindent
Suppose $U$ is a finite population of known size $N<\infty$, which is partitioned into $H$ homogeneous strata with $M_h$ clusters nested within stratum $h$ $(h=1, 2, ... , H)$ and $M=\sum_{h=1}^H M_h$ the total number of clusters in $U$. Also, let $N_{hj}$ be the size of cluster $j$ in stratum $h$, and thus $N=\sum_{h=1}^H\sum_{j=1}^{N_h}N_{hj}$. One can identify the stratum and cluster of unit $i$ in $U$ through an $(H-1)$-dimensional vector of dummies denoted as $d_i$ and an $(M-H)$-dimensional vector of dummies denoted as $c_i$, respectively. For each $i=1, 2, ... , N$, we consider $y_i$ to be the realized values of a scalar outcome variable, $Y$, and $x_{i}=[x_{i1}, x_{i2}, ... , x_{ip}]^T$ the values of a $p$-dimensional set of relevant auxiliary variables, $X$, in $U$.\par

Let $S_A$ be a non-probability sample selected from $U$ with $[y_i, x^T_i, d^T_i, c^T_i]$ observed and $n_A$ being the sample size. Descriptive inference aims at learning about an unknown parameter of $U$ that is a function of $Y$. The goal here is to draw point and interval estimates for the population mean, i.e. $Q(y)= \bar y_U=\sum_{i=1}^Ny_i/N$. Suppose $\delta^A_{i}=I(i\in S_A)$ represents the inclusion indicator variable of $S_A$ for $i\in U$ whose distribution can be explained by $x_i$. To make valid inference for $S_A$ possible, we assume that conditions \textbf{C1-C3} determined in \cite{rafei2022robust} hold.\par

Consider $S_R$ to be a parallel reference survey of size $n_R$, for which the same set of covariates, $X$, has been measured, though $Y$ has to be unobserved. Also, let $\delta^R_i=I(i\in S_R)$ denote the inclusion indicator variable associated with $S_R$ for $i\in U$. As highlighted in the introduction, we concentrate on a setting where units of $S_R$ are selected via a stratified two-stage clustered sampling design. Note that probability samples often involve more than two stages of clustering, but the majority of existing theories for complex sample survey analysis are built upon the ``ultimate clustering principle'' where the second and higher stages of clustering are ignored, and the sample is treated as if it follows a single-stage clustering design \citep{wolter2007introduction}. From each given stratum $h$ ($h=1, 2, ... , H$), we draw $m_h$ clusters as the primary sampling units (PSU) with first-order inclusion probabilities of $\pi^R_{j|h}$ $(j=1, 2, ... , M_h)$, and $n_{hj}$ subjects from the $j$\textsuperscript{th} cluster as the secondary sampling units (SSU) with second-order inclusion probabilities of $\pi^R_{i|hj}$ $(i=1, 2, ... , N_{hj})$. Note that $n_R=\sum_{h=1}^H\sum_{j=1}^{m_h}n_{hj}$.\par

While the proposed methods in this article impose no restriction on the number of PSUs per stratum, the reference surveys across both the simulation and application parts involve a two-PSU-per-stratum design, i.e. $m_h=2$ for all $h=1, 2, ... , H$. As discussed earlier, such a sampling design is widely used among U.S. government statistical agencies. One common practical way to randomly select the PSUs within strata is probability proportional-to-size (PPS). If $v_{hj}$ denotes the measure of size (MOH) associated with PSU $j$ of stratum $h$, under a PPS technique, $\pi^R_{hj}=v_{hj}/\sum_{j=1}^{M_h} v_{hj}$. In practice, probability survey data are only accompanied by an external set of sampling weights that are supposed to be inversely proportional to the joint selection probabilities, i.e. $w^R_i\propto 1/\pi^R_{hji}$, where $\pi^R_{hij}=\pi^R_{hj}\pi^R_{i|hj}$, which are only observed for $i\in S_R$. It is likely that the sampling weights already involve post-survey adjustments for ineligibility, non-response, and non-coverage errors as well \citep{Valliant2018}. Throughout this article, we assume that the given weights correctly adjust for all these potential sources of bias. Defining $Z=\{D, C, w^R\}$ as the set of all sampling design variables in $S_R$, we also assume $Z$ is fully observed in both $S_R$ and $S_A$ with no item missingness.\par

Despite the possible overlap or correlation between $X$ and $Z$, in addition to the aforementioned conditions, we also assume:
\begin{enumerate}
  \item[\textbf{C4.}] \textbf{Independence of Samples}--- conditional on $\{X, Z\}$, $S_R$ and $S_A$ are selected independently, i.e. $\delta^A\indep\delta^R|X,Z$. 
\end{enumerate}
Then, considering \textbf{C1-C4}, the joint density of $y_i$, $\delta^A_i$ and $\delta^R_i$ can be factored as follows:
\begin{equation}\label{eq:5.1}
p(y_i, \delta^A_i, \delta^R_i| x_i, z_i; \theta, \beta)=p(y_i|x_i, z_i; \theta)p(\delta^A_i|x_i; \beta)p(\delta^R_i| z_i), \hspace{4mm} \forall i\in U
\end{equation}
where $\eta=(\theta, \beta)$ are unknown parameters indexing the conditional distribution of $Y|X, Z$ and $\delta^A|X$, respectively. In the following subsection, we describe a fully model-based strategy for inference based on $S_A$ that models both $p(y_i|x_i, z_i; \theta)$ and $p(\delta^A_i|x_i; \beta)$.\par 





\subsection{Model-based inference}\label{S:5.2.2}
\noindent 
Consider $\hat\pi^A$ and $\hat y$ to be consistent estimates of the pseudo-selection probabilities in $S_A$ and outcome variable in $S_C=S_A\cup S_R$, respectively. The proposed robust estimators by \cite{chen2019doubly} and \cite{rafei2021robust} take the following form: 
\begin{equation}\label{eq:5.2.1}
\hat{\bar y}_U = \left(\sum_{i\in S_A}\frac{\left(y_i-\hat y_i\right)}{\hat\pi^A_i} + \sum_{i\in S_R}\frac{\hat y_i}{\pi^R_i}\right)/N
\end{equation}
which involves two Horvitz-Thompson (HT) estimators from $S_A$ and $S_R$. While preserving double robustness, \cite{rafei2022robust} propose to drop the denominator from the first term as below:
\begin{equation}\label{eq:5.2.2}
\hat{\bar y}_U = \left(\sum_{i\in S_A}\left(y_i-\hat y_i\right) + \sum_{i\in S_R}\frac{\hat y_i}{\pi^R_i}\right)/N
\end{equation}
The main goal here was to reduce the computational burden under a Bayesian framework by limiting the imputations to the combined sample. However, Eq.~\ref{eq:5.2.2} still relies on a \emph{HT} term which can yield inefficient estimates if there are extremely small values of $\pi^R_i$. To get rid of this design-based term, we use the estimator proposed by \cite{liu2021inference}:
\begin{equation}\label{eq:5.2}
\hat{\bar y}_U = \left(\sum_{i\in S_A}\left(y_i-\hat y_i\right) + \sum_{i\in U}\hat y_i\right)/N
\end{equation}
where $\hat y_i$ is the prediction of $y_i$ for $i\in U$. This is a fully model-based estimator in the sense that the unobserved outcome, $Y$, is imputed for the entire population units. \par 

Analogous to \cite{rafei2022robust}, the goal is to specify a model for the joint distribution of $(y_i, \delta^A_i)$ across the units of $U$, which can be formulated by
\begin{equation}\label{eq:5.3}
p(y_i, \delta^A_i | x_i, z_i; \theta, \beta)=p(y_i|x_i, z_i, \delta^A_i; \theta)p(\delta^A_i|x_i; \beta), \hspace{4mm} i\in U
\end{equation}
In a Bayesian setting, descriptive inference about $\bar y_U$ is attained by deriving its posterior predictive distribution given the observed data. By integrating out the unknown parameters, we have
\begin{equation}\label{eq:5.6}
p(\bar y_U|y_A, \delta^A_U, x_U, z_U)=\int \int p(\bar y_U|y_A, \delta^A_U, x_U, z_U, \theta, 
\beta)p(\theta,\beta|y_A, \delta^A_U, x_U, z_U)d\theta d\beta
\end{equation}
where subscripts $_U$ and $_A$ denote a vector of the variables defined in $U$ and $S_A$, respectively. 
Thus $\{X, Z\}$ have to be observed for all of the units of $U$ in order to simulate the posterior predictive distribution of $\bar y_U$.\par

In a practical setting, the measurement of $\{X, Z\}$ is often confined to the pooled sample. To circumvent this issue, we choose to generate a finite set of synthetic populations as the first step by undoing the sampling mechanism in $S_R$, which can be performed non-parametrically through the idea of finite population Bayesian bootstrapping (FPBB) \citep{little2007bayesian, dong2014nonparametric}. Theoretical details of such a method adapted to reference surveys with a complex design, are provided in the following subsection. For any given synthetic population, we then predict the pseudo-selection probabilities of $S_R$, i.e. $\pi^A_i$, as the second step via a QR model. Having $\pi^A_i$ imputed for the entire units of the synthetic population, the third step involves predicting $y_i$ for the entire synthetic population by fitting a PM on $S_A$ with a flexible function of $\hat\pi^A_i$ as a predictor in the model. As demonstrated in \cite{rafei2022robust}, this leads to double robustness in the adjusted estimator.\par

\begin{figure}[hbt!]
\centering\includegraphics[width=1.02\linewidth]{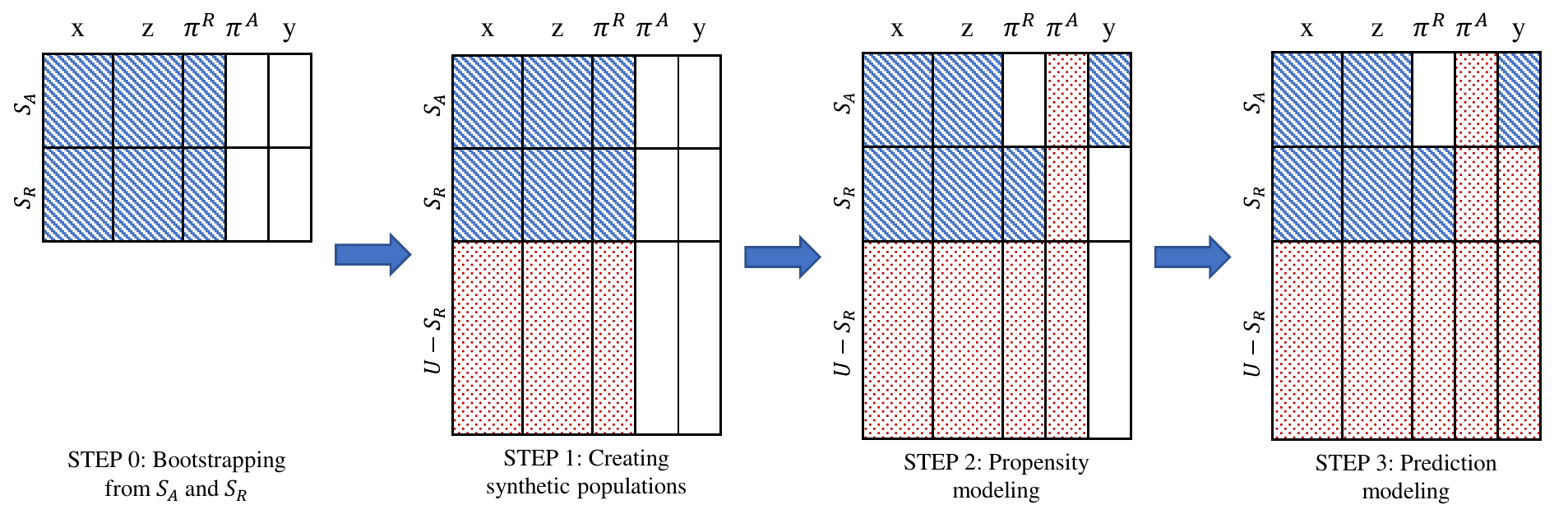}
\caption{Data structure in the population and the combined sample}\label{fig:5.1}
\end{figure}

Once $y_i$ is imputed for all units of a synthetic population, one can compute Eq.~\ref{eq:5.2} and derive point and interval estimates for the population mean by combining the estimates across synthetic populations based on Rubin's multiple imputation rules \citep{rubin1976inference}. This research implements a frequentist approach joint with resampling techniques to train the model parameters based on maximum likelihood estimation (MLE), and to suitably capture the uncertainties due to modeling and complex sampling.  We do this for two reasons:  (1) fitting Bayesian models repeatedly on synthetic populations is demanding computationally for the joint estimation of $(y_i, \pi^A_i)$ especially when $N$ is large, and (2) public-use survey data usually lack detailed information about the selection probabilities across different stages that are essential for reverting the selection mechanism in multi-stage designs. Figure~\ref{fig:5.1} visualizes these steps for the proposed bias adjustment procedure. The shaded blue areas show the observed data while the shaded red areas display the imputed data at each step.\par


\subsubsection{STEP 1: Weighted finite population Bayesian bootstrap}\label{S:5.2.2.1}
\noindent
The first step involves generating synthetic populations using Finite Population Bayesian Bootstrapping (FPBB). \cite{rafei2022robust} discussed in detail how FPBB operates in a probability sample setting, where the goal is to derive the posterior distribution of a given set of variables, such as $\{X, Z\}$, for the non-sampled units of $U$ under an exchangeability assumption. Recently, \cite{zhou2016synthetic} have expanded the weighted variant of FPBB based on P\'olya posterior \citep{cohen1997bayesian, dong2014nonparametric} to create synthetic populations based on a probability sample with a stratified two-stage clustering design. However, their method requires the first and second-order probabilities of inclusion to be known to the analyst such that sampling stages can be reverted hierarchically. Most often, public-use survey data contain merely a variable representing the sampling weights whose reciprocal can be regarded as marginal selection probabilities, but lack selection probabilities at different stages. Because of this constraint, \cite{zhou2016two} propose an alternative two-step approach, where design-adjusted bootstrap resamples are initially selected, and a weighed p\'olya posterior is used to create synthetic populations as the second step.\par

In this study, we employ a similar method using \cite{rao1988resampling}'s bootstrap technique for $S_R$, which is complex in design. It basically uses the idea of ``with replacement'' bootstrap proposed by \cite{mccarthy1985bootstrap}, with an additional rescaling factor applied to the original sampling weights. The asymptotic properties of the bootstrap technique, including consistency, have been well-established for both smooth and non-smooth statistics, such as quantiles \citep{shao1998bootstrapping}. Besides, a bootstrap method can be easily adapted to designs with more than two PSUs per stratum, unlike some other resampling techniques. Compared to the conventional bootstrap, the Rao-Wu variant yields more stable variance estimates when the frequency of PSUs within strata is small at the population level. This is often the case in two-PSU-per-stratum designs where a large pool of strata exists in the sample. For $S_A$, whose elements are assumed to be selected independently according to \textbf{C3}, we use a regular bootstrap method where with replacement random samples of size $n_A$ are drawn repeatedly from the original sample \citep{efron1992bootstrap}.\par 

For Rao-Wu's bootstrap technique, the simplest procedure involves drawing random samples of $m^*_h=m_h-1$ PSUs with replacement from the stratum $h$ ($h=1, 2, ... , H$) of the original sample. These re-samples are then pooled to construct a bootstrap sample of size $m^*=m-H$, and the sampling weights are then rescaled to $w^*_{hji}=w^R_{hji}m_h/(m_h-1)$. This procedure is replicated $B$ times for the $S_R$ to create $S^{(b)}_R$'s ($b=1, 2, ..., B$) with $n^*_R=n_R-H$ being the new sample size. \cite{zhou2016synthetic} note that because the new weights implicitly carry over the information in strata, PSUs, and unequal selection probabilities, one can treat the bootstrapped samples as if units are selected independently with unequal weights, and accordingly, the subscript $hj$ can be dropped, and $w^*_i$'s are normalized such that $N=\sum_{i=1}^{n-H} w^*_i$. In the following, we show how one can create a synthetic population based on a given $S^{(b)}_R$ using a P\'olya posterior distribution. We also denote the $b$\textsuperscript{th} bootstrap replicate of $S_A$ by $S^{(b)}_A$ $(b=1, 2, ... , B)$.\par

Let $t=\{t_1, t_2, ... , t_J\}$ represents a $J$-dimensional vector containing all distinct values of $(x_i, z_i)$ in $S^{(b)}_R$, and $\xi=\{\xi_1, \xi_2, ... , \xi_J\}$ denotes the vector of probabilities $p(x_i=t_j|\xi)=\xi_j$ for $i=1, 2, ... , n^*_R$, $j=1, 2, ... , J$, where $\sum_{j=1}^J\xi_j=1$. Now, suppose $n^*_j$ and $r^*_j$ are the frequencies of units taking the vector $t_j$ in $S^{(b)}_R$ and $\bar S^{(b)}_R=U-S^{(b)}_R$, respectively, for $j=1, 2, ... , J$. It is clear that $\sum_{j=1}^J n^*_j=n^*_R$, and $\sum_{j=1}^J r^*_j=N-n^*_R$. Considering a non-informative Haldane prior, i.e. $\xi\sim Dirichlet(0, ... , 0)$, with a multinomial likelihood function of $p(n^*_1, n^*_2, ... , n^*_J|\xi)\propto\prod_{j=1}^J\xi_j^{n^*_j}$, the posterior distribution of $\xi$ in $\bar S^{(b)}_R$ is given by $(\xi|n^*_1, n^*_2, ... , n^*_J)\sim Dirichlet(n^*_1-1, n^*_2-1, ... , n^*_J-1)$. Using a weighted P\'olya Urn Scheme, \cite{dong2014nonparametric} prove that the posterior predictive draws of $r^*_j$ can be directly generated by
\begin{equation}\label{eq:5.7}
    p(r^*_1, r^*_2, ... , r^*_J | w^*_1, w^*_2, ... , w^*_J)=\frac{\prod_{j=1}^K\Gamma(w^*_j+r^*_j)/\Gamma(w^*_j)}{\Gamma(2N-n^*_R)/\Gamma(N)}
\end{equation}
where the $w^*_j$'s are the normalized rescaled weights of $S^{(b)}_R$ such that $\sum_{j=1}^Kw^*_j=N$.\par 

Finally, one can use \cite{cohen1997bayesian}'s computational algorithm to obtain the weighted-Pol\'ya posterior draws of $\{X, Z\}$ based on $S^{(b)}_R$. This involves taking a Pol\'ya sample of size $N-n^*_R$ denoted by $(X^{(ls)}, Z^{(ls)})$ from the urn $(X^{(l)}, Z^{(l)})$ using the following selection probabilities:
\begin{equation}\label{eq:5.8}
    \zeta_i=\frac{l_{i,k-1}\left(\frac{N-n^*_R}{n^*_R}\right)+w^{(l)^*}-1}{(k-1)\left(\frac{N-n^*_R}{n^*_R}\right)+N-n^*_R}, \hspace{5mm} k=1, 2, ..., N-n^*_R+1
\end{equation}
where $w^{(l)^*}$ denotes the bootstrap weights associated with the $i$-th unit in the $l$-th replicate BB sample, and setting $l_{i,0}=0$, $l_{i,k-1}$ denotes the frequency of unit $i$ selected up to $(k-1)$-th selection \citep{zhou2016two}. For the $l$-th posterior predictive draw, $n^*_j+r^*_j$ represents the frequency of $(x_i, d_i)$ in the synthetic population whose value is $t_j$. Therefore, for any given $S^{(b)}_R$ $(b=1, 2, ... , B)$, one can use this approach to generate $L$ synthetic populations, denoted by $\hat U^{(b,l)}$ $(l=1, 2, ... , L)$. Note that the total number of generated synthetic populations will be $B\times L$.\par

\subsubsection{STEP 2: Imputing the pseudo-selection probabilities in $S_A$}\label{S:5.2.2.2}
\noindent
The second step involves modeling $p(\delta^A_i=1|x_i)$ with the aim to predict $\pi^A_i$ for $i\in\hat U^{(b,l)}$ ($b=1, 2, ... , B; 1, 2, ... , L$). Assuming that $S_A$ has a Poisson sampling design with $n_A$ being random, one can formulate the likelihood function across the population units by
\begin{equation}\label{eq:5.9}
L(\beta|\delta_U^A, x_U)= \prod_{i=1}^N p(\delta^A_i=1|x_i, \beta)^{\delta^A_i}\left[1-p(\delta^A_i=1|x_i, \beta)\right]^{1-\delta^A_i}
\end{equation}
where $\beta$ the vector of parameters associated with the QR model. Note that although the population units are autocorrelated due to clustering, $L$ is written as if units of $U$ are independent. Considering a logistic regression model, we have
\begin{equation}\label{eq:5.10}
\pi^A_i=p(\delta^A_i=1|x_i;\beta)=\frac{exp\{\beta_0+x^T_i\beta_1\}}{1+exp\{\beta_0+x^T_i\beta_1\}}
\end{equation}
where $\beta^T=[\beta_0,\beta^T_1]$. A consistent estimate for $\beta$ can be attained through an MLE method by solving a set of estimating equations. However, one major obstacle is that $\delta^A_i$ is unobserved in a generated synthetic population $\hat U^{(b,l)}$, and imputing this missing variable would require one to match units of $S_A$ to units of $\hat U^{(b,l)}$ \citep{liu2021inference}. To avoid this cumbersome task, we utilize a duplicate removal trick proposed by \cite{wang2020adjusted}.

For any given $(b,l)$ $(b=1, 2, ..., B; l=1, 2, ... , L)$, we initially append $S^{(b)}_A$ to $\hat U^{(b,l)}$, and define $\hat U^{(b,l)*} =S^{(b)}_A\cup^*\hat U^{(b,l)}$, where $\cup^*$ denotes the union of $S_A$ and $\hat U^{(b,l)}$ while allowing for duplicated units of $S^{(b)}_A$ in $\hat U^{(b,l)}$. We then define a new indicator variable $\delta_i^*$ in $\hat U^{(b,l)*}$ taking the value of $1$ for $i\in S^{(b)}_A$ and $0$ for $i\in\hat U^{(b,l)}$. According to \cite{wang2020adjusted}, for a given $(b,l)$, one can show that
\begin{equation}\label{eq:5.12}
\begin{aligned}
\pi^A_i&=p\left(i\in S^{(b)}_A|x_i, i\in \hat U^{(b,l)}\right)\\
&=\frac{p\left(i\in S^{(b)}_A|x_i\right)}{p\left(i\in\hat U^{(b,l)}|x_i\right)}\\
&=\frac{p\left(i\in S^{(b)}_A|x_i\right)/p\left(i\in\hat U^{(b,l)*}|x_i\right)}{p\left(i\in\hat U^{(b,l)}|x_i\right)/p\left(i\in\hat U^{(b,l)*}|x_i\right)}\\
&=\frac{p\left(i\in S^{(b)}_A|x_i, i\in\hat U^{(b,l)*}\right)}{p\left(i\in\hat U^{(b,l)}|x_i, i\in\hat U^{(b,l)*}\right)}\\
&=\frac{p_i(\beta^*)}{1-p_i(\beta^*)}
\end{aligned}
\end{equation}
where $\beta^*$ is a new vector of parameters associated with $\hat U^{(b,l)*}$ and $p_i(\beta^*)=p\left(\delta^*_i=1|x_i, \beta^*\right)$. Note that
\begin{equation}\label{eq:5.13}
    p_i(\beta^*)=\frac{p\left(i\in S^{(b)}_A|\hat U^{(b,l)}\right)}{p\left(i\in S^{(b)}_A|\hat U^{(b,l)}\right)+p\left(i\in\hat U^{(b,l)}|\hat U^{(b,l)}\right)}\leq\frac{1}{2}
\end{equation}
and the equality holds only if $p\left(i\in S_A|\hat U^{(b,l)}\right)=1$. This ensures that $0<\hat\pi^A_i<1$.\par 

Again, under the logistic regression model, we have
\begin{equation}\label{eq:5.14}
    p_i(\beta^*)= p(\delta^*_i=1|x_i;\beta^*_i)=\frac{exp\{x^T_i\beta^*_i\}}{1+exp\{x^T_i\beta^*_i\}}
\end{equation}
where $\beta^*$ can be estimated again using the standard MLE method. Considering (\ref{eq:5.3}), the estimate of $\pi^A_i$ for $i\in\hat U^{(b,l)}$ is given by
\begin{equation}\label{eq:5.15}
    \hat\pi^A_i=\frac{\hat p_i(\beta^*)}{1-\hat p(\beta^*_i)}=exp\{x^T_i\hat\beta^*_i\}
\end{equation}
Hence, for a given $\hat U^{(b,l)}$, to obtain the posterior distribution $p(\pi^A_U|\delta^A_U, x_U)$, all one needs is to append $S^{(b)}_A$ to $\hat U^{(b,l)}$ and fit a Bayesian logistic regression model on the pooled data. Note that the term odds of PS was also seen in the Propensity-adjusted Probability Prediction (PAPP) method in \cite{rafei2020big}. However, its first term, which involved modeling the sampling weights, is dropped by expanding the reference survey to the population.\par

\subsubsection{STEP 3: Prediction modeling using a partially linear Gaussian process regression}\label{S:5.2.2.3}
\noindent
The goal of PM here is to impute $y_i$ for $i\in U^{(b,l)}$ $(b=1, 2, ... , B; l=1, 2, ... , L)$ by estimating $E(y_i|x_i, z_i; \theta)$, which is essential for the calculation of Eq.~\ref{eq:5.2}. To this end, one has to know about the true functional form of the response surface that governs the superpopulation structure from which $U$ has been selected. Considering the fact that $U$ is clustered, suppose the true PM is given by
\begin{equation}\label{eq:5.17}
g\left(E\left[y_{ij}|x_{ij}, d_{ij}, c_i, w^R_{ij}; \theta\right]\right)=\theta_0 + x^T_{ij}\theta_1 + d^T_{ij}\theta_2 + log\left(w^{R}_{ij}\right)\theta_3 + u_i
\end{equation}
where the subscript $ij$ indicates the observed variable in the $j$\textsuperscript{th} unit of the $i$\textsuperscript{th} cluster, $\theta$ is a $(p+H+1)$-dimensional vector of unknown model parameters, $g$ is an appropriate \emph{link} function, and $u_i\sim N(0, \sigma^2_u)$ denotes a random effect term associated associated with the cluster $i$ $(i=1, 2, ... , M)$. Note that the notational convention used in Eq.~\ref{eq:5.17} suppresses showing the cluster indicator, $c_i$, on the right side of the equation.\par

In a non-probability sample setting, since $y_i$ is only observed for $i\in S_A$ the PM has to be trained based on $S^{(b)}_A$ $(b=1, 2, ... , B)$. Considering \textbf{C2} and by the use of the Bayes rule, one can show that $p(y_i|x_i, z_i; \theta)=p(y_i|\delta^A_i=1, x_i, z_i; \theta)$. This implies that a consistent estimate for $\theta$ can be derived from $S_A$. Therefore, one can use a marginal likelihood function, restricted to $S_A$, to obtain a consistent estimate for $\theta$ as below:
\begin{equation}\label{eq:5.18}
L(\theta|y_A, x_A, z_A)=\prod_{i=1}^m\prod_{j=1}^{n^A_i} \int_u p(y_{ij}|x_{ij},z_{ij}, u_i;\theta)\phi(u_i;0,\sigma^2_u)d u_i
\end{equation}
where $\phi(.;\mu,\sigma^2)$ denotes a normal density function with mean $\mu$ and variance $\sigma^2$. To obtain an MLE for $\theta$ when the outcome is non-normal, one can utilize adaptive Gaussian quadrature to assess the integral in Eq.~\ref{eq:5.17} \citep{pinheiro1995approximations}. It is important to note that there must be adequate sample from all $H$ strata observed in $S_A$ and $(d_i, w^R_i)$ must be known for all $i\in S_A$ such that one can fit the model in Eq.~\ref{eq:5.17} on $S_A$. This introduces a big challenge for inference based on a non-probability sample, which will be further elaborated in the discussion.\par

In reality, however, the true functional form of the response surface in $U$ is not known to the analyst. It is likely that the auxiliary variables, especially MOS, are non-linearly associated with the conditional expectation of the outcome variable or multi-way interaction effects are present among the covariates. Similar to \cite{rafei2022robust}, one can add the estimated $\pi^A_i$ from Step II as a predictor to the PM with the aim to further protect against the misspecification of the PM. This leads to a doubly robust (DR) estimator, which preserves consistency if the underlying model for either QR or PM is correctly specified \citep{robins1994estimation}. Linear-in-weight (LWP) prediction is the most common type of a model-based DR estimator, which uses the inverse of $\hat\pi^A_i$ as a linear predictor in the PM in addition to $\{x_i,z_i\}$, and is equivalent to the earliest class of DR methods, i.e. augmented inverse propensity weighting (AIPW) \citep{scharfstein1999adjusting, bang2005doubly}. \cite{zheng2003penalized} realized that the use of a more flexible function of $\hat\pi^A_i$ can yield improved efficiency in the ultimate DR estimator. The improvement is more tangible when extremely low values of $\hat\pi^A_i$ are present \citep{zhang2011comparative}. This problem often arises when there is a partial lack of common distributional support in $x_A$.\par 

While \cite{zheng2003penalized} suggest using a penalized spline structure to non-parametrically link $\hat\pi^A_i$ to the response surface, \cite{rafei2022robust} proposed to use a partially linear Gaussian process (GP) regression as the PM. By choosing a stationary isotropic covariance function, we showed that GP behaves as an optimal kernel matching technique based on $\hat\pi^A_i$. The interest in this study is to use GP for matching the units of $S^{(b)}_A$ to those of $U^{(b,l)}$ for any given $b=1, 2, ... , B; l=1, 2, ... , L$. Therefore, we update the PM given in Eq.~\ref{eq:5.18} as below:
\begin{equation}\label{eq:5.19}
g\left(E\left[y_{ij}|x_{ij}, d_{ij}, w^R_{ij}, \hat\pi_{ij}; \theta\right]\right) = \theta_0 + x^T_{ij}\theta_1 + d^T_{ij}\theta_2 + log(\pi^R_{ij})\theta_3 + f\left(log(\hat\pi^A_{ij})\right)
\end{equation}
where $f(x)\sim GP\left(\mu(x), K\left(x, x'\right)\right)$ is a GP with mean $\mu(x)$ and covariance matrix $K\left(x, x'\right)$. As illustrated, we propose to use a $log$ transformation of $\hat\pi^A_i$ as the GP inputs. This is mainly because $log\left(\hat\pi^A_i\right)$ tends to be normally distributed as $log(\hat\pi^A_i)=x^T_i\hat\beta^*$ according to Step II. Note that to minimize the computational burden, we dropped the random intercept term associated with the cluster effects in Eq.~\ref{eq:5.14}, as the mean of random effects is expected to be $zero$, and we already accounted for the cluster-level variance through the Rao-Wu's bootstrapping method.\par

\cite{rafei2022robust} dealt with a fully Bayesian framework where we aimed to jointly estimate $(\pi^A_i, y_i)$. The proposed GP was characterized by a two-parameter M\'atern covariance function added to an inharmonious polunomial kernel. It is well-understood that the joint posterior distribution of the length-scale and marginal standard error parameters are weakly identifiable while their ratio is well-identifiable \citep{zhang2004inconsistent}. Hence, such a method may not be computationally efficient in this study, as the PM in Eq.~\ref{eq:5.19} has to be fitted $BL$ we employ a frequentist variant of GP with a single smoothing parameter where model parameters are estimated through a maximum a posteriori (MAP) technique. Suppose $K(x,x_i)=k(||x,x_i||)$ is a non-negative function with $||.||$ being the Euclidean norm, such that $k(0)=1$ and $k(d)\rightarrow 0$ monotically as $d\rightarrow\infty$. Now, we define $f$ as
\begin{equation}\label{eq:5.20}
f(log(\hat\pi^A))=log(\hat\pi^A)\theta_4+\sum_{i=1}^{n^{(b)}_A} v_iK\left(log(\hat\pi^A), log(\hat\pi^A_i)\right)
\end{equation}
where $v\sim N(0,(\lambda K)^{-1})$ with $\lambda$ being a smoothing parameter. Note that the covariance matrix of $f$ is $K/\lambda$ \citep{wood2017generalized}. The MAP estimate for $\theta$ is then given by
\begin{equation}\label{eq:5.20}
\hat\theta=argmin_\theta||y-x^T\theta_1-log(w^R)\theta_2-d^T\theta_3-\theta_4log(\hat\pi^A)-Kv||^2/\sigma^2+\lambda v^TKv
\end{equation}
For the structure of $K$, we use a simplified 
version of Mat\'ern covariance function as below:
\begin{equation}\label{eq:5.20}
K(x, x_i)=\left(1+\frac{||x-x_i||}{\rho}\right)exp\bigg\{-\frac{||x-x_i||}{\rho}\bigg\}
\end{equation}
where $\rho=\max_{1\leq i,j\leq n_A}||x_i-x_j||$ as recommended by \cite{kammann2003geoadditive}. This ensures scale invariance and numerical stability in finding the parameter estimates. This partially linear GP regression model can be implemented in R using the package `\emph{mgcv}'. 

Having $y_i$ imputed for all $i\in U^{(b, l)*}$, we estimate the population mean of $Y$ for a given $(b, l)$ by
\begin{equation}\label{eq:5.21}
\hat{\bar y}^{(b, l)}_U = \left(\sum_{i\in S^{(b)}_A}\left(y_i-\hat y_i\right) + \hat y^{(b, l)}_U\right)/N
\end{equation}
where $\hat y^{(b, l)}_U=\sum_{i\in U^{(b, l)}} \hat y_i$. Upon calculating the expression~\ref{eq:5.21} for all $b=1, 2, ... , B$ and $l=1, 2, ... , L$, one can build a DR estimate for the population mean via Rubin's combining rule
\begin{equation}\label{eq:5.22}
\hat{\bar y}_U=\frac{1}{BL}\sum_{b=1}^B\sum_{l=1}^L\hat{\bar y}^{(b,l)}_U
\end{equation}
and the corresponding variance estimator is given by
\begin{equation}\label{eq:5.23}
\widehat{Var}(\hat{\bar y}_{U})=\frac{B+1}{B(B-1)}\sum_{b=1}^B\left(\hat{\bar y}^{(b)}_{U}-\hat{\bar y}_{U}\right)^2
\end{equation}
where $\hat{\bar y}^{(b)}_{U}=\sum_{l=1}^L\hat{\bar y}^{(b, l)}_{U}/L$. It is important noting that the within-imputation variance in Rubin's combining rule, i.e. $Var(\hat{\bar y}^{(b)}_{U})$, is $zero$ as we impute the PS and outcome for the entire population \citep{zhou2016two}. One can construct the 95\% CI for the population mean of $Y$ based on a $t$-student reference distribution with degrees of freedom equal to $min\{m-H, B-1\}$ \citep{zhou2016jos}.\par

Note that the population size $N$ was assumed to be known. Most often, this population parameter is unknown to the analyst. Although an estimate of the population size can always be obtained from the public-use weights of the reference survey by $\hat N=\sum_h\sum_j\sum_i w^R_{hji}$, one has to account for the uncertainty of $\hat N$ as well. For this purpose, we propose to replace $N$ in Eq.~\ref{eq:5.21} with $N^{(b)}_R=\sum_h\sum_j\sum_i w^*_{hji}$ where $w^*_{hji}$'s denote the rescaled sampling weights in $S^{(b)}_R$ $(b=1, 2, ... , B)$ based on Rao-Wu's method. Algorithm~\ref{alg:1} summarizes the entire procedure of the proposed GPPP method.\par

\vspace{5mm}

\begin{algorithm}[H]
\SetAlgoLined
\raggedright

\footnotesize{\KwResult{Adjusted point and variance estimates of finite population mean}
 Combine $S_A$ with $S_R$\;
 \For{b from 1 to B}{
 Draw $S^{(b)}_A$, a regular bootstrap sample of size $n_A$, from $S_A$\;
  \For{h from 1 to H}{
  Draw $m_h-1$ bootstrap PSUs from $h$\textsuperscript{th} stratum of $S_R$\;
  }
  Construct $S^{(b)}_R$ by combining bootstrapped PSUs across $H$ strata\;
  Apply \cite{rao1988resampling} rescaling method to sampling weights of $S^{(b)}_R$\;
  \For{l from 1 to L}{
     Generate $\hat U^{(b, l)}$, a synthetic population of size $N$, based on $S^{(b)}_R$\;
     Attach $S^{(b)}_A$ to $\hat U^{(b, l)}$ to construct $\hat U^{(b, l)*}$, and define $\delta^{A*}_i=I(i\in S^{(b)}_A)$\;
     Estimate $\hat\pi^A_i$ for $i\in \hat U^{(b, l)*}$ by modeling $p(\delta^{A*}_i=1|x_i)$ on $\hat U^{(b, l)*}$\;
     Predict $\hat y_i$ for $i\in \hat U^{(b, l)*}$ by modeling $E[y_i|\hat\pi^A_i,x_i, z_i]$ on $S^{(b)}_A$\;
     Compute $\hat{\bar y}^{(b, l)}_U = \left(\sum_{i\in S^{(b)}_A}\left(y_i-\hat y_i\right) + \sum_{i\in U^{(b, l)}} \hat y_i\right)/N$\;
  }
  Compute $\hat{\bar y}^{(b)}=\left(\sum_{l=1}^L \hat{\bar y}^{(b, l)}_U \right)/L$\;
 }
 Compute $\hat{\bar y}_U=\sum_{b=1}^B \hat{\bar y}^{(b)}_U/B$ and $\widehat{Var}(\hat{\bar y}_U)=(1+B^{-1})\sum_{b=1}^B \left(\hat{\bar y}^{(b)}_U-\hat{\bar y}_U\right)^2/(B-1)$\;}
 \caption{Fully model-based inference for non-probability samples}\label{alg:1}
\end{algorithm}



\section{Simulation study}\label{S:5.3}
\noindent
We design an extensive simulation study to gauge how the proposed GPPP method operates with respect repeated sampling properties, including unbiasedness and efficiency, and contrast the results with those based on the PSPP, LWP, AIPW and IPSW methods. While all the competing methods are recognized as DR except for the IPSW, the two latter constitute a design-based form unlike the rest that are fully model-based. For any given pair of bootstrap samples, $S^{(b)}_A$ and $S^{(b)}_R$ $(b=1, 2, ... , B)$ and a synthetic population $\hat U^{(b,l)}$ $(l=1, 2, ... L)$ created based on $S^{(b)}_R$, an AIPW estimator for the population mean is given by
\begin{equation}\label{eq:5.23}
\hat{\bar y}^{(b, l)}_U = \frac{1}{N}\sum_{i\in S^{(b)}_A}\frac{(y_i-\hat y_i)}{\hat\pi^A_i} + \frac{1}{N}\sum_{i\in U^{(b, l)}} \hat y_i
\end{equation}
where $\hat y^{(b, l)}_i$ is the prediction of $y_i$ for $i\in \hat U^{(b, l)}$ based the PM presented in Eq.~\ref{eq:5.18}. Note that the main difference between Eq.~\ref{eq:5.23} and Eq.~\ref{eq:5.2} is in how $\hat\pi^A_i$ is applied. In the earlier, $\hat\pi^A_i$ is used to inversely weight the PM residuals, whereas the latter utilizes $\hat\pi^A_i$ as a model covariate to predict $y_i$ in PM.\par

Regarding the PSPP method, we use a P-spline smoothing technique proposed by \cite{eilers1996flexible} in the PM, which specifies a discrete penalty to the basis functions. This is equivalent to a situation where knot coefficients are treated as random effects \citep{ruppert2003semiparametric}. Here a cubic spline is employed under a ridge penalty with $10$ evenly spaced knots. Setting $B=50$ and $L=10$, we then utilize Rubin's combining rules to obtain point and interval estimates of the population mean for all the competing methods. Our focus here is on a situation where $S_R$ is complex in design, but units of $S_A$ are selected independently with unequal probabilities of selection. Once $S_A$ and $S_R$ are drawn from $U$, we assume that $\pi^A_i$ for $i\in S_A$ and $y_j$ for $j\in S_R$ are unobserved, and the aim is to adjust for the selection bias in $S_A$ based on the combined sample. The simulation is then iterated $K=1,008$ times (which is a multiple of $36$, the number of cores we employed for parallel computing), where the bias-adjusted point estimates, SE, and associated 95\% confidence intervals (CI) are estimated for $\bar y_U$ in each iteration.\par

\subsection{Simulation design}\label{S:5.3.1}
\noindent
Initially, we generate a hypothetical population of size $N=50,000$ with $H=50$ strata and $M_h=20$ clusters, each with a fixed size of $N_{hj}=50$ ($j=1, 2, ... , 20$), within the stratum $h$ ($h=1, 2, ... , 50$). Then, two random variables, $v_1$ and $v_2$, are created at the cluster and unit levels, respectively with the following distributions
\begin{equation}\label{eq:5.24}
V_1\sim EXP(\mu=1), \hspace{15mm} V_2 \sim Unif(1, 5)
\end{equation}
which are supposed to be the MOS associated with first- and second-order probabilities of inclusion in $S_R$. The first-order selection probabilities are produced by $\pi^R_{j|h}\propto c+ v_1$ where $c$ is defined such that $max\{\pi^R_{j|h}\}/min\{\pi^R_{j|h}\}=30$. we then normalize $\pi^R_{j|h}$ such that $\sum_{h=1}^H\sum_{j=1}^{M_h}\pi^R_{j|h}=1$. Also, for each $i$\textsuperscript{th} unit of the $j$\textsuperscript{th} cluster in the stratum $h$, the second-order probabilities of inclusion are obtained by $\pi^R_{i|hj}=v_{2hji}/\sum_{j=1}^{M_h}\sum_{i=1}^{N_{hj}}v_{2hji}$. As a result, the joint selection probabilities and sampling weights in $S_R$ are then calculated by $\pi^R_{hji}=\pi^R_{j|h}\pi^R_{i|hj}$ and $w^R_{hji}\propto 1/\pi^R_{hji}$. Finally, we normalize the weights such that $\sum_{h=1}^H\sum_{j=1}^{M_h}\sum_{i=1}^{N_{hj}}w^R_{hji}=N$.

Associated with the selection mechanism of $S_A$, we also create a random variable $X$, which is assumed to be correlated with $w^R_{hji}$ through $x_{hji}=-7+log(w^R_{hji})+\rho\epsilon_{hji}$ where $\epsilon_{hji}\sim N(0, 1)$ and $\rho$ is determined such that $cor(x_{hji}, log(w^R_{hji}))=0.5$. A \emph{logistic} function is then employed to produce the selection probabilities related to $S_A$ by
\begin{equation}\label{eq:5.25}
\pi^A(x_{hji})=p(\delta^A_{hji}=1|x_{hji}) = \frac{exp\{\gamma_0+\gamma_1x_i\}}{1+exp\{\gamma_0+\gamma_1x_i\}}
\end{equation}
We pick two values for $\gamma_1$, $0.3$ and $0.6$, where the latter produces extremely low selection probabilities. The goal here is to assess how the competing adjustment methods perform in the presence of influential weights. For a given $\gamma_1$, the value of $\gamma_0$ is then found such that $\sum_{h=1}^H\sum_{j=1}^{M_h}\sum_{i=1}^{N_{hj}}\pi^A_{hji}=n_A$.

\begin{figure}[hbt!]
\centering\includegraphics[scale=0.22]{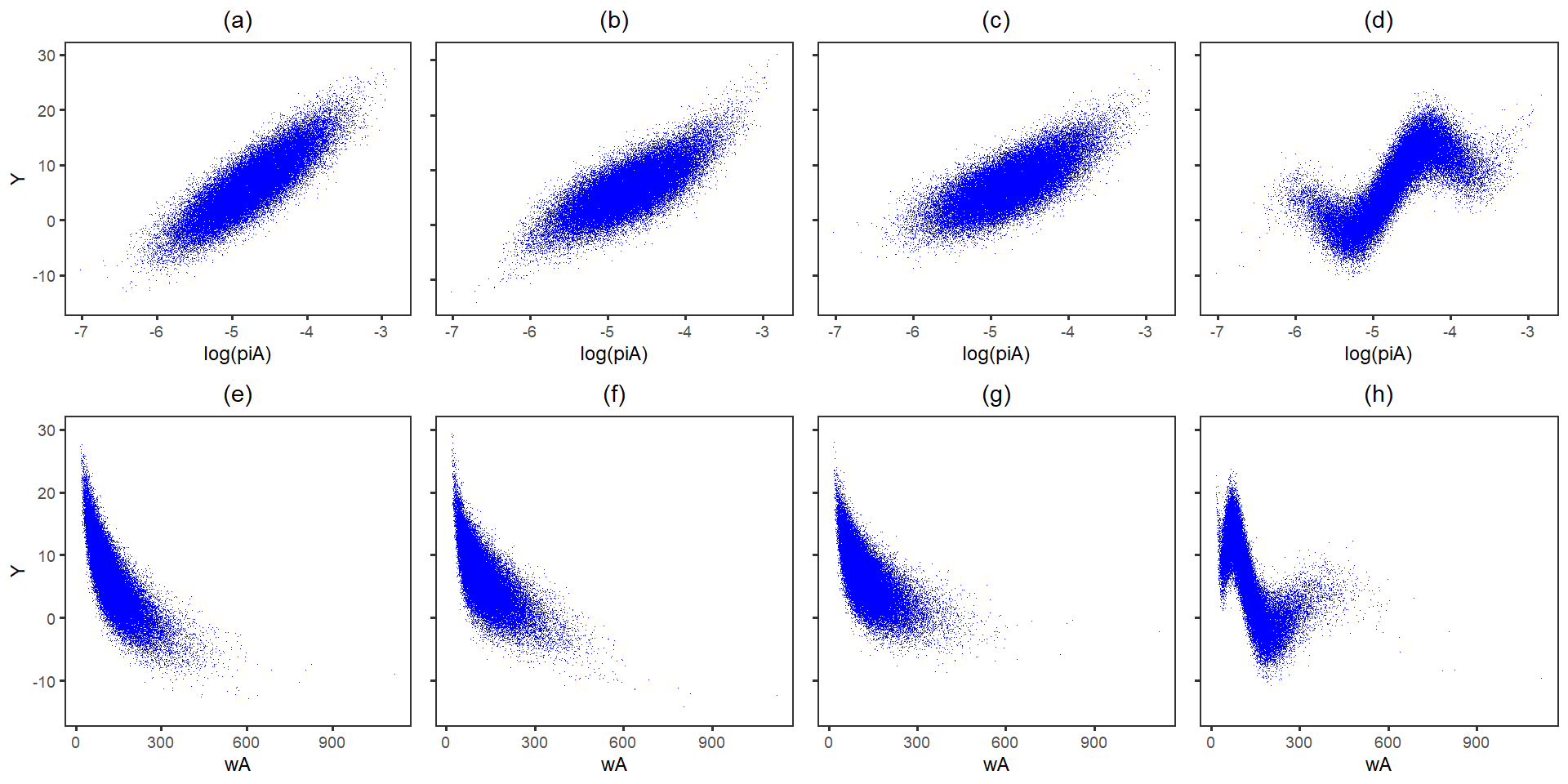}
\caption{The proposed relationships between the outcome variable $Y$ and $log(\pi^A)$ in $U$ for (a) $LIN$, (b) $CUB$, (c) $EXP$ and (d) $SIN$ scenarios, and between the outcome $Y$ and sampling weights $w^A$ for (e) $LIN$, (f) $CUB$, (g) $EXP$ and (h) $SIN$ scenarios.}\label{fig:5.2}
\end{figure}

Using a non-linear mixed effect model, as the next step, we generate a continuous outcome variable denoted by $y^c_{ij}$ and a binary outcome variable denoted by $y^b_{ij}$ by
\begin{equation}\label{eq:5.26}
\small{\begin{aligned}
y^c_{kij}|x_{ij}, w^R_{ij}, d_{ij} & \sim N(\mu=1+f_k(x_{ij})+log(w^R_{ij})-0.1d_{ij}+0.2x_{ij}log(w^R_{ij})+u_i, \sigma^2=4)\\
y^b_{kij}|x_{ij}, w^R_{ij}, d_{ij} & \sim Ber\left(\frac{exp\{-7+f_k(x_{ij})+log(w^R_{ij})-0.1d_{ij}+0.2x_{ij}log(w^R_{ij})+u_i\}}{1+exp\{-7+f_k(x_{ij})+log(w^R_{ij})-0.1d_{ij}+0.2x_{ij}log(w^R_{ij})+u_i\}}\right)
\end{aligned}}
\end{equation}
where $d_{ij}$ is a categorical variable indicating the strata in $U$ which takes values of $0, 1, ... , 49$, $u_i\sim N(0, \sigma^2_u)$, and $\sigma^2_u$ is determined such that the intraclass correlation (ICC) equals $0.2$ \citep{oman2001modelling, hunsberger2008testing}. The function $f_k(.)$ is assumed to take one of the following forms:
\begin{equation}\label{eq:5.28}
\begin{aligned}
LIN: f_1(x)&=x \hspace{32mm} CUB:f_2(x)=(x/3)^3\\ EXP: f_3(x)&=exp(x/2)/5 \hspace{15mm} SIN: f_4(x)=5sin(\pi x/2)
\end{aligned}
\end{equation}
Figure~\ref{fig:5.2} depicts the association between $y_i$ and $\pi^A_i$ and also between $y_i$ and $w^A_i=1/\pi^A_i$ in the generated hypothetical population under each of these scenarios.

Associated with $S_R$ and $S_A$, independent samples of size $n_R=1,000$ and $n_A=500$ were selected randomly from $U$. Regarding $S_R$, we consider a stratified two-stage clustering design with two PSU selected per stratum with $m_h=2$ and $n_{hj}=10$ ($h=1, 2, ..., 50$) and ($j=1, 2$). At each stage, without replacement random samples are selected with unequal inclusion probabilities from PSUs and SSUs. In addition, a Poisson sampling design is used to select units of $S_A$ with unequal inclusion probabilities $\pi^A_{hji}$ directly from $U$.  Finally, to test the DR property of the proposed methods, we investigate different scenarios regarding whether models for QR and PM are correctly specified or not. To misspecify a model, we only include $x^2_{ij}$ as a predictor in the model.\par

To evaluate the repeated sampling properties of the competing method, relative bias (rBias), relative root mean square error (rMSE), the nominal coverage rate of 95\% CIs (crCI), relative length of 95\% CIs (rlCI) and SE ratio (rSE) are calculated as below:
\begin{align}
rbias\left(\hat{\bar y}_{U}\right) &=100 \times\frac{1}{K}\sum_{k=1}^K \left(\hat{\bar y}^{(k)}_{U}-\bar y_U\right) /\bar y_U\\
rMSE\left(\hat{\bar y}_{U}\right) &=100 \times\sqrt{\frac{1}{K}\sum_{k=1}^K\left(\hat{\bar y}^{(k)}_{U}-\bar y_U\right)^2} /\bar y_U\\
crCI\left(\hat{\bar y}_{U}\right) &=100 \times \frac{1}{K}\sum_{k=1}^K I\left(\big|\hat{\bar y}^{(k)}_{U} - \bar y_U\big| <z_{0.975}\sqrt{var\left(\hat{\bar y}^{(k)}_{U}\right)}\right)\\
rlCI\left(\hat{\bar y}_{U}\right) & = 100\times \frac{2}{K}\sum_{k=1}^K z_{0.975}\sqrt{var\left(\hat{\bar y}^{(k)}_{U}\right)}/\bar y_U\\
rSE\left(\hat{\bar y}_{U}\right) &= \frac{1}{K}\sum_{k=1}^K \sqrt{var(\hat{\bar y}^{(k)}_{U})}/\sqrt{\frac{1}{K-1}\sum_{k=1}^K \left(\hat{\bar y}^{(k)}_{U}-\bar{\bar y}_{U}\right)^2}
\end{align}
where $\hat{\bar y}^{(k)}_{U}$ denotes the adjusted sample mean from iteration $k$, $\bar{\bar y}_{U}=\sum_{k=1}^K \hat{\bar y}^{(k)}_{U}/K$, $\bar y_U$ is the finite population true mean, and $var(.)$ represents the variance estimate of the adjusted mean based on the sample.\par

\subsection{Simulation results}\label{S:5.3.2}
\begin{figure}[hbt!]
\centering\includegraphics[scale=0.28]{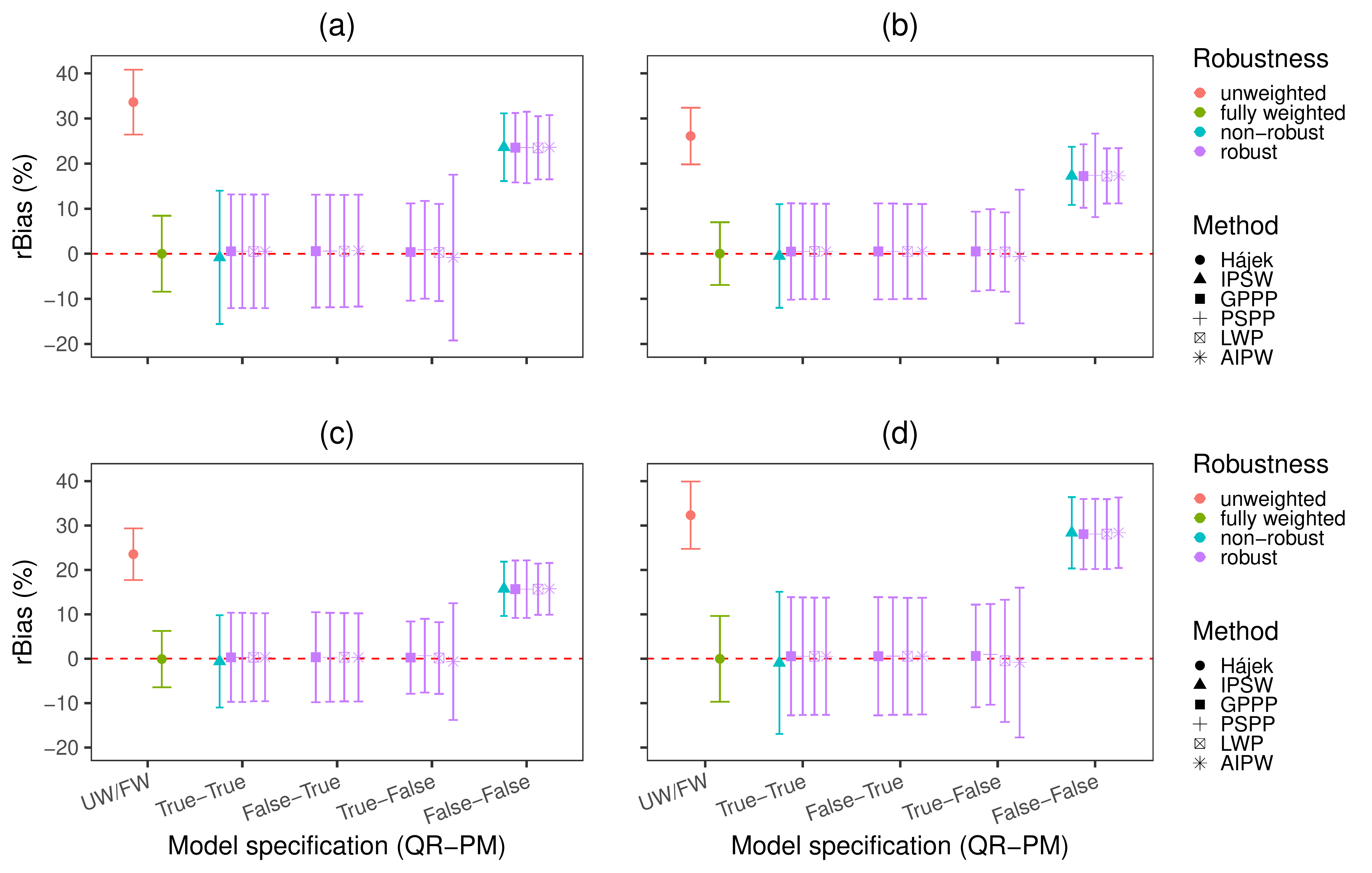}
\caption{Comparing the performance of the adjusted estimators for the \emph{continuous} outcome variable across different model-specification scenarios under (a) LIN, (b) CUB, (c) EXP, and (d) SIN when $\gamma_1=0.3$. The error bars point out the relative length of 95\% CIs. UW: unweighted; FW: Fully weighted; IPSW: Inverse Propensity Score Weighting; GPPP: Gaussian Processes of Propensity Prediction; PSPP: Penalized Spline of Propensity Prediction; LWP: Linear-in-weight Prediction; AIPW: Augmented Inverse Propensity Weighting}\label{fig:5.3}
\end{figure}

\noindent
Figure~\ref{fig:5.3} visualizes the magnitude of bias and efficiency of the competing methods for various types of response surface structures and across different scenarios of model specification when $\gamma_1=0.3$. Note that the error bars point out the expected rlCI over the simulation iterations with narrower bands indicating higher efficiency. As illustrated, naive estimates are severely biased while adjustment based on the true sampling weight (if known to the analyst) removes the bias entirely. Generally, we observe that all the competing methods produce unbiased estimates as long as the working model is valid for either QR or PM. This finding holds across all the four outcome variables when there is no evidence of influential pseudo-weights. When both working models are wrong, not surprisingly, estimates based on all the applied methods are biased.\par

\begin{figure}[hbt!]
\centering\includegraphics[scale=0.28]{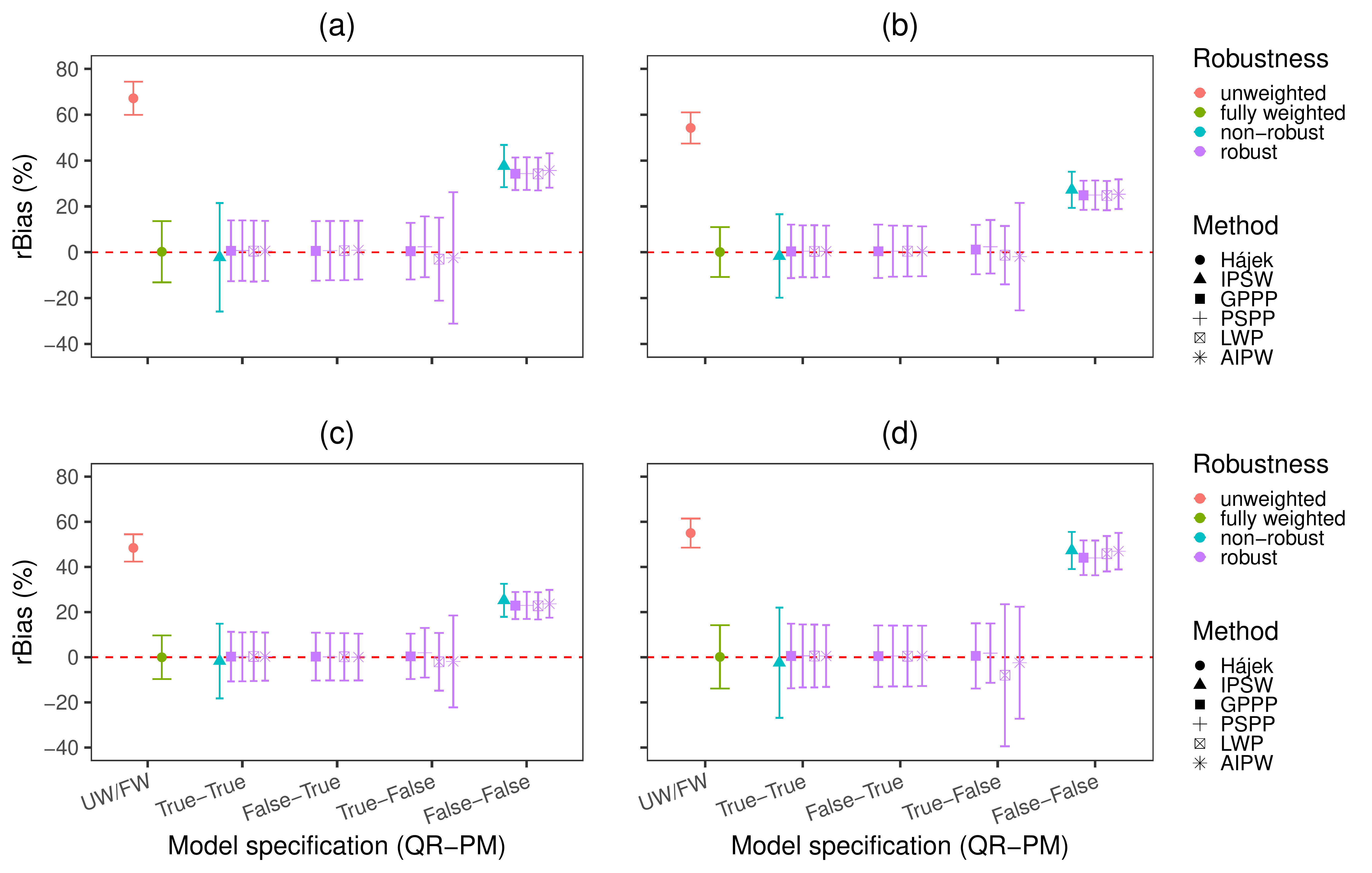}
\caption{Comparing the performance of the adjusted estimators for the \emph{continuous} outcome variable across different model-specification scenarios under (a) LIN, (b) CUB, (c) EXP, and (d) SIN when $\gamma_1=0.6$. The error bars point out the relative length of 95\% CIs. UW: unweighted; FW: Fully weighted; IPSW: Inverse Propensity Score Weighting; GPPP: Gaussian Processes of Propensity Prediction; PSPP: Penalized Spline of Propensity Prediction; LWP: Linear-in-weight Prediction; AIPW: Augmented Inverse Propensity Weighting}\label{fig:5.4}
\end{figure}

When it comes to efficiency, the adjustment methods perform almost the same, except for the design-based estimators, i.e. IPSW and AIPW, that consistently show lower efficiency across the four plots in situations where it is only the QR model that is correctly specified. In addition, when extreme non-linearity is present, such as in Figure~\ref{fig:5.3}d, the LWP estimator is slightly less efficient than those based on the GPPP and PSPP methods when the PM is invalid. While non-parametric terms in GPPP and PSPP improves the efficiency under a wrongly specified PM, it turns out that no gain is reached with respect to efficiency when the PM is correctly specified regardless of how the QR model is specified. Analogous results are depicted in Figure~\ref{fig:5.4} but for $\gamma_1=0.6$, which leads to generating influential pseudo-weights. In this circumstance, slight residual bias is evident for the PSPP, LWP, and AIPW methods where the underlying model is valid just for the QR approach. Furthermore, the LWP method loses more efficiency when outlying pseudo-weights are present, especially for higher degrees of non-linearity in the outcome variable.\par 

\begin{figure}[hbt!]
\centering\includegraphics[scale=0.22]{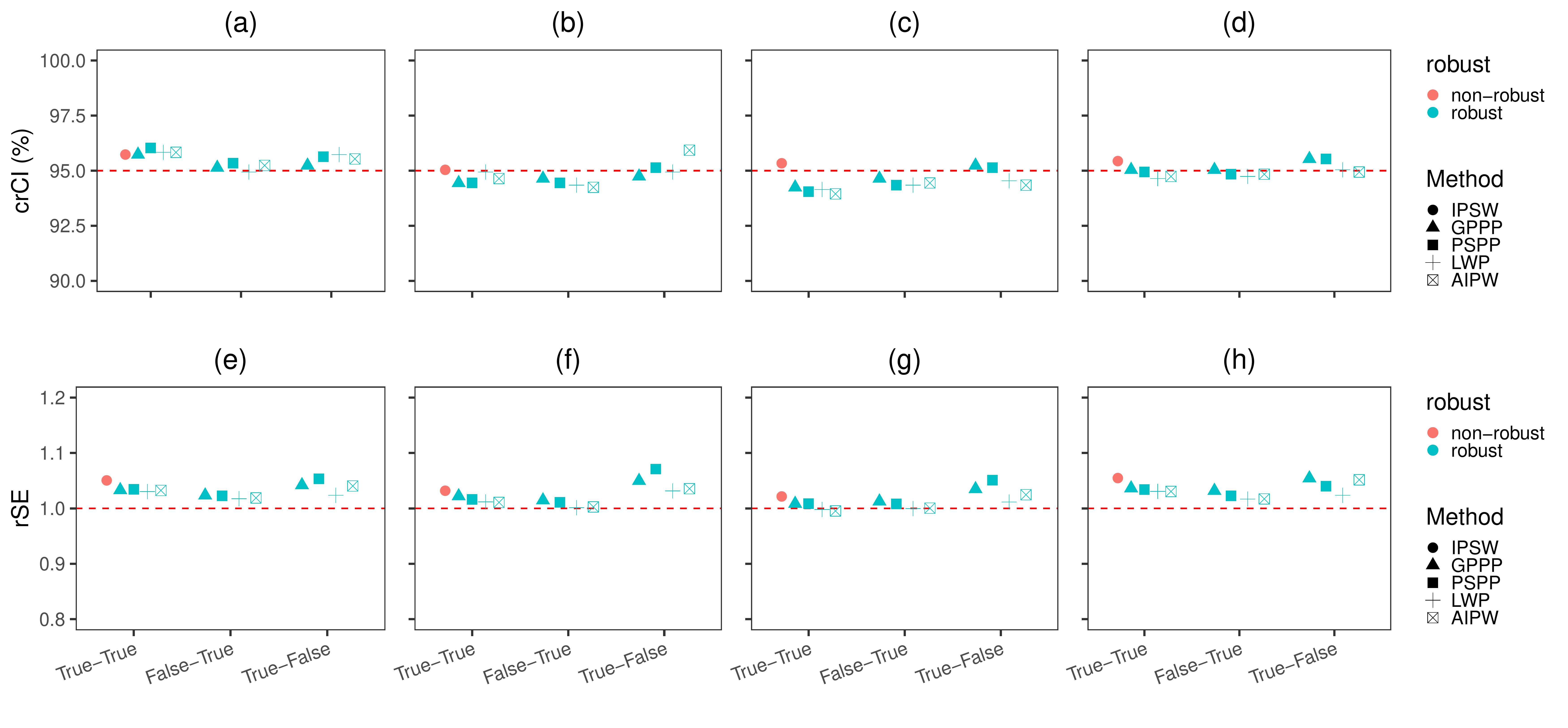}
\caption{Comparing the 95\% CI coverage rates (crCI) of the adjusted means for the \emph{continuous} outcome variable under (a) LIN, (b) CUB, (c) EXP, and (d) SIN, and SE ratios (rSE) under (e) LIN, (f) CUB, (g) EXP, and (h) SIN, across different DR methods under different model specification scenarios when $\gamma_1=0.3$. UW: unweighted; FW: Fully weighted; PAPP: Propensity Adjusted Probability Prediction; GPPP: Gaussian Processes of Propensity Prediction; LWP: Linear-in-weight Prediction; AIPW: Augmented Inverse Propensity Weighting}\label{fig:5.5}
\end{figure}

\begin{figure}[hbt!]
\centering\includegraphics[scale=0.22]{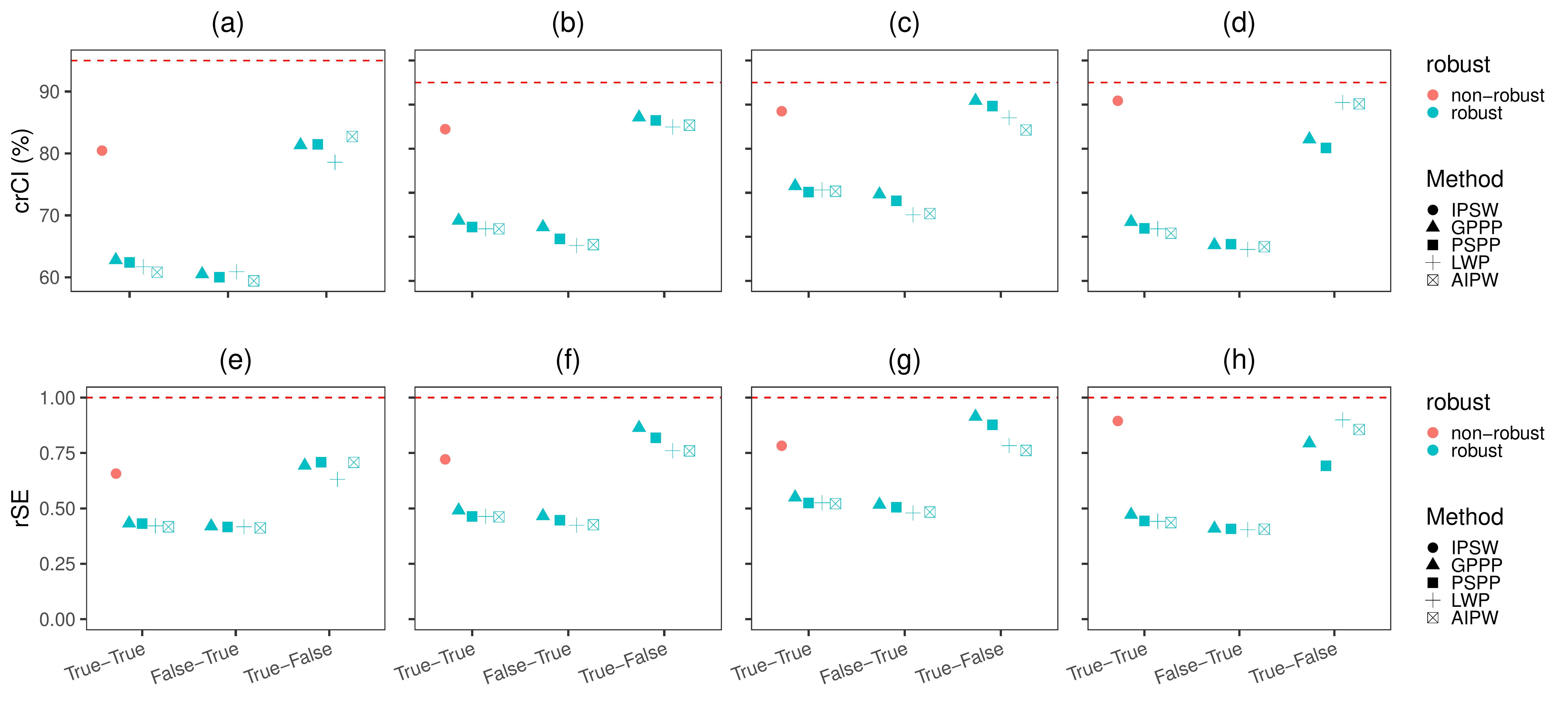}\caption{Comparing the 95\% CI coverage rates (crCI) of the DR adjusted means for the \emph{continuous} outcome variable under (a) LIN, (b) CUB, (c) EXP, and (d) SIN, and SE ratios (rSE) under (e) LIN, (f) CUB, (g) EXP, and (h) SIN, across different DR methods under different model specification scenarios when $\gamma_1=0.3$ and the complexity in the design of $S_R$ is ignored. UW: unweighted; FW: Fully weighted; PAPP: Propensity Adjusted Probability Prediction; GPPP: Gaussian Processes of Propensity Prediction; LWP: Linear-in-weight Prediction; AIPW: Augmented Inverse Propensity Weighting}\label{fig:5.6}
\end{figure}

We have summarized the simulation results for rSE and crCI with $\gamma_0=0.3$ in Figure~\ref{fig:5.5}. At first glance, it can be seen that the values of rSE are very close to one for all the competing methods and outcome variables, indicating accurate estimation of the variance based on the proposed bootstrap-Pol\'ya posterior method as long as at least one of the models for QR or PM holds. For higher degrees of non-linearity in the outcome, i.e. in the plot (h), there is evidence of slight overestimation in the variance. In addition, plots (a) to (h) reflect that the percentages of crCI are close to the nominal value, $95\%$. An exception includes plot (b) CUB, where 95\% CIs slightly undercover the true population mean when the PM is valid. In Figure~\ref{fig:5.6}, we show how ignoring the effects of strata and clusters in $S_R$ affects the values of rSE and crCI. To this end, we skip the step of bootstrapping from $S_R$ and directly generate synthetic populations using $w^R_i$'s. The new values of rSE indicate that the variance is substantially underestimated if the complex design of $S_R$ is ignored. In addition, the estimated 95\% CIs tend to significantly undercover the true population mean. Note that ignoring the stratification and clustering effects does not affect the magnitude of bias of point estimates.\par

\begin{figure}[hbt!]
\centering\includegraphics[scale=0.22]{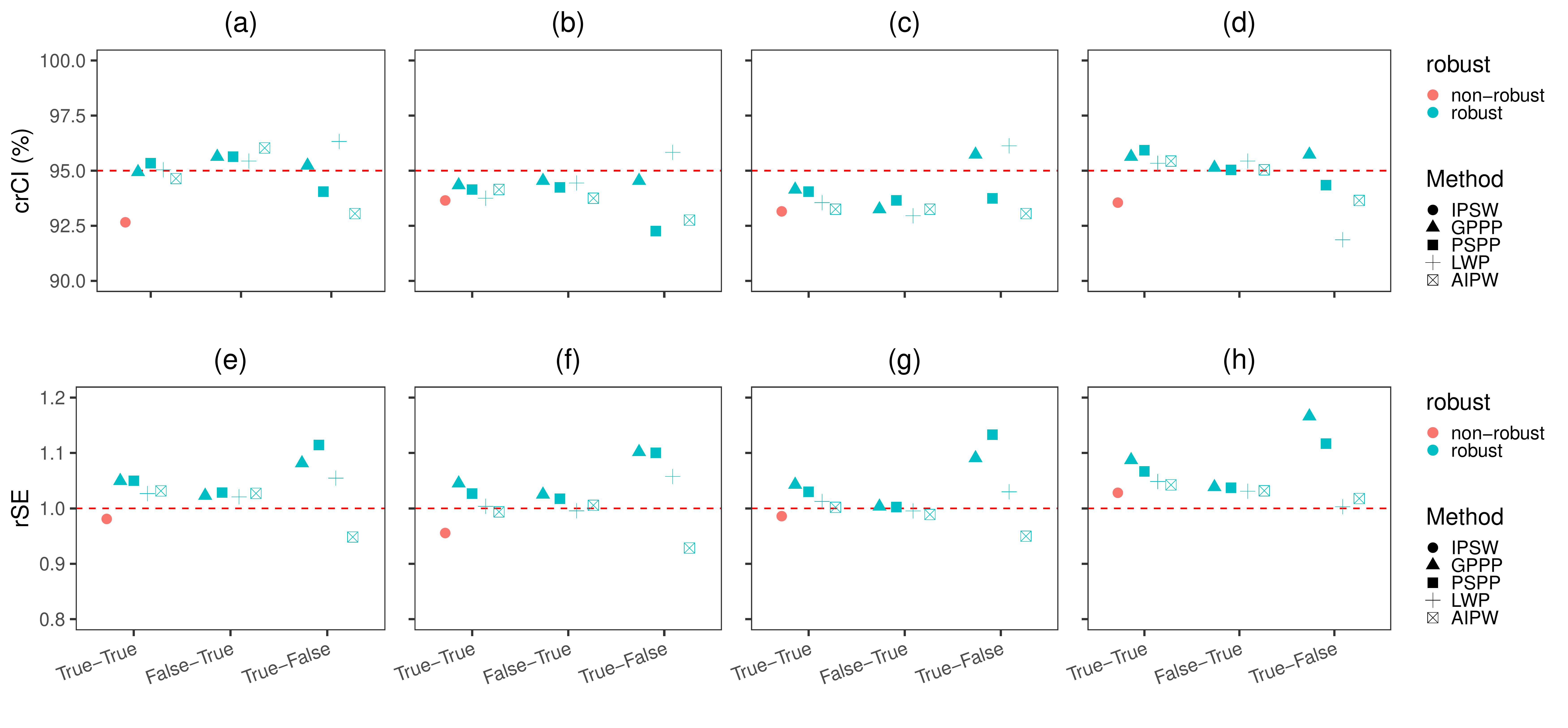}
\caption{Comparing the 95\% CI coverage rates (crCI) of the DR adjusted means for the \emph{continuous} outcome variable under (a) LIN, (b) CUB, (c) EXP, and (d) SIN, and SE ratios (rSE) under (e) LIN, (f) CUB, (g) EXP, and (h) SIN, across different DR methods under different model specification scenarios when $\gamma_1=0.6$. UW: unweighted; FW: Fully weighted; PAPP: Propensity Adjusted Probability Prediction; GPPP: Gaussian Processes of Propensity Prediction; LWP: Linear-in-weight Prediction; AIPW: Augmented Inverse Propensity Weighting}\label{fig:5.7}
\end{figure}

In Figure~\ref{fig:5.7}, we re-evaluate the simulation results for rSE and crCI when outlying pseudo-weights are present, i.e. under $\gamma_1=0.6$. According to the values of rSE, the overestimation of variance tends to be higher under this circumstance, especially for the GPPP and PSPP methods. While we observe no consistent outperformance across the competing methods, it seems the worst situation is associated with non-linear PM and where the working PM is invalid. The AIPW method exceptionally underestimates the variance for the outcomes associated with LIN, CUB, and EXP. Furthermore, it turns out that the presence of influential pseudo-weights leads to higher variability in the percentages of crCI, though deviations from the nominal 95\% seem to be negligible.\par

In Appendix~\ref{S:5.6}, we replicate the simulation results for the binary outcome variable with $\pi^A_i$'s generated based on $\gamma_1=0.3$ and $\gamma_1=0.6$, respectively. Regarding the bias magnitude and efficiency, the plots reveal almost no difference between the binary and continuous outcome variables. However, unlike the continuous outcome variable where improvements in efficiency were observed when the PM is incorrectly specified, this improvement no longer holds for the binary outcome. In addition, we realized that the estimated 95\% CIs consistently undercover the population true mean across all the competing methods. In addition to the simulation results for the binary outcome, Appendix~\ref{S:5.6} involves the numerical tables associated with all the plots in the current section as well as extensions to simulations with $(n_A, n_R)=(1,000, 500)$ and $(n_A, n_R)=(500, 500)$.\par


\section{Application}\label{S:5.4}
\noindent
Using Crash Injury Research Engineering Network (CIREN) data from 2005 to 2015, we aim at estimating crash injury rates in various body regions, including head, abdomen, thorax and spine. To correct for the selection bias in CIREN using differnet methods, we consider pooled data of Crashworthiness Data System (CDS) over the same time period as the reference survey. A brief description of these two datasets is given in the following subsections.

\subsection{Crashworthiness Data System}\label{S:5.4.1}
\noindent
The CDS has served as one of the key components of the National Automotive Sampling System (NASS) for about three decades, investigating severe traffic accidents involving passenger cars in the U.S. The CDS entails a series of stratified three-stage and multi-phase probability samples conducted annually with the primary aim to learn about the crashworthiness of vehicles and its consequences to the affected occupants. A PSU consists of geographical locations with a minimum population size of $50,000$ that are grouped into $12$ strata. Two PSUs are selected randomly from each stratum in the first stage. A secondary sampling unit (SSU) involves police jurisdictions processing reports of motor vehicle crashes within the selected PSUs. Finally, crashes are sampled randomly on a weekly basis from the police jurisdictions selected in the second stage. As the chief inclusion criteria, a police-reported crash must have caused at least one towed passenger vehicle to be eligible for selection in the final stage of sampling. (Note that the selected PSUs and SSUs are fixed across the years of study \citep{fleming2015auto}.) Thorough information about vehicle crash profiles, restraint system performance, and injury mechanisms are then collected from the selected crashes via face-to-face interviews, medical records, vehicle inspections, and scene investigations. The 2015 round of the survey involves 2,633 randomly selected police-reported accidents with 5,479 occupants investigated \citep{zhang2019crash}.\par

\subsection{Crash Injury Research and Engineering Network}\label{S:5.4.2}
\noindent
Launched by the National Highway Traffic Safety Administration (NHTSA), CIREN collects prospective clinical and biomechanical data about motor vehicle crashes with severe injury across the U.S. More specifically, CIREN investigates crash causality and consequences with the aim to tailor enhanced strategies for preventing, treating, and rehabilitating injured occupants in traffic accidents. Subjects in CIREN are selected from severely injured occupants due to motor vehicle crashes who are admitted to one of the CIREN-assigned medical centers through a set of specific inclusion criteria. Medical data are gathered by expert trauma/emergency physicians in level one trauma centers, whereas biomechanical data are gathered by trained engineers in academic engineering laboratories. While CIREN collects more detailed information than CDS, the sampling mechanism in CIREN is non-probabilistic. The trauma centers are located in urban areas and there is a tendency to recruit patients with more severe injury and with newer model-year vehicles \citep{flannagan2009comparison, elliott2010appropriate}. To adjust for the potential selection bias in CIREN, we analyze the data gathered from 2005 through 2015, which consist of 2,104 subjects.\par

\subsection{Data harmonization and auxiliary variables}\label{S:5.4.3}
\noindent
Before performing any statistical analysis, extensive effort was invested to harmonize the data of the two studies at the occupant level such that both datasets represent potentially the same target population. This involves occupants of any non-derivable motor vehicle crash between 2005 and 2015 in the U.S. who have either been fatally injured or hospitalized. As the first step, we excluded all the individuals participating in CDS who failed to meet CIREN-specific inclusion criteria. Details of these criteria have been reported in the supplemental appendix of \cite{elliott2010appropriate}. We also restricted the samples to those who were seated in the first two rows of the crashed vehicle with known crash direction and restraint status. In addition, for both datasets, individuals who survived without an urgent need for hospitalization or whose treatment status was unknown were filtered out. This finally left us a sample of size $n_R=7,721$ for CDS and a sample of size $n_A=1,738$ for CIREN.\par

\begin{figure}[hbt!]
\centering\includegraphics[scale=0.17]{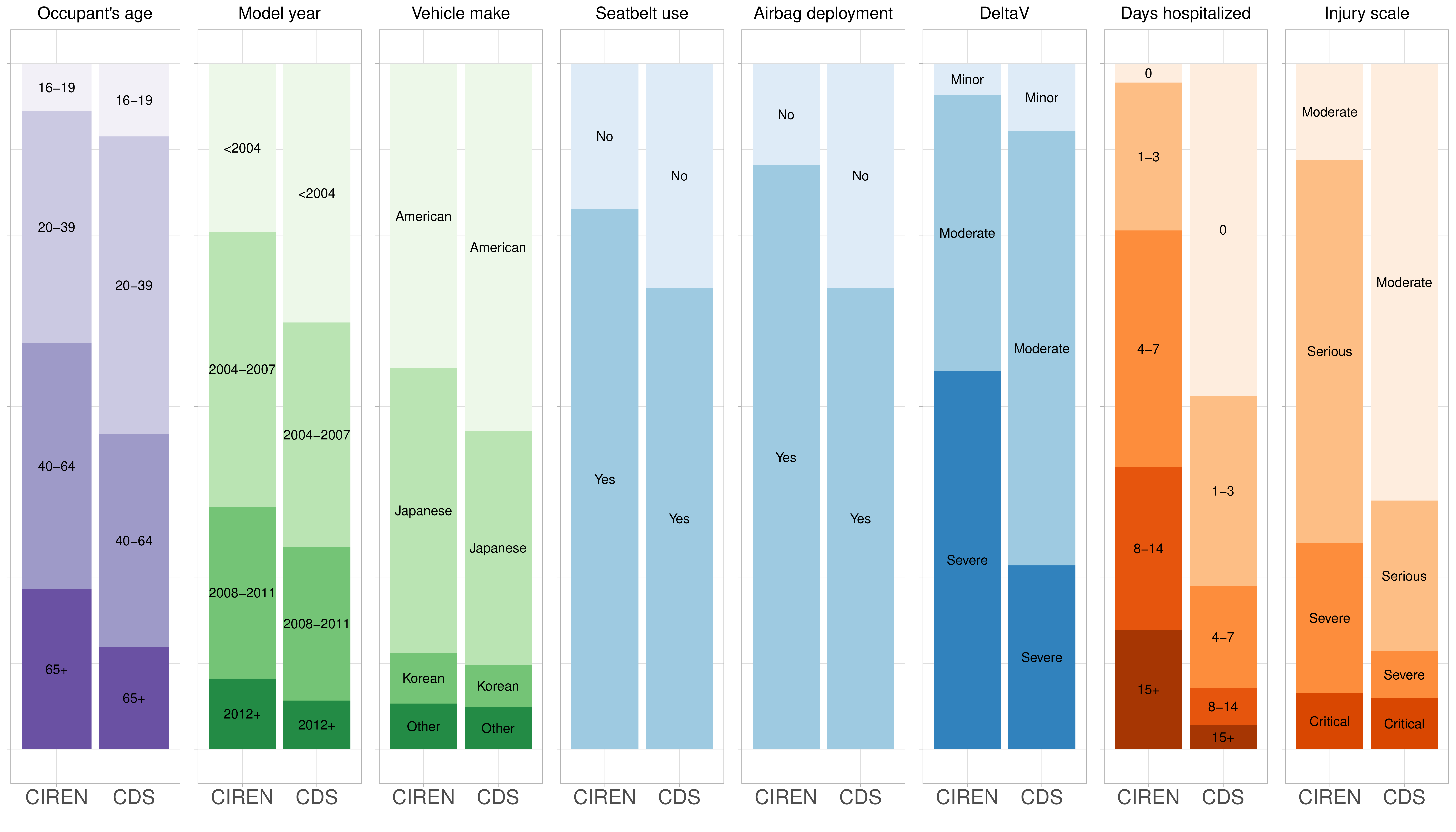}
\caption{Comparing the distributions of common auxiliary variables in CIREN with weighted CDS}\label{fig:5.8}
\end{figure}

We identified and harmonized $20$ common auxiliary variables between the two datasets, mainly describing occupants' demographic, position, and behavior at the incident, vehicle characteristics, and the intensity of crash and resulting injuries. Two variables associated with race/ethnicity were set aside due to a high rate of item-level missingness in CDS. Among the others, we chose the variables whose distribution was significantly imbalanced in CIREN compared to weighted CDS. To this end, we performed stepwise variable selection based on Bayesian Information Criterion (BIC) in a weighted logistic regression whose outcome was the indicator of being in CIREN given the combined sample. Note that the sampling weights were set to $1$ for the units of CIREN \citep{wang2020efficient}. We specifically found eight variables that are significant in the model. Figure~\ref{fig:5.8} depicts these variables, comparing their sample distribution in CIREN with the target population using weighted CDS data. \par

The bias adjustment procedure in the next step of analysis is restricted to this set of auxiliary variables for modeling the response indicator in CIREN. To be able to evaluate the performance of the adjustment methods on the actual data, in addition to the common auxiliary variables, we recognize a set of injury-level variables that were available in both datasets as the outcome variables. This entailed the indicators of being diagnosed with a $3+$ level injury (according to AIS codes) to the following body regions: head, thorax, abdomen, and spine. One would expect that unweighted estimates of the prevalence of such injuries using CIREN-only data are overestimated, as CIREN tends to recruit severely injured occupants. To construct the PM, we conduct the BIC-based stepwise procedure of variable selection for each outcome variable separately. Once the bias-adjusted estimates of the prevalence are obtained for these outcome variables, we compare them with the corresponding weighted estimates using CDS as the benchmark.\par

\subsection{Results}\label{S:5.4.4}
\begin{figure}[hbt!]
\centering\includegraphics[scale=0.35]{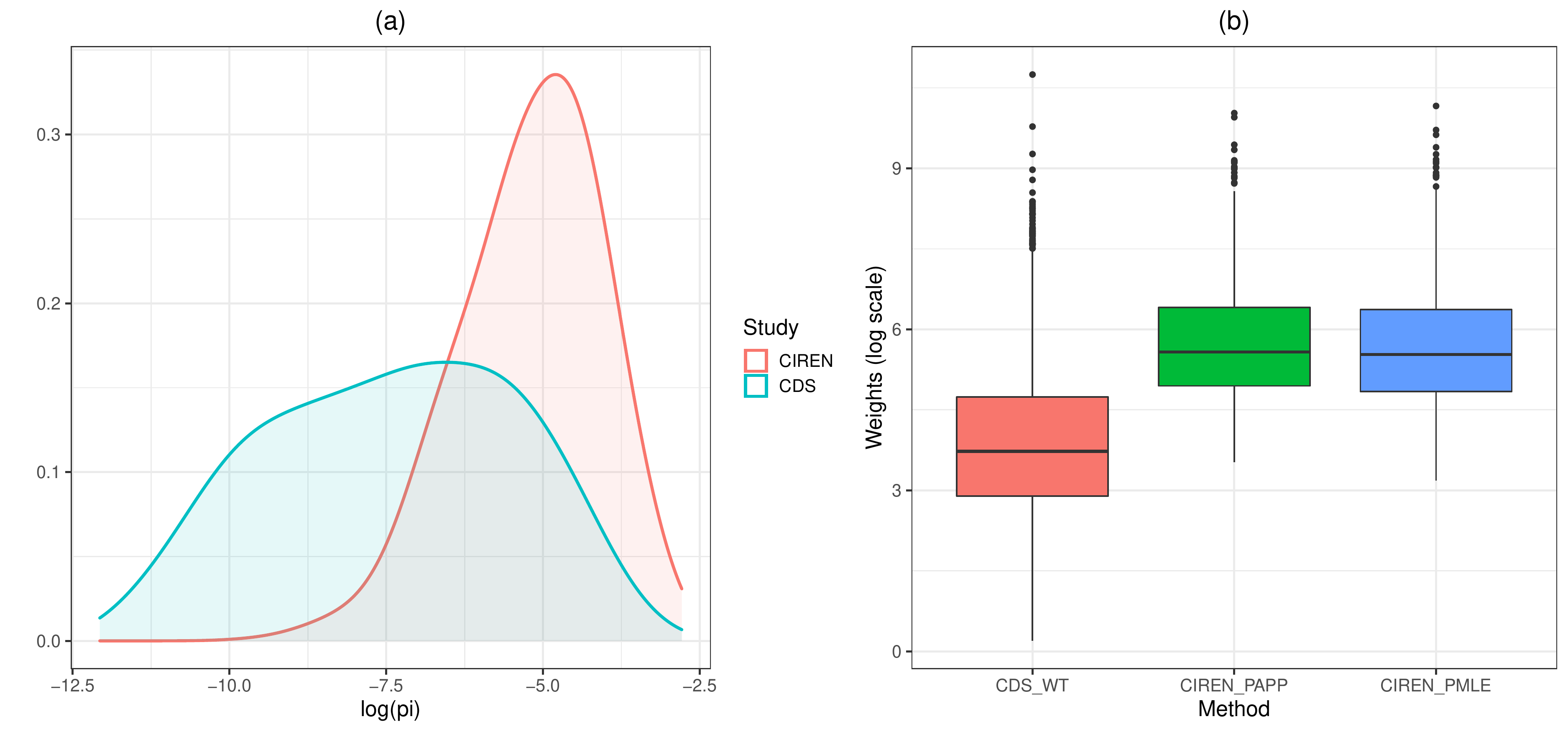}
\caption{Comparing the empirical density of (a) estimated propensity scores between CIREN and CDS and (b) estimated pseudo-weights in CIREN across the applied quasi-randomization methods}\label{fig:5.9}
\end{figure}

\noindent
As the first step of the analysis, we visually compared the empirical density of the estimated propensity scores (PS) in CIREN data with that of the target population to see if there is any lack of common support in the distribution of estimated PS in CIREN. The estimates of PS were obtained through a weighted logistic regression fitted on the combined sample with the weights equal to $1$ for the CIREN units. This comparison has been displayed in Figure~\ref{fig:5.9}a. As illustrated there is evidence of a lack of common support in the left tail of the distribution. This may suggest that further filtering should be done on the CDS data such that the undercoverage of the common representing population by CIREN is minimized. We then estimated the pseudo-weights associated with the CIREN units based on both the proposed PAPP method and the PMLE method proposed by \cite{wang2020adjusted}.\par

\begin{figure}[hbt!]
\centering\includegraphics[scale=0.17]{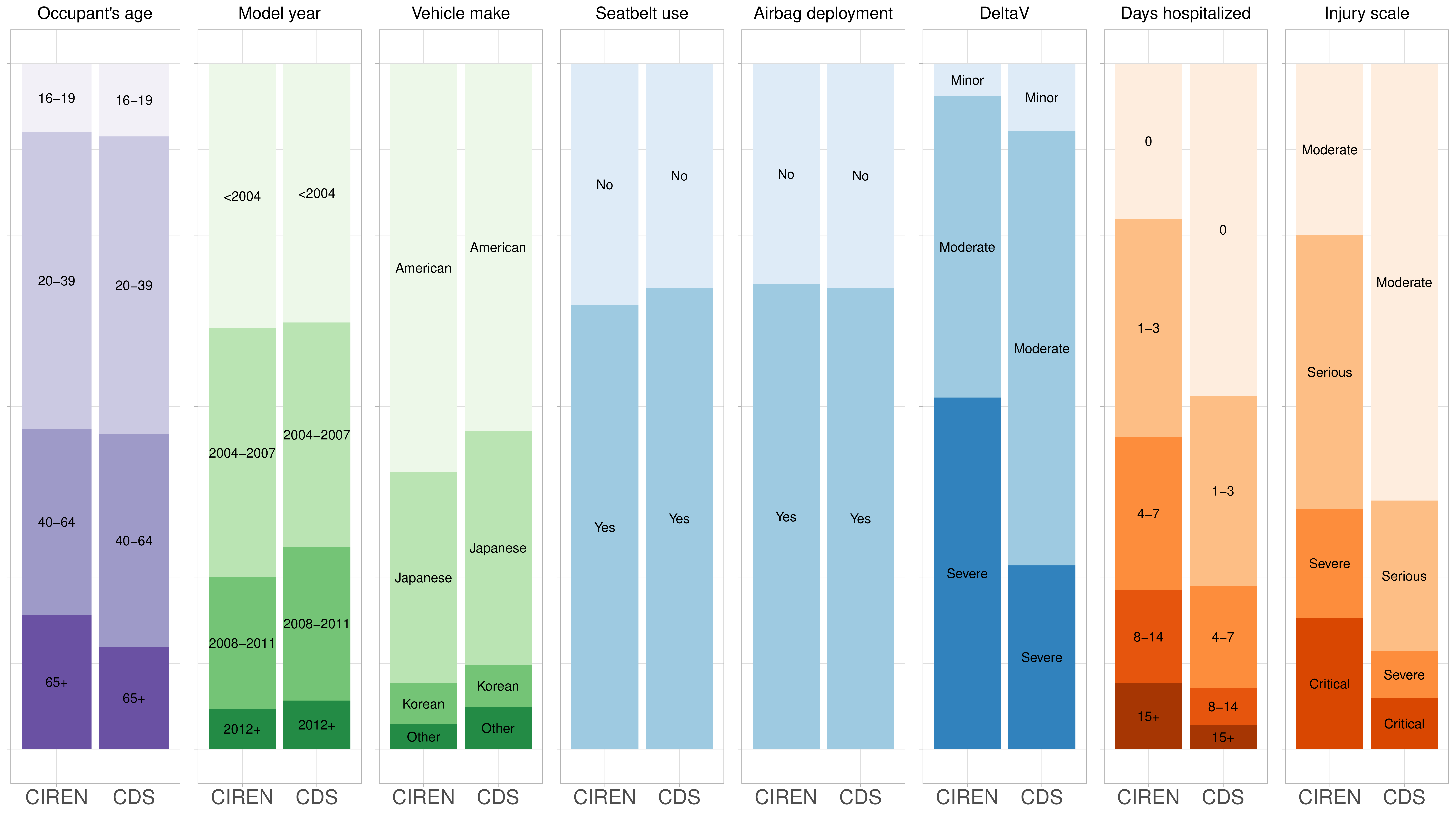}
\caption{Comparing the distribution of common auxiliary variables in pseudo-weighted CIREN based on the PAPP method with weighted CDS}\label{fig:5.10}
\end{figure}

Figure~\ref{fig:5.9}b compares the distribution of the estimated pseudo-weights between the two methods through a boxplot. As illustrated, there is evidence of influential pseudo-weights. However, we believe that the proposed GPPP method is not widely affected by this issue. This mainly originates from the specific structure of the GP covariance, which is a function of $||\hat\pi^A_i-\hat\pi^A_j||$, not $\hat\pi^A_i\hat\pi^A_j$ as in the LWP method. In addition, we showed in \cite{rafei2022robust} how GP behaves like a matching technique, and it is well-understood that PS-based matching reduces the adverse effects of influential pseudo-weights. We observed a Pearson correlation of $\rho=0.69$ in the estimates of pseudo-weights between the two PAPP and PMLE methods. Now, we compare the pseudo-weighted distribution of auxiliary variables in CIREN with their weighted distribution in CDS based on the PAPP method to see how well it mitigates the previously observed discrepancies in Figure~\ref{fig:5.8}. As illustrated in Figure~\ref{fig:5.10}, many of these discrepancies are removed or improved after assigning the pseudo-weights to the CIREN units. If the association of these variables is strong enough with the outcome variables of interest, one would expect that pseudo-weighted estimates would improve the potential bias in those variables to a significant extent as well.\par

For the PM, as was concluded from the joint likelihood factorization in Eq.~\ref{eq:5.1}, one has to use both $X$ and $Z$ as the predictors in the model. However, one major obstacle is that $Z$, the design features of CDS including strata indicators and sampling weights are not known or calculable for units of CIREN. Even if the CDS strata was identifiable for units of CIREN, it is most likely that for some strata, there is no observation in CIREN. This may imply a lack of coverage by the CIREN sample, and if the strata indicators are associated with the outcome variable of interest, final estimates may be biased. However, there is still a hope that this association is at least partially explained by the observed set of $X$. Since the chosen outcome variables in this study are available for both samples, one can check the role of $Z$ in addition to $X$ in the goodness-of-fit of the PM by fitting it on the CDS data.\par 

\begin{figure}[hbt!]
\centering\includegraphics[scale=0.5]{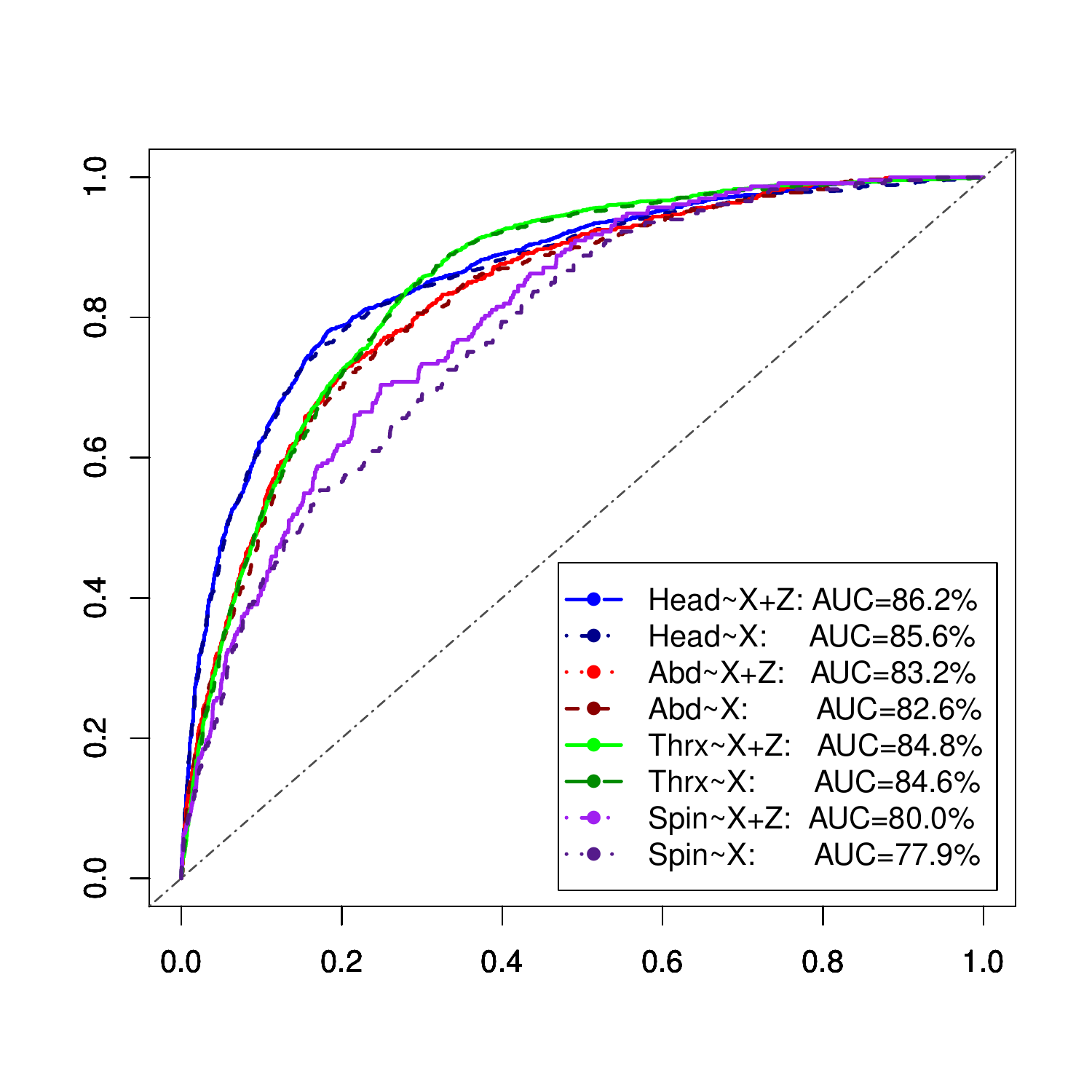}
\caption{Assessing the predictive power of the design features of CDS (Z) in the prediction models using the area under ROC}\label{fig:5.110}
\end{figure}

Table~\ref{tab:5.10} compares the model's goodness-of-fit before and after including the design features $Z$ across different outcome variables. Note that the models under $X+Z$ involve two GP terms, one linking $\pi_i^R$ and the other linking $\hat\pi^A_i$ to the response surface non-linearly. As the values of adjusted-$R^2$ and explained deviance reflect, the improvement in the PM's goodness-of-fit after including the $Z$ variables, i.e. $GP(\pi^R_i)$ and $d_i$, is small for all the outcome variables. Figure~\ref{fig:5.110} visualizes the predictive power of $Z$ in the PM associated with different outcome variables using the Area Under the Curve (AUC) of receiver operating characteristic (ROC) curve. The values of AUC suggest that $Z$ does not improve the predictive power of $X$ in the PM substantially.\par

\begin{table}[hbt!]
\centering
\caption{Assessing goodness-of-fit of the prediction model in CDS when the design features (Z) are excluded}\label{tab:5.10}
\begin{threeparttable}
\begin{tabular}{l l l l l l }
\toprule
 & \multicolumn{2}{c}{Adjusted-$R^2$}  &  & \multicolumn{2}{c}{Deviance (\%)}\\\cline{2-3}\cline{5-6}
Outcome & X & X+Z &  & X & X+Z\\
\midrule
\textbf{CDS} &  &  &  &  & \\
\hline
\hspace{2mm}Head & 0.355 & 0.360 &  & 31.00 & 31.90\\
\hspace{2mm}Abdomen & 0.164 & 0.172 &  & 21.10 & 22.10\\
\hspace{2mm}Thorax & 0.334 & 0.336 &  & 29.20 & 29.70\\
\hspace{2mm}Spine & 0.050 & 0.056 &  & 12.70 & 14.70\\
 \midrule
\textbf{CIREN} &  &  &  &  & \\
 \midrule
\hspace{2mm}Head & 0.187 & --- &  & 20.40 & ---\\
\hspace{2mm}Abdomen & 0.180 & --- &  & 21.10 & ---\\
\hspace{2mm}Thorax & 0.261 & --- &  & 24.30 & ---\\
\hspace{2mm}Spine & 0.137 & --- &  & 17.10 & ---\\
\bottomrule
\end{tabular}
  \begin{tablenotes}
   \footnotesize
   \item 
  \end{tablenotes}
 \end{threeparttable}
\end{table}

\begin{figure}[hbt!]
\centering\includegraphics[scale=0.27]{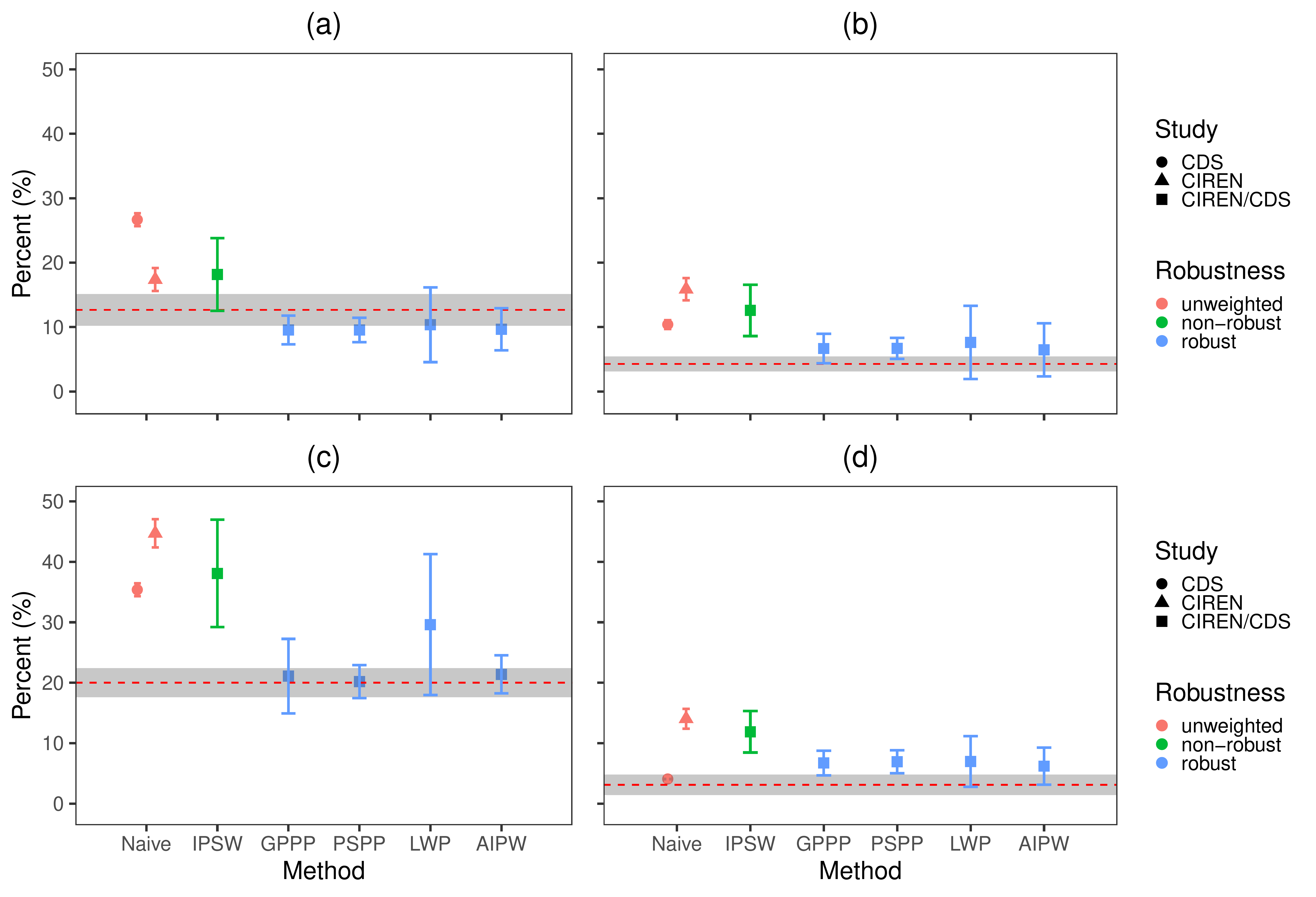}
\caption{Comparing the performance of adjustment methods for estimating the percentage of occupants with severe injury in (a) head, (b) abdomen, (c) thorax, and (d) spine based on CIREN/CDS. The dashed line and surrounding shadowed area represent weighted estimates and 95\% CIs in CDS, respectively. UW: unweighted; FW: Fully weighted; IPSW: Inverse Propensity Score Weighting; GPPP: Gaussian Processes of Propensity Prediction; PSPP: Penalized Spline of Propensity Prediction; LWP: Linear-in-weight Prediction; AIPW: Augmented Inverse Propensity Weighting.}\label{fig:5.11}
\end{figure}

As the last step, we apply the adjustment methods on the outcome variables of interest and compare the bias-adjusted estimates based on the GPPP approach with those based on the AIPW method and the weighted CDS. The results are shown in Figure~\ref{fig:5.11}. The red dashed line and the surrounding shaded area illustrate the true population estimates using the CDS weighted data and associated $95\%$CI. As expected, naive percentages based on the CDS-only and CIREN-only data are overestimated substantially across the four outcome variable. An exception includes the outcome variable associated with the spinal injury where the CDS-only unweighted estimate is unbiased. At first glance, one can infer that all DR bias adjustment methods shift the prevalence towards the true population estimates, except for the PAPP method. It turns out that estimates based on the IPSW method are biased, and the magnitude of bias is even larger than that in naive estimates for injuries in the head and thorax. The minimum amount of residual bias seems to be associated with the GPPP and PSPP methods.\par

Regarding the efficiency, the widest 95\% CIs are associated with the IPSW and LWP methods, and the narrowest 95\% CI is associated with the PSPP method. This is contrary to the simulation results where we observed better efficiency in GPPP estimator than the PSPP estimator. A notable discrepancy between the application and simulation studies is that, unlike the simulation, all the auxiliary variables used for predicting both PS and outcome variable in the application were categorical. However, in terms of bias, we found no consistent differences between the GPPP and PSPP methods. Further numerical comparisons across different outcome variables and different DR methods by levels of auxiliary variables are provided in tables~\ref{tab:5.13}-\ref{tab:5.16} in Appendix~\ref{S:5.6.2}.\par

\section{Discussion}\label{S:5.5}
\noindent
This article developed a fully model-based approach for finite population inference using a non-probability sample. Our focus was on a situation where the available reference sample is complex in design. To address this, we particularly used a bootstrap-Pol\'ya posterior technique proposed by~\cite{zhou2016two} to generate synthetic populations based on the reference survey. Generally, the simulation results demonstrate that the proposed GPPP estimator is DR, producing the least degree of residual bias with improved efficiency among the competing methods. The superiority of the GPPP was more evident when influential pseudo-weights are present. On the contrary, the empirical study showed that adjustments based on PSPP lead to more efficient estimates than those based on the GPPP method. This specific contradiction may have arisen from the fact that all the auxiliary variables used in the application were categorical.\par

Unlike the proposed estimator in prior literature, the one introduced by this study involved no design-based term, because the outcome variable was predicted for all non-sampled units. This along with the use of a flexible function of estimated propensity based on GP not only yielded double robustness in the consistency of the ultimate estimator but reduced the sensitivity to the presence of influential pseudo-weights. More importantly, the new proposed method eliminated the need for modeling the sampling weights of the reference survey as the outcome in the process of estimating the PS, which can be challenging to the analyst. In fact, one can get rid of the complexity in the design of the reference survey, especially the unequal sampling weights, by generating synthetic populations as the first step. This is advantageous for Bayesian inference and for the use of alternative algorithmic predictive tools such as BART. \cite{mercer2018selection, liu2021inference} uses BART for prediction modeling through a similar model-based idea with the estimated PS used as a predictor in the model. In his comparison with the PSPP method, the author shows by simulation that estimates based on the PSPP method are more efficient than those based on BART.\par

One can easily expand the proposed method to a fully Bayesian setting where the outcome and PS are jointly estimated. However, unlike the proposed method in \cite{rafei2022robust}, developing a unified Bayesian framework may not be possible under a fully model-based approach given the available Bayesian platforms. This is mainly because one has to generate the synthetic population as the beginning step, while software like winBUGS and Stan does not allow for using the simulated posterior predictive distribution as the input for the following steps. Therefore, one has to use a two-step approach where initially synthetic populations are created repeatedly, and then for any given synthetic population, one can fit the models for QR and PM jointly under a Bayesian setting. Therefore, simulating the posterior predictive distribution of the population's unknown quantity will be prohibitive as one would have to rely on Rubin's combining rules for the variance estimation. In addition, such a method can turn out to be very intensive computationally, as for any given synthetic population, one has to fit the QR model on the entire population and predict the outcome for all units of the population repeatedly.\par

Another major challenge with the PM, which was argued in the results section of the application, is the fact that design features of the reference sample are often unobserved for units of the non-probability sample. In the application part, because we had picked outcome variables that were available in both samples, we could check the improvements in the PM's goodness-of-fit using the reference survey where both $X$ and $Z$ are available. The good news is that our proposed method is still DR. Even if the PM lacks the reference survey's design features as predictors, one can still rely on the QR model for producing unbiased estimates. As a DR method, if the QR method is correctly specified, one would expect that the final estimates are unbiased.\par

\section{Acknowledgement}
The present study was part of the doctoral research of the first author of this article at the Michigan Program in Survey and Data Science. Therefore, we would like to thank the respected members of the dissertation committee, Professors Brady T. West, Roderick J. Little, and Philip S. Boonstra at the University of Michigan, who have continuously supported this research with their excellent comments and critical feedback. Our gratitude also goes to Professors Katharine Abraham, Stanley Presser, and Joseph Sedarski at the University of Maryland who have significantly contributed to the development of the main idea of this paper with their valuable comments and feedback over the doctoral seminar course. Last but not least, the authors would like to thank all the researchers and staff who have been involved in collecting the data of CIREN and CDS from 2005 to 2015.

\section{Conflict of Interest} The authors declare that there was no conflict of interest in the current research.


\bibliographystyle{chicago}
\bibliography{arXiv-Rafei-BA-paper}


\setcounter{page}{1}

\newpage
\clearpage

\section{Appendix}\label{S:5.6}

\subsection{Extension of the simulation study}\label{S:5.6.1}

\begin{figure}[hbt!]
\centering\includegraphics[scale=0.28]{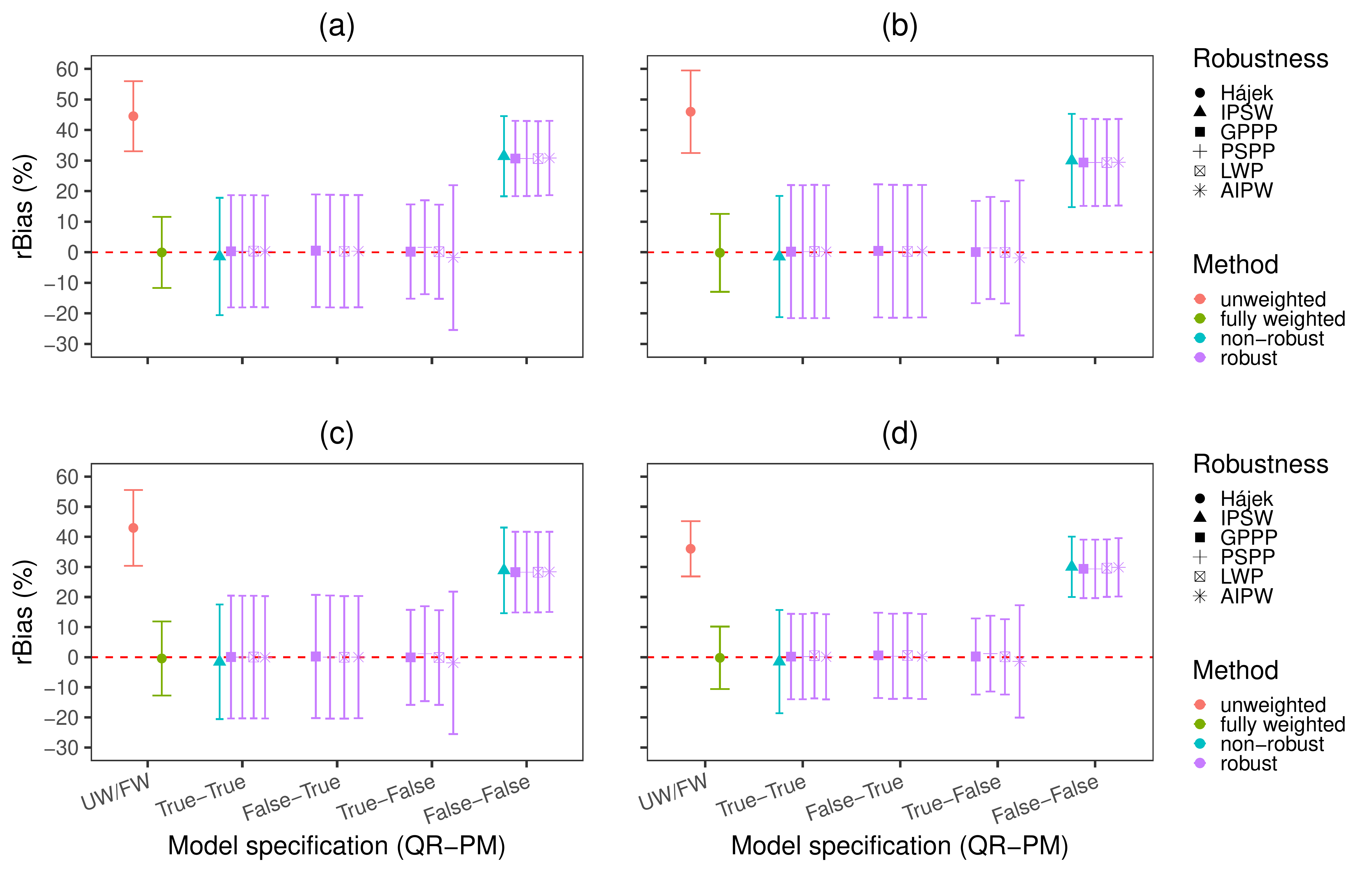}
\caption{Comparing the performance of the DR estimators under different model-specification scenarios for the \emph{binary} outcome variable under (a) LIN, (b) CUB, (c) EXP, and (d) SIN when $\gamma_1=0.3$. The error bars point out the relative length of 95\% CIs. UW: unweighted; FW: Fully weighted; GPPP: Gaussian Processes of Propensity Prediction; PSPP: Penalized Spline of Propensity Prediction; LWP: Linear-in-weight Prediction; AIPW: Augmented Inverse Propensity Weighting}\label{fig:5.12}
\end{figure}

\begin{figure}[hbt!]
\centering\includegraphics[scale=0.28]{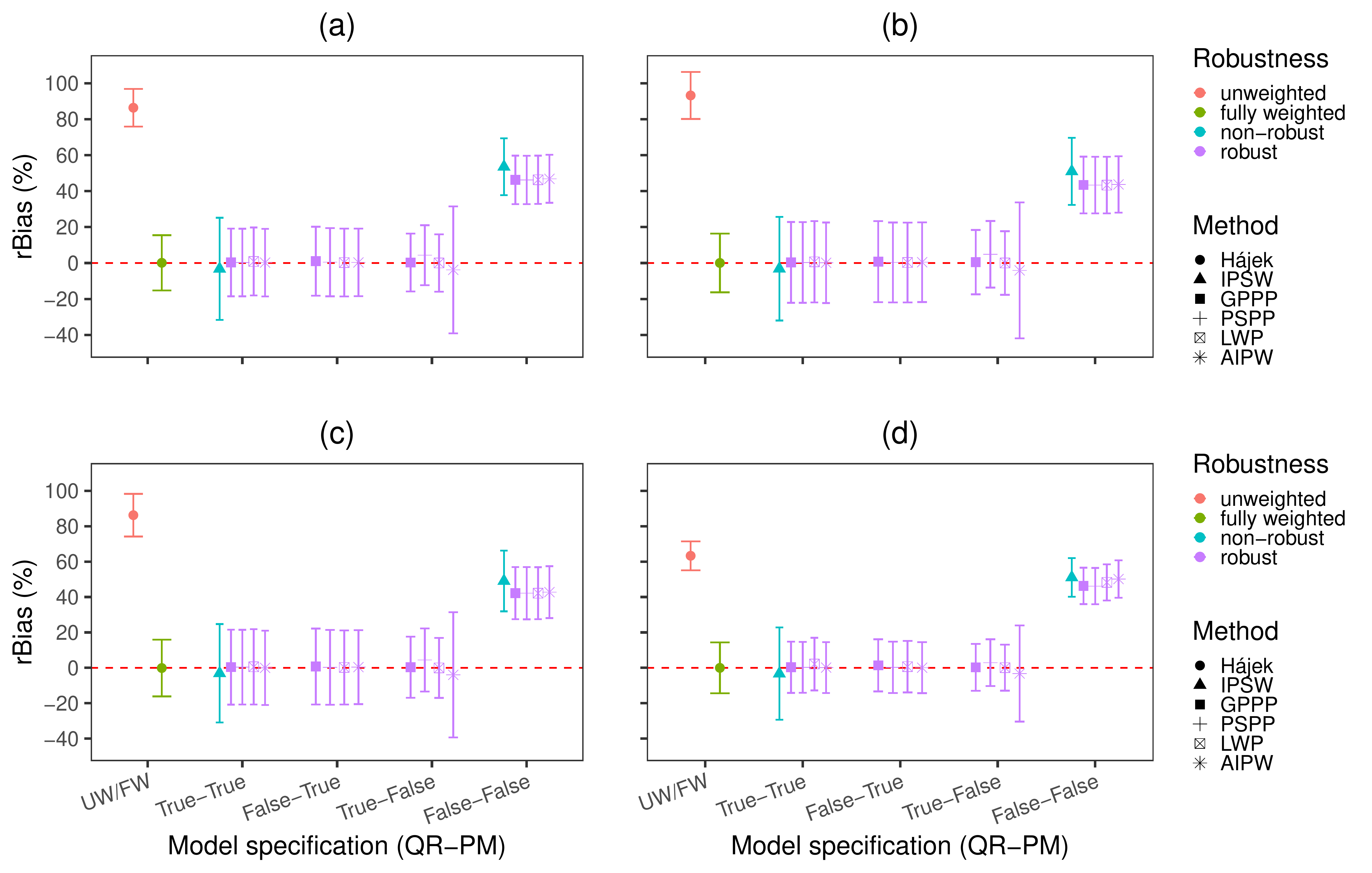}
\caption{Comparing the performance of the DR estimators under different model-specification scenarios for the \emph{binary} outcome variable under (a) LIN, (b) CUB, (c) EXP, and (d) SIN when $\gamma_1=0.6$. The error bars point out the relative length of 95\% CIs. UW: unweighted; FW: Fully weighted; GPPP: Gaussian Processes of Propensity Prediction; PSPP: Penalized Spline of Propensity Prediction; LWP: Linear-in-weight Prediction; AIPW: Augmented Inverse Propensity Weighting}\label{fig:5.13}
\end{figure}

\begin{figure}[hbt!]
\centering\includegraphics[scale=0.22]{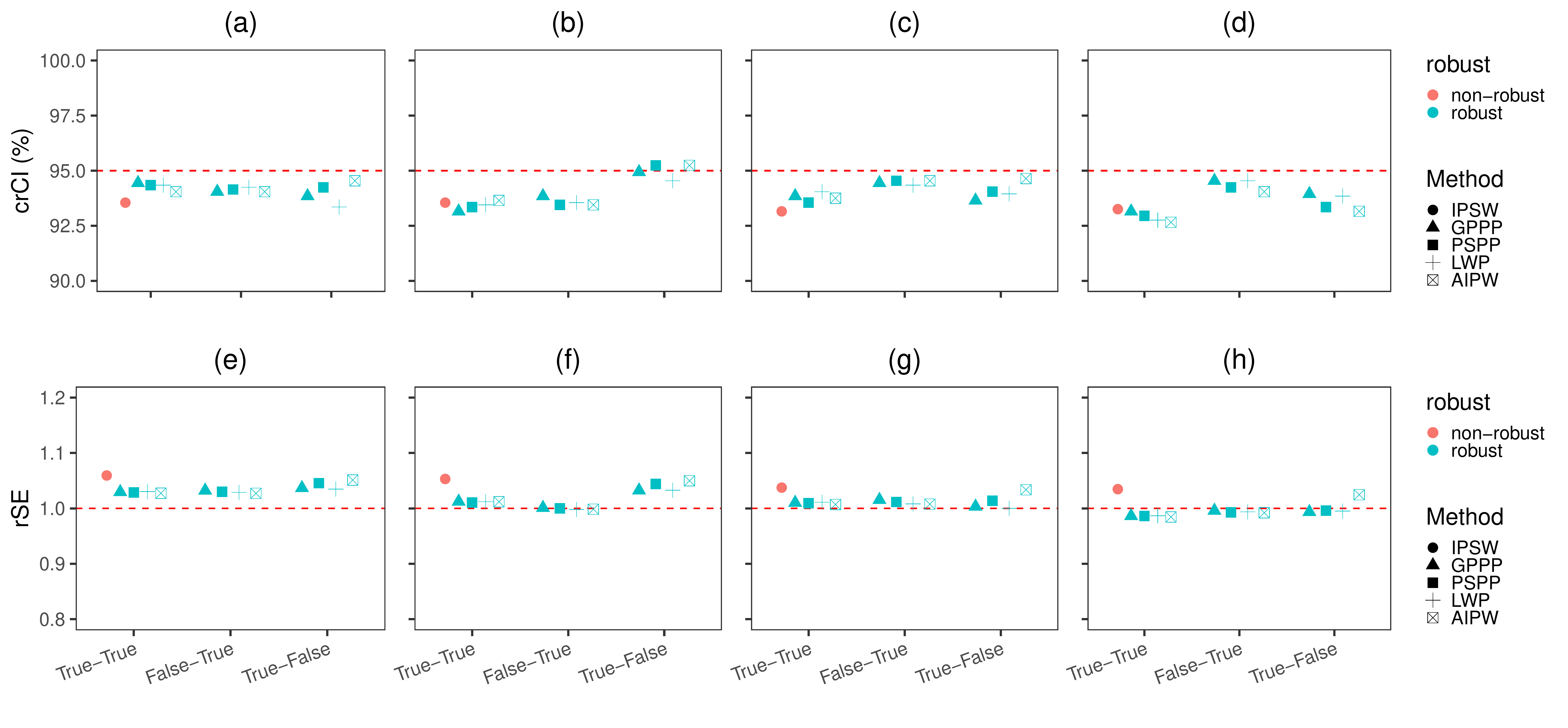}
\caption{Comparing the 95\% CI coverage rates (crCI) of the DR adjusted means for the \emph{continuous} outcome variable under (a) LIN, (b) CUB, (c) EXP, and (d) SIN, and SE ratios (rSE) under (e) LIN, (f) CUB, (g) EXP, and (h) SIN, across different DR methods under different model specification scenarios when $\gamma_1=0.3$. UW: unweighted; FW: Fully weighted; PAPP: Propensity Adjusted Probability Prediction; GPPP: Gaussian Processes of Propensity Prediction; LWP: Linear-in-weight Prediction; AIPW: Augmented Inverse Propensity Weighting}\label{fig:5.14}
\end{figure}

\begin{figure}[hbt!]
\centering\includegraphics[scale=0.22]{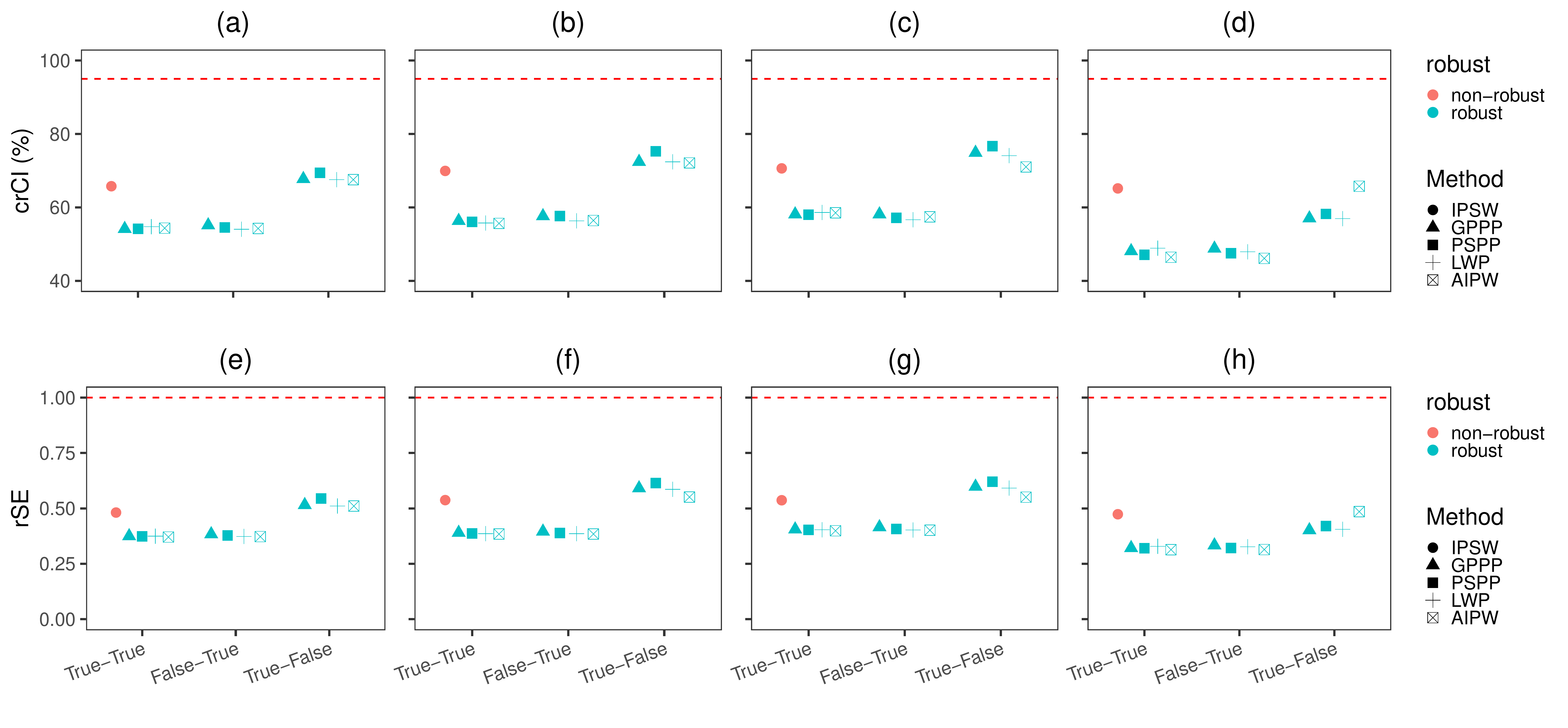}
\caption{Comparing the 95\% CI coverage rates (crCI) of the DR adjusted means for the \emph{continuous} outcome variable under (a) LIN, (b) CUB, (c) EXP, and (d) SIN, and SE ratios (rSE) under (e) LIN, (f) CUB, (g) EXP, and (h) SIN, across different DR methods under different model specification scenarios when $gamma_1=0.3$ and the complexity in the design of $S_R$ is ignored. UW: unweighted; FW: Fully weighted; PAPP: Propensity Adjusted Probability Prediction; GPPP: Gaussian Processes of Propensity Prediction; LWP: Linear-in-weight Prediction; AIPW: Augmented Inverse Propensity Weighting}\label{fig:5.15}
\end{figure}

\begin{figure}[hbt!]
\centering\includegraphics[scale=0.22]{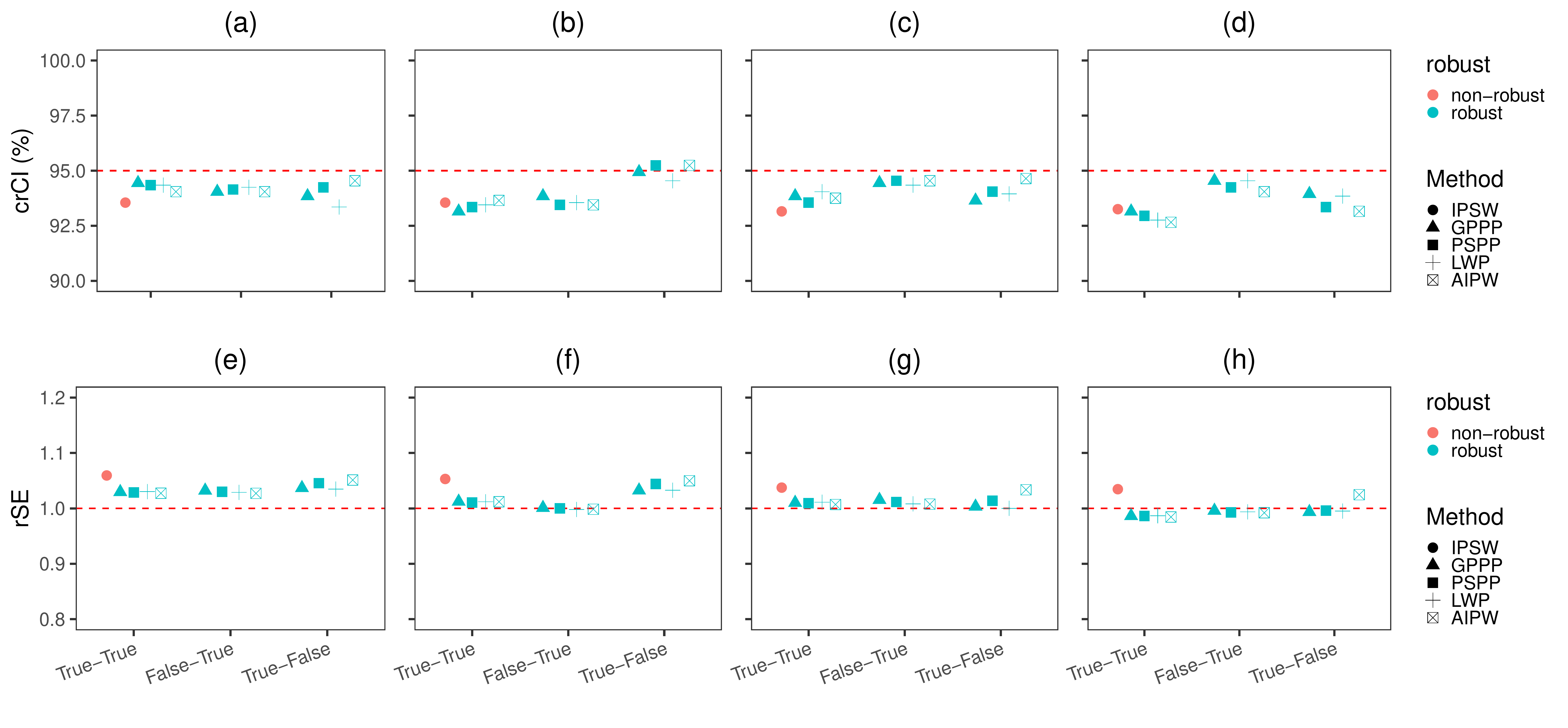}
\caption{Comparing the 95\% CI coverage rates (crCI) of the DR adjusted means for the \emph{continuous} outcome variable under (a) LIN, (b) CUB, (c) EXP, and (d) SIN, and SE ratios (rSE) under (e) LIN, (f) CUB, (g) EXP, and (h) SIN, across different DR methods under different model specification scenarios when $\gamma_1=0.6$. UW: unweighted; FW: Fully weighted; PAPP: Propensity Adjusted Probability Prediction; GPPP: Gaussian Processes of Propensity Prediction; LWP: Linear-in-weight Prediction; AIPW: Augmented Inverse Propensity Weighting}\label{fig:5.16}
\end{figure}

Figures~\ref{fig:5.12}-\ref{fig:5.16} show the simulation results for the binary outcome variable with respect to rBias, lCI, rSE and crCI. In addition, tables~\ref{tab:5.1}-\ref{tab:5.12} provides numerical results of the simulation extended to sample sizes of $(n_A, n_R)=(1,000, 500)$ and $(n_A, n_R)=(500, 500)$.

\newpage

\setlength{\tabcolsep}{3pt}
\begin{sidewaystable}[hbt!]
\centering
\caption{Comparing the performance of the bias adjustment methods in the simulation study for the \emph{continuous} outcome with $(n_A, n_R)=(500, 1,000)$ and $\gamma_1=0.3$}\label{tab:5.1}
\begin{threeparttable}
\scriptsize{\begin{tabular}{l l l l l l l l l l l l l l l l l l l l l l l l l}
\toprule
& \multicolumn{5}{c}{\textbf{$LIN$}} & & \multicolumn{5}{c}{\textbf{$CUB$}} & & \multicolumn{5}{c}{\textbf{$EXP$}} & & \multicolumn{5}{c}{\textbf{$SIN$}}\\\cline{2-6}\cline{8-12}\cline{14-18}\cline{20-24}
\textbf{Measure} & rBias & rMSE & crCI & lCI & rSE & & rBias & rMSE & crCI & lCI & rSE & & rBias & rMSE & crCI & lCI & rSE & & rBias & rMSE & crCI & lCI & rSE \\
\midrule
\multicolumn{24}{l}{\textbf{Probability sample ($S_R$)}}\\ 
\hline
\hspace{2mm} UW & -31.577 & 31.829 & 0.000 & 9.476 & 0.604 & & -25.314 & 25.534 & 0.000 & 7.895 & 0.603 & & -23.741 & 23.950 & 0.000 & 7.280 & 0.589 & & -33.737 & 34.038 & 0.000 & 11.272 & 0.636\\
\hspace{2mm} FW & 0.575 & 6.592 & 94.345 & 25.873 & 1.005 & & 0.400 & 5.581 & 93.552 & 21.858 & 1.001 & & 0.333 & 5.279 & 94.444 & 20.565 & 0.995 & & 0.545 & 6.820 & 94.147 & 26.873 & 1.008\\
\hline 
\multicolumn{24}{l}{\textbf{Non-probability sample ($S_A$)}}\\ 
\hline 
\hspace{2mm} UW & 33.632 & 33.837 & 0.000 & 14.373 & 0.985 & & 26.107 & 26.300 & 0.000 & 12.565 & 1.007 & & 23.527 & 23.712 & 0.000 & 11.612 & 1.001 & & 32.309 & 32.543 & 0.000 & 15.189 & 0.994\\
\hspace{2mm} FW & -0.002 & 4.433 & 94.444 & 16.857 & 0.970 & & 0.032 & 3.599 & 94.841 & 13.921 & 0.986 & & -0.088 & 3.197 & 95.437 & 12.691 & 1.013 & & -0.033 & 4.947 & 95.337 & 19.326 & 0.996\\
\hline 
\multicolumn{24}{l}{\textbf{Non-robust method}}\\ 
\hline 
\multicolumn{24}{l}{Model specification: QR--True}\\ 
\hline 
\hspace{2mm} IPSW & -0.807 & 7.220 & 95.734 & 29.561 & 1.051 & & -0.495 & 5.698 & 95.040 & 22.974 & 1.032 & & -0.608 & 5.222 & 95.337 & 20.781 & 1.022 & & -0.945 & 7.794 & 95.437 & 32.004 & 1.055\\
\hline
\multicolumn{24}{l}{Model specification: QR--False}\\ 
\hline 
\hspace{2mm} IPSW & 23.624 & 23.936 & 0.000 & 15.020 & 0.995 & & 17.277 & 17.587 & 0.000 & 12.880 & 1.000 & & 15.739 & 16.043 & 0.099 & 12.204 & 1.001 & & 28.363 & 28.657 & 0.000 & 16.066 & 1.000\\
\hline 
\multicolumn{24}{l}{\textbf{Doubly robust methods}}\\ 
\hline 
\multicolumn{24}{l}{Model specification: QR--True, PM--True}\\ 
\hline 
\hspace{2mm} GPPP & 0.547 & 6.244 & 95.734 & 25.204 & 1.033 & & 0.503 & 5.352 & 94.444 & 21.362 & 1.022 & & 0.302 & 5.082 & 94.246 & 20.065 & 1.008 & & 0.553 & 6.573 & 95.040 & 26.622 & 1.036\\
\hspace{2mm} PSPP & 0.559 & 6.238 & 96.032 & 25.210 & 1.035 & & 0.521 & 5.352 & 94.444 & 21.228 & 1.016 & & 0.287 & 5.078 & 94.048 & 20.057 & 1.009 & & 0.557 & 6.554 & 94.940 & 26.483 & 1.034\\
\hspace{2mm} LWP & 0.531 & 6.256 & 95.833 & 25.188 & 1.030 & & 0.498 & 5.358 & 94.940 & 21.168 & 1.012 & & 0.328 & 5.082 & 94.147 & 19.853 & 0.998 & & 0.552 & 6.556 & 94.643 & 26.412 & 1.031\\
\hspace{2mm} AIPW & 0.542 & 6.246 & 95.833 & 25.199 & 1.033 & & 0.511 & 5.360 & 94.643 & 21.153 & 1.011 & & 0.322 & 5.088 & 93.948 & 19.823 & 0.995 & & 0.550 & 6.554 & 94.742 & 26.397 & 1.031\\
\hline 
\multicolumn{24}{l}{Model specification: QR--True, PM--False}\\ 
\hline 
\hspace{2mm} GPPP & 0.580 & 6.258 & 95.139 & 25.016 & 1.024 & & 0.514 & 5.369 & 94.643 & 21.271 & 1.015 & & 0.320 & 5.110 & 94.643 & 20.254 & 1.013 & & 0.558 & 6.601 & 95.040 & 26.620 & 1.032\\
\hspace{2mm} PSPP & 0.603 & 6.245 & 95.337 & 24.934 & 1.023 & & 0.524 & 5.380 & 94.444 & 21.232 & 1.011 & & 0.319 & 5.072 & 94.345 & 20.012 & 1.008 & & 0.574 & 6.617 & 94.841 & 26.441 & 1.023\\
\hspace{2mm} LWP & 0.595 & 6.261 & 94.940 & 24.871 & 1.018 & & 0.515 & 5.374 & 94.345 & 21.007 & 1.001 & & 0.322 & 5.081 & 94.345 & 19.881 & 1.000 & & 0.557 & 6.615 & 94.742 & 26.288 & 1.017\\
\hspace{2mm} AIPW & 0.690 & 6.246 & 95.238 & 24.808 & 1.019 & & 0.519 & 5.366 & 94.246 & 21.004 & 1.003 & & 0.286 & 5.071 & 94.444 & 19.865 & 1.000 & & 0.578 & 6.608 & 94.841 & 26.260 & 1.017\\
\hline 
\multicolumn{24}{l}{Model specification: QR--False, PM--True}\\ 
\hline 
\hspace{2mm} GPPP & 0.382 & 5.287 & 95.238 & 21.551 & 1.042 & & 0.517 & 4.321 & 94.742 & 17.664 & 1.050 & & 0.242 & 4.027 & 95.238 & 16.317 & 1.035 & & 0.618 & 5.628 & 95.536 & 23.133 & 1.054\\
\hspace{2mm} PSPP & 0.877 & 5.321 & 95.635 & 21.682 & 1.053 & & 0.915 & 4.376 & 95.139 & 17.975 & 1.071 & & 0.665 & 4.082 & 95.139 & 16.602 & 1.051 & & 0.958 & 5.647 & 95.536 & 22.702 & 1.040\\
\hspace{2mm} LWP & 0.281 & 5.371 & 95.734 & 21.534 & 1.024 & & 0.384 & 4.367 & 94.940 & 17.599 & 1.032 & & 0.140 & 4.067 & 94.544 & 16.124 & 1.012 & & -0.486 & 6.869 & 95.040 & 27.506 & 1.024\\
\hspace{2mm} AIPW & -0.844 & 9.046 & 95.536 & 36.755 & 1.041 & & -0.606 & 7.323 & 95.933 & 29.641 & 1.036 & & -0.659 & 6.581 & 94.345 & 26.312 & 1.025 & & -0.869 & 8.225 & 94.940 & 33.733 & 1.052\\
\hline 
\multicolumn{24}{l}{Model specification: QR--False, PM--False}\\ 
\hline 
\hspace{2mm} GPPP & 23.523 & 23.802 & 0.893 & 15.401 & 1.079 & & 17.230 & 17.533 & 1.786 & 14.060 & 1.105 & & 15.657 & 15.936 & 2.083 & 12.941 & 1.111 & & 28.044 & 28.317 & 0.198 & 15.841 & 1.031\\
\hspace{2mm} PSPP & 23.569 & 23.851 & 1.091 & 15.829 & 1.104 & & 17.375 & 17.900 & 2.480 & 18.527 & 1.097 & & 15.670 & 15.950 & 1.389 & 12.967 & 1.112 & & 28.080 & 28.352 & 0.099 & 15.816 & 1.029\\
\hspace{2mm} LWP & 23.483 & 23.746 & 0.000 & 14.020 & 1.014 & & 17.250 & 17.523 & 0.000 & 12.217 & 1.011 & & 15.655 & 15.921 & 0.000 & 11.549 & 1.015 & & 28.059 & 28.330 & 0.000 & 15.767 & 1.029\\
\hspace{2mm} AIPW & 23.625 & 23.902 & 0.000 & 14.239 & 1.001 & & 17.290 & 17.568 & 0.000 & 12.276 & 1.006 & & 15.737 & 16.013 & 0.000 & 11.660 & 1.004 & & 28.366 & 28.651 & 0.000 & 15.857 & 1.004\\
\bottomrule
\end{tabular}}
  \begin{tablenotes}
   \footnotesize
   \item UW: Unweighted; FW: Fully weighted; GPPP: Gaussian Process of Propensity Prediction; PSPP: Penalized Spline of Propensity Prediction; LWP: Linear-in-weight prediction; AIPW: Augmented Inverse Propensity Weighting;
  \end{tablenotes}
 \end{threeparttable}
\end{sidewaystable}

\begin{sidewaystable}[hbt!]
\centering
\caption{Comparing the performance of the bias adjustment methods in the simulation study for the \emph{continuous} outcome with $(n_A, n_R)=(1,000, 500)$ and $\gamma_1=0.3$}\label{tab:5.2}
\begin{threeparttable}
\scriptsize{\begin{tabular}{l l l l l l l l l l l l l l l l l l l l l l l l l}
\toprule
& \multicolumn{5}{c}{\textbf{$LIN$}} & & \multicolumn{5}{c}{\textbf{$CUB$}} & & \multicolumn{5}{c}{\textbf{$EXP$}} & & \multicolumn{5}{c}{\textbf{$SIN$}}\\\cline{2-6}\cline{8-12}\cline{14-18}\cline{20-24}
\textbf{Measure} & rBias & rMSE & crCI & lCI & rSE & & rBias & rMSE & crCI & lCI & rSE & & rBias & rMSE & crCI & lCI & rSE & & rBias & rMSE & crCI & lCI & rSE \\
\midrule
\multicolumn{24}{l}{\textbf{Probability sample ($S_R$)}}\\ 
\hline 
\hspace{2mm} UW & -32.193 & 32.514 & 0.000 & 13.346 & 0.747 & & -25.780 & 26.036 & 0.000 & 11.118 & 0.780 & & -24.168 & 24.422 & 0.000 & 10.264 & 0.746 & & -34.387 & 34.793 & 0.000 & 15.911 & 0.766\\
\hspace{2mm} FW & -0.415 & 7.362 & 92.659 & 28.223 & 0.979 & & -0.342 & 6.150 & 92.163 & 24.037 & 0.998 & & -0.306 & 5.941 & 92.460 & 22.876 & 0.983 & & -0.296 & 7.867 & 93.452 & 30.374 & 0.985\\
\hline 
\multicolumn{24}{l}{\textbf{Non-probability sample ($S_A$)}}\\ 
\hline 
\hspace{2mm} UW & 33.130 & 33.229 & 0.000 & 10.159 & 1.009 & & 25.641 & 25.734 & 0.000 & 8.875 & 1.035 & & 23.202 & 23.296 & 0.000 & 8.197 & 1.004 & & 32.069 & 32.176 & 0.000 & 10.744 & 1.044\\
\hspace{2mm} FW & -0.143 & 3.036 & 95.040 & 11.958 & 1.005 & & -0.114 & 2.478 & 94.940 & 9.882 & 1.018 & & -0.093 & 2.289 & 94.147 & 8.969 & 1.000 & & 0.048 & 3.346 & 94.841 & 13.613 & 1.037\\
\hline 
\multicolumn{24}{l}{\textbf{Non-robust method}}\\ 
\hline 
\multicolumn{24}{l}{Model specification: QR--True}\\ 
\hline 
\hspace{2mm} IPSW & -2.670 & 9.146 & 91.766 & 33.905 & 0.988 & & -2.075 & 7.056 & 92.560 & 26.030 & 0.984 & & -1.770 & 6.239 & 92.857 & 23.366 & 0.996 & & -2.672 & 9.423 & 92.063 & 35.748 & 1.009\\
\hline 
\multicolumn{24}{l}{Model specification: QR--False}\\ 
\hline 
\hspace{2mm} IPSW & 22.769 & 23.070 & 0.000 & 14.501 & 0.995 & & 16.574 & 16.873 & 0.000 & 12.370 & 0.998 & & 15.171 & 15.451 & 0.000 & 11.407 & 0.993 & & 27.891 & 28.054 & 0.000 & 12.288 & 1.037\\
\hline 
\multicolumn{24}{l}{\textbf{Doubly robust methods}}\\ 
\hline 
\multicolumn{24}{l}{Model specification: QR--True, PM--True}\\ 
\hline 
\hspace{2mm} GPPP & -0.403 & 7.024 & 92.956 & 26.771 & 0.973 & & -0.432 & 5.810 & 92.460 & 22.190 & 0.977 & & -0.375 & 5.426 & 92.361 & 20.812 & 0.980 & & -0.213 & 7.489 & 93.353 & 29.064 & 0.990\\
\hspace{2mm} PSPP & -0.391 & 7.023 & 92.857 & 26.763 & 0.973 & & -0.429 & 5.815 & 92.262 & 22.168 & 0.975 & & -0.386 & 5.428 & 92.361 & 20.817 & 0.980 & & -0.213 & 7.485 & 93.254 & 29.036 & 0.990\\
\hspace{2mm} LWP & -0.424 & 7.036 & 92.857 & 26.752 & 0.971 & & -0.441 & 5.807 & 92.063 & 22.151 & 0.975 & & -0.365 & 5.420 & 92.163 & 20.789 & 0.980 & & -0.218 & 7.478 & 93.254 & 29.030 & 0.990\\
\hspace{2mm} AIPW & -0.434 & 7.056 & 92.659 & 26.828 & 0.971 & & -0.435 & 5.805 & 92.361 & 22.179 & 0.977 & & -0.356 & 5.400 & 92.163 & 20.781 & 0.983 & & -0.229 & 7.492 & 93.056 & 29.064 & 0.990\\
\hline 
\multicolumn{24}{l}{Model specification: QR--True, PM--False}\\ 
\hline 
\hspace{2mm} GPPP & -0.449 & 7.035 & 92.560 & 26.762 & 0.972 & & -0.389 & 5.785 & 92.361 & 22.169 & 0.979 & & -0.375 & 5.435 & 92.262 & 20.817 & 0.979 & & -0.207 & 7.577 & 93.651 & 29.008 & 0.977\\
\hspace{2mm} PSPP & -0.423 & 7.025 & 92.560 & 26.754 & 0.973 & & -0.392 & 5.789 & 92.361 & 22.146 & 0.978 & & -0.387 & 5.438 & 92.063 & 20.817 & 0.979 & & -0.198 & 7.571 & 93.750 & 28.994 & 0.977\\
\hspace{2mm} LWP & -0.450 & 7.032 & 92.659 & 26.725 & 0.971 & & -0.402 & 5.787 & 92.361 & 22.144 & 0.978 & & -0.385 & 5.437 & 92.163 & 20.797 & 0.978 & & -0.206 & 7.566 & 93.651 & 29.004 & 0.978\\
\hspace{2mm} AIPW & -0.335 & 7.001 & 92.659 & 26.656 & 0.972 & & -0.400 & 5.810 & 92.063 & 22.126 & 0.973 & & -0.428 & 5.438 & 91.964 & 20.802 & 0.979 & & -0.183 & 7.552 & 93.849 & 28.981 & 0.979\\
\hline 
\multicolumn{24}{l}{Model specification: QR--False, PM--True}\\ 
\hline 
\hspace{2mm} GPPP & -0.370 & 6.080 & 91.567 & 23.011 & 0.967 & & -0.220 & 4.731 & 92.262 & 18.194 & 0.982 & & -0.238 & 4.359 & 92.262 & 16.587 & 0.972 & & 0.005 & 6.556 & 92.956 & 25.445 & 0.990\\
\hspace{2mm} PSPP & -0.141 & 6.049 & 92.063 & 23.040 & 0.971 & & -0.052 & 4.706 & 93.056 & 18.231 & 0.988 & & -0.019 & 4.343 & 91.865 & 16.635 & 0.977 & & 0.140 & 6.536 & 92.956 & 25.376 & 0.990\\
\hspace{2mm} LWP & -0.247 & 5.957 & 91.766 & 22.571 & 0.967 & & -0.221 & 4.717 & 92.063 & 18.143 & 0.982 & & -0.186 & 4.274 & 92.262 & 16.322 & 0.975 & & -0.026 & 6.598 & 93.948 & 25.603 & 0.989\\
\hspace{2mm} AIPW & -3.276 & 11.468 & 92.659 & 42.552 & 0.987 & & -2.575 & 9.175 & 94.048 & 34.192 & 0.990 & & -2.202 & 8.033 & 92.460 & 29.757 & 0.982 & & -2.820 & 10.042 & 92.063 & 37.905 & 1.003\\
\hline 
\multicolumn{24}{l}{Model specification: QR--False, PM--False}\\ 
\hline 
\hspace{2mm} GPPP & 23.085 & 23.346 & 0.000 & 13.851 & 1.015 & & 16.852 & 17.118 & 0.099 & 12.023 & 1.020 & & 15.413 & 15.651 & 0.099 & 10.979 & 1.030 & & 27.846 & 28.017 & 0.000 & 12.644 & 1.044\\
\hspace{2mm} PSPP & 23.093 & 23.353 & 0.397 & 14.099 & 1.035 & & 16.899 & 17.211 & 0.794 & 13.589 & 1.063 & & 15.425 & 15.663 & 0.298 & 11.093 & 1.038 & & 27.841 & 28.012 & 0.000 & 12.695 & 1.048\\
\hspace{2mm} LWP & 23.068 & 23.330 & 0.000 & 13.694 & 1.002 & & 16.837 & 17.103 & 0.000 & 11.764 & 1.000 & & 15.399 & 15.637 & 0.000 & 10.846 & 1.017 & & 28.010 & 28.159 & 0.000 & 12.121 & 1.068\\
\hspace{2mm} AIPW & 23.053 & 23.323 & 0.000 & 13.735 & 0.990 & & 16.837 & 17.106 & 0.000 & 11.776 & 0.993 & & 15.393 & 15.639 & 0.000 & 10.861 & 1.001 & & 27.990 & 28.146 & 0.000 & 12.059 & 1.039\\
\bottomrule
\end{tabular}}
  \begin{tablenotes}
   \footnotesize
   \item UW: Unweighted; FW: Fully weighted; GPPP: Gaussian Process of Propensity Prediction; PSPP: Penalized Spline of Propensity Prediction; LWP: Linear-in-weight prediction; AIPW: Augmented Inverse Propensity Weighting;
  \end{tablenotes}
 \end{threeparttable}
\end{sidewaystable}

\begin{sidewaystable}[hbt!]
\centering
\caption{Comparing the performance of the bias adjustment methods in the simulation study for the \emph{continuous} outcome with $(n_A, n_R)=(500, 500)$ and $\gamma_1=0.3$}\label{tab:5.3}
\begin{threeparttable}
\scriptsize{\begin{tabular}{l l l l l l l l l l l l l l l l l l l l l l l l l}
\toprule
& \multicolumn{5}{c}{\textbf{$LIN$}} & & \multicolumn{5}{c}{\textbf{$CUB$}} & & \multicolumn{5}{c}{\textbf{$EXP$}} & & \multicolumn{5}{c}{\textbf{$SIN$}}\\\cline{2-6}\cline{8-12}\cline{14-18}\cline{20-24}
\textbf{Measure} & rBias & rMSE & crCI & lCI & rSE & & rBias & rMSE & crCI & lCI & rSE & & rBias & rMSE & crCI & lCI & rSE & & rBias & rMSE & crCI & lCI & rSE \\
\midrule
\multicolumn{24}{l}{\textbf{Probability sample ($S_R$)}}\\ 
\hline 
\hspace{2mm} UW & -32.193 & 32.514 & 0.000 & 13.346 & 0.747 & & -25.780 & 26.036 & 0.000 & 11.118 & 0.780 & & -24.168 & 24.422 & 0.000 & 10.264 & 0.746 & & -34.387 & 34.793 & 0.000 & 15.911 & 0.766\\
\hspace{2mm} FW & -0.395 & 7.335 & 92.956 & 28.305 & 0.985 & & -0.338 & 6.109 & 92.262 & 24.115 & 1.008 & & -0.331 & 5.907 & 92.659 & 22.926 & 0.991 & & -0.318 & 7.858 & 93.056 & 30.282 & 0.983\\
\hline 
\multicolumn{24}{l}{\textbf{Non-probability sample ($S_A$)}}\\ 
\hline 
\hspace{2mm} UW & 33.538 & 33.739 & 0.000 & 14.374 & 0.996 & & 25.970 & 26.160 & 0.000 & 12.566 & 1.019 & & 23.492 & 23.681 & 0.000 & 11.610 & 0.992 & & 32.315 & 32.539 & 0.000 & 15.176 & 1.016\\
\hspace{2mm} FW & -0.097 & 4.370 & 94.048 & 16.894 & 0.986 & & -0.082 & 3.616 & 94.742 & 13.925 & 0.982 & & -0.097 & 3.310 & 94.246 & 12.694 & 0.978 & & 0.055 & 4.904 & 95.238 & 19.299 & 1.003\\
\hline 
\multicolumn{24}{l}{\textbf{Non-robust method}}\\ 
\hline 
\multicolumn{24}{l}{Model specification: QR--True}\\ 
\hline 
\hspace{2mm} IPSW & -2.816 & 9.640 & 92.857 & 36.078 & 0.998 & & -2.150 & 7.463 & 93.849 & 27.889 & 0.995 & & -1.945 & 6.626 & 93.254 & 25.123 & 1.011 & & -2.840 & 9.985 & 92.560 & 38.102 & 1.015\\
\hline 
\multicolumn{24}{l}{Model specification: QR--False}\\ 
\hline 
\hspace{2mm} IPSW & 22.879 & 23.267 & 0.000 & 16.832 & 1.015 & & 16.608 & 17.015 & 0.000 & 14.371 & 0.991 & & 15.199 & 15.579 & 0.198 & 13.450 & 1.002 & & 28.020 & 28.331 & 0.000 & 16.501 & 1.006\\
\hline 
\multicolumn{24}{l}{\textbf{Doubly robust methods}}\\ 
\hline 
\multicolumn{24}{l}{Model specification: QR--True, PM--True}\\ 
\hline 
\hspace{2mm} GPPP & -0.416 & 7.151 & 92.163 & 27.480 & 0.982 & & -0.422 & 5.928 & 92.063 & 22.941 & 0.989 & & -0.433 & 5.549 & 92.560 & 21.670 & 0.999 & & -0.220 & 7.633 & 93.651 & 29.602 & 0.989\\
\hspace{2mm} PSPP & -0.397 & 7.145 & 92.361 & 27.469 & 0.982 & & -0.420 & 5.922 & 91.865 & 22.818 & 0.985 & & -0.450 & 5.557 & 92.659 & 21.623 & 0.995 & & -0.224 & 7.616 & 93.750 & 29.444 & 0.986\\
\hspace{2mm} LWP & -0.459 & 7.170 & 92.063 & 27.418 & 0.977 & & -0.429 & 5.931 & 91.964 & 22.791 & 0.982 & & -0.415 & 5.543 & 92.460 & 21.513 & 0.992 & & -0.252 & 7.604 & 93.750 & 29.361 & 0.985\\
\hspace{2mm} AIPW & -0.453 & 7.178 & 92.163 & 27.478 & 0.978 & & -0.415 & 5.914 & 92.063 & 22.809 & 0.986 & & -0.431 & 5.525 & 92.560 & 21.510 & 0.996 & & -0.253 & 7.612 & 93.651 & 29.377 & 0.985\\
\hline 
\multicolumn{24}{l}{Model specification: QR--True, PM--False}\\ 
\hline 
\hspace{2mm} GPPP & -0.476 & 7.200 & 92.163 & 27.494 & 0.976 & & -0.408 & 5.939 & 91.964 & 22.918 & 0.986 & & -0.421 & 5.540 & 92.262 & 21.423 & 0.989 & & -0.246 & 7.618 & 93.254 & 29.664 & 0.993\\
\hspace{2mm} PSPP & -0.441 & 7.199 & 91.964 & 27.484 & 0.975 & & -0.404 & 5.928 & 91.667 & 22.886 & 0.987 & & -0.458 & 5.557 & 92.262 & 21.455 & 0.988 & & -0.237 & 7.618 & 93.452 & 29.639 & 0.992\\
\hspace{2mm} LWP & -0.478 & 7.200 & 91.964 & 27.413 & 0.973 & & -0.420 & 5.931 & 91.865 & 22.781 & 0.982 & & -0.446 & 5.541 & 92.163 & 21.331 & 0.985 & & -0.251 & 7.606 & 93.651 & 29.567 & 0.992\\
\hspace{2mm} AIPW & -0.349 & 7.171 & 91.964 & 27.329 & 0.973 & & -0.410 & 5.931 & 91.667 & 22.760 & 0.981 & & -0.514 & 5.557 & 92.361 & 21.342 & 0.984 & & -0.231 & 7.609 & 93.552 & 29.506 & 0.989\\
\hline 
\multicolumn{24}{l}{Model specification: QR--False, PM--True}\\ 
\hline 
\hspace{2mm} GPPP & -0.442 & 6.204 & 92.163 & 23.930 & 0.986 & & -0.147 & 4.945 & 93.254 & 19.356 & 0.998 & & -0.304 & 4.567 & 92.857 & 18.001 & 1.007 & & -0.021 & 6.805 & 93.651 & 26.639 & 0.998\\
\hspace{2mm} PSPP & 0.045 & 6.185 & 93.056 & 24.170 & 0.996 & & 0.231 & 4.955 & 93.651 & 19.624 & 1.011 & & 0.159 & 4.593 & 93.750 & 18.225 & 1.012 & & 0.298 & 6.739 & 93.552 & 26.163 & 0.991\\
\hspace{2mm} LWP & -0.474 & 6.189 & 93.452 & 23.853 & 0.986 & & -0.219 & 4.974 & 93.254 & 19.331 & 0.992 & & -0.343 & 4.532 & 93.452 & 17.846 & 1.007 & & -0.890 & 7.763 & 94.643 & 30.292 & 1.002\\
\hspace{2mm} AIPW & -3.371 & 12.232 & 94.246 & 45.300 & 0.982 & & -2.445 & 9.765 & 94.345 & 36.292 & 0.979 & & -2.309 & 8.690 & 94.544 & 32.432 & 0.987 & & -3.081 & 10.672 & 91.865 & 40.608 & 1.013\\
\hline 
\multicolumn{24}{l}{Model specification: QR--False, PM--False}\\ 
\hline 
\hspace{2mm} GPPP & 23.208 & 23.555 & 1.091 & 16.675 & 1.056 & & 16.864 & 17.241 & 0.893 & 14.527 & 1.033 & & 15.449 & 15.799 & 1.091 & 13.411 & 1.035 & & 27.977 & 28.284 & 0.099 & 16.494 & 1.012\\
\hspace{2mm} PSPP & 23.239 & 23.592 & 0.992 & 17.006 & 1.067 & & 16.895 & 17.282 & 2.183 & 15.411 & 1.080 & & 15.470 & 15.819 & 1.687 & 13.592 & 1.050 & & 28.014 & 28.319 & 0.000 & 16.495 & 1.016\\
\hspace{2mm} LWP & 23.169 & 23.511 & 0.000 & 15.785 & 1.007 & & 16.854 & 17.222 & 0.000 & 13.679 & 0.986 & & 15.412 & 15.751 & 0.099 & 12.734 & 0.998 & & 28.074 & 28.358 & 0.000 & 16.187 & 1.030\\
\hspace{2mm} AIPW & 23.159 & 23.512 & 0.000 & 15.863 & 0.997 & & 16.862 & 17.238 & 0.000 & 13.662 & 0.974 & & 15.417 & 15.767 & 0.099 & 12.785 & 0.988 & & 28.121 & 28.420 & 0.000 & 16.220 & 1.005\\
\bottomrule
\end{tabular}}
  \begin{tablenotes}
   \footnotesize
   \item UW: Unweighted; FW: Fully weighted; GPPP: Gaussian Process of Propensity Prediction; PSPP: Penalized Spline of Propensity Prediction; LWP: Linear-in-weight prediction; AIPW: Augmented Inverse Propensity Weighting;
  \end{tablenotes}
 \end{threeparttable}
\end{sidewaystable}

\begin{sidewaystable}[hbt!]
\centering
\caption{Comparing the performance of the bias adjustment methods in the simulation study for the \emph{continuous} outcome with $(n_A, n_R)=(500, 1,000)$ and $\gamma_1=0.6$}\label{tab:5.4}
\begin{threeparttable}
\scriptsize{\begin{tabular}{l l l l l l l l l l l l l l l l l l l l l l l l l}
\toprule
& \multicolumn{5}{c}{\textbf{$LIN$}} & & \multicolumn{5}{c}{\textbf{$CUB$}} & & \multicolumn{5}{c}{\textbf{$EXP$}} & & \multicolumn{5}{c}{\textbf{$SIN$}}\\\cline{2-6}\cline{8-12}\cline{14-18}\cline{20-24}
\textbf{Measure} & rBias & rMSE & crCI & lCI & rSE & & rBias & rMSE & crCI & lCI & rSE & & rBias & rMSE & crCI & lCI & rSE & & rBias & rMSE & crCI & lCI & rSE \\
\midrule
\multicolumn{24}{l}{\textbf{Probability sample ($S_R$)}}\\ 
\hline 
\hspace{2mm} UW & -31.577 & 31.829 & 0.000 & 9.476 & 0.604 & & -25.314 & 25.534 & 0.000 & 7.895 & 0.603 & & -23.741 & 23.950 & 0.000 & 7.280 & 0.589 & & -33.737 & 34.038 & 0.000 & 11.272 & 0.636\\
\hspace{2mm} FW & 0.544 & 6.601 & 93.552 & 25.637 & 0.994 & & 0.365 & 5.619 & 93.750 & 21.558 & 0.980 & & 0.323 & 5.261 & 93.948 & 20.656 & 1.003 & & 0.606 & 6.848 & 93.948 & 26.869 & 1.004\\
\hline 
\multicolumn{24}{l}{\textbf{Non-probability sample ($S_A$)}}\\ 
\hline 
\hspace{2mm} UW & 67.162 & 67.257 & 0.000 & 14.469 & 1.030 & & 54.231 & 54.329 & 0.000 & 13.571 & 1.058 & & 48.409 & 48.503 & 0.000 & 12.142 & 1.024 & & 54.984 & 55.080 & 0.000 & 12.872 & 1.014\\
\hspace{2mm} FW & 0.214 & 7.308 & 91.667 & 26.736 & 0.933 & & 0.087 & 5.907 & 91.171 & 21.731 & 0.938 & & -0.041 & 5.175 & 93.056 & 19.335 & 0.953 & & 0.127 & 7.239 & 93.552 & 28.085 & 0.989\\
\hline 
\multicolumn{24}{l}{\textbf{Non-robust method}}\\ 
\hline 
\multicolumn{24}{l}{Model specification: QR--True}\\ 
\hline 
\hspace{2mm} IPSW & -2.198 & 12.496 & 92.659 & 47.329 & 0.981 & & -1.641 & 9.851 & 93.651 & 36.404 & 0.956 & & -1.691 & 8.734 & 93.155 & 33.136 & 0.986 & & -2.445 & 12.350 & 93.552 & 48.815 & 1.028\\
\hline 
\multicolumn{24}{l}{Model specification: QR--False}\\ 
\hline 
\hspace{2mm} IPSW & 37.613 & 37.906 & 0.000 & 18.408 & 0.997 & & 27.228 & 27.529 & 0.000 & 15.746 & 0.989 & & 25.215 & 25.501 & 0.000 & 14.666 & 0.984 & & 47.301 & 47.483 & 0.000 & 16.469 & 1.012\\
\hline 
\multicolumn{24}{l}{\textbf{Doubly robust methods}}\\ 
\hline 
\multicolumn{24}{l}{Model specification: QR--True, PM--True}\\ 
\hline 
\hspace{2mm} GPPP & 0.634 & 6.481 & 94.940 & 26.557 & 1.050 & & 0.365 & 5.701 & 94.345 & 23.324 & 1.045 & & 0.270 & 5.380 & 94.147 & 21.977 & 1.043 & & 0.555 & 6.730 & 95.635 & 28.602 & 1.087\\
\hspace{2mm} PSPP & 0.727 & 6.448 & 95.337 & 26.385 & 1.050 & & 0.442 & 5.634 & 94.147 & 22.616 & 1.027 & & 0.154 & 5.352 & 94.048 & 21.614 & 1.030 & & 0.572 & 6.690 & 95.933 & 27.883 & 1.067\\
\hspace{2mm} LWP & 0.484 & 6.623 & 95.040 & 26.601 & 1.027 & & 0.371 & 5.810 & 93.750 & 22.827 & 1.004 & & 0.318 & 5.478 & 93.552 & 21.715 & 1.013 & & 0.516 & 6.792 & 95.337 & 27.849 & 1.049\\
\hspace{2mm} AIPW & 0.588 & 6.517 & 94.643 & 26.259 & 1.032 & & 0.397 & 5.764 & 94.147 & 22.412 & 0.994 & & 0.266 & 5.436 & 93.254 & 21.336 & 1.002 & & 0.537 & 6.719 & 95.437 & 27.381 & 1.042\\
\hline 
\multicolumn{24}{l}{Model specification: QR--True, PM--False}\\ 
\hline 
\hspace{2mm} GPPP & 0.574 & 6.496 & 95.635 & 25.962 & 1.023 & & 0.408 & 5.807 & 94.544 & 23.289 & 1.025 & & 0.270 & 5.402 & 93.254 & 21.233 & 1.003 & & 0.457 & 6.715 & 95.139 & 27.287 & 1.039\\
\hspace{2mm} PSPP & 0.692 & 6.448 & 95.635 & 25.861 & 1.029 & & 0.467 & 5.599 & 94.246 & 22.261 & 1.017 & & 0.170 & 5.336 & 93.651 & 20.972 & 1.003 & & 0.531 & 6.656 & 95.040 & 26.988 & 1.037\\
\hspace{2mm} LWP & 0.726 & 6.506 & 95.437 & 25.888 & 1.021 & & 0.471 & 5.677 & 94.444 & 22.094 & 0.996 & & 0.167 & 5.384 & 92.956 & 21.010 & 0.996 & & 0.474 & 6.692 & 95.437 & 26.992 & 1.031\\
\hspace{2mm} AIPW & 0.926 & 6.432 & 96.032 & 25.640 & 1.027 & & 0.428 & 5.538 & 93.750 & 21.776 & 1.006 & & 0.059 & 5.347 & 93.254 & 20.739 & 0.989 & & 0.590 & 6.629 & 95.040 & 26.723 & 1.032\\
\hline 
\multicolumn{24}{l}{Model specification: QR--False, PM--True}\\ 
\hline 
\hspace{2mm} GPPP & 0.473 & 5.841 & 95.238 & 24.700 & 1.082 & & 1.173 & 5.124 & 94.544 & 21.555 & 1.102 & & 0.377 & 4.722 & 95.734 & 20.135 & 1.091 & & 0.583 & 6.340 & 95.734 & 28.878 & 1.166\\
\hspace{2mm} PSPP & 2.367 & 6.515 & 94.048 & 26.531 & 1.114 & & 2.392 & 5.917 & 92.262 & 23.361 & 1.101 & & 1.968 & 5.317 & 93.750 & 21.951 & 1.133 & & 1.783 & 6.260 & 94.345 & 26.286 & 1.117\\
\hspace{2mm} LWP & -3.023 & 9.265 & 96.329 & 36.224 & 1.055 & & -1.289 & 6.272 & 95.833 & 25.461 & 1.058 & & -2.038 & 6.653 & 96.131 & 25.586 & 1.030 & & -7.996 & 17.881 & 91.865 & 62.917 & 1.003\\
\hspace{2mm} AIPW & -2.472 & 15.606 & 93.056 & 57.317 & 0.948 & & -1.936 & 13.033 & 92.758 & 46.933 & 0.928 & & -1.871 & 11.081 & 93.056 & 40.685 & 0.950 & & -2.444 & 12.657 & 93.651 & 49.579 & 1.018\\
\hline 
\multicolumn{24}{l}{Model specification: QR--False, PM--False}\\ 
\hline 
\hspace{2mm} GPPP & 34.240 & 34.426 & 0.000 & 14.190 & 1.012 & & 24.837 & 25.044 & 0.000 & 12.727 & 1.010 & & 22.870 & 23.079 & 0.000 & 11.997 & 0.987 & & 44.075 & 44.248 & 0.000 & 15.383 & 1.005\\
\hspace{2mm} PSPP & 34.287 & 34.473 & 0.000 & 14.242 & 1.015 & & 24.940 & 25.146 & 0.000 & 12.766 & 1.013 & & 22.967 & 23.175 & 0.000 & 12.035 & 0.991 & & 43.983 & 44.155 & 0.000 & 15.419 & 1.011\\
\hspace{2mm} LWP & 34.158 & 34.350 & 0.000 & 14.344 & 1.009 & & 24.704 & 24.917 & 0.000 & 12.785 & 1.002 & & 22.762 & 22.976 & 0.000 & 12.053 & 0.981 & & 45.852 & 46.020 & 0.000 & 15.687 & 1.019\\
\hspace{2mm} AIPW & 35.685 & 35.890 & 0.000 & 15.018 & 1.001 & & 25.334 & 25.549 & 0.000 & 12.935 & 0.998 & & 23.645 & 23.860 & 0.000 & 12.313 & 0.982 & & 46.967 & 47.144 & 0.000 & 16.223 & 1.013\\
\bottomrule
\end{tabular}}
  \begin{tablenotes}
   \footnotesize
   \item UW: Unweighted; FW: Fully weighted; GPPP: Gaussian Process of Propensity Prediction; PSPP: Penalized Spline of Propensity Prediction; LWP: Linear-in-weight prediction; AIPW: Augmented Inverse Propensity Weighting;
  \end{tablenotes}
 \end{threeparttable}
\end{sidewaystable}

\begin{sidewaystable}[hbt!]
\centering
\caption{Comparing the performance of the bias adjustment methods in the simulation study for the \emph{continuous} outcome with $(n_A, n_R)=(1,000, 500)$ and $\gamma_1=0.6$}\label{tab:5.5}
\begin{threeparttable}
\scriptsize{\begin{tabular}{l l l l l l l l l l l l l l l l l l l l l l l l l}
\toprule
& \multicolumn{5}{c}{\textbf{$LIN$}} & & \multicolumn{5}{c}{\textbf{$CUB$}} & & \multicolumn{5}{c}{\textbf{$EXP$}} & & \multicolumn{5}{c}{\textbf{$SIN$}}\\\cline{2-6}\cline{8-12}\cline{14-18}\cline{20-24}
\textbf{Measure} & rBias & rMSE & crCI & lCI & rSE & & rBias & rMSE & crCI & lCI & rSE & & rBias & rMSE & crCI & lCI & rSE & & rBias & rMSE & crCI & lCI & rSE \\
\midrule
\multicolumn{24}{l}{\textbf{Probability sample ($S_R$)}}\\ 
\hline 
\hspace{2mm} UW & -32.193 & 32.514 & 0.000 & 13.346 & 0.747 & & -25.780 & 26.036 & 0.000 & 11.118 & 0.780 & & -24.168 & 24.422 & 0.000 & 10.264 & 0.746 & & -34.387 & 34.793 & 0.000 & 15.911 & 0.766\\
\hspace{2mm} FW & -0.525 & 7.302 & 92.659 & 28.356 & 0.993 & & -0.341 & 6.133 & 92.560 & 23.982 & 0.999 & & -0.328 & 5.973 & 92.560 & 22.907 & 0.979 & & -0.308 & 7.900 & 93.254 & 30.372 & 0.981\\
\hline 
\multicolumn{24}{l}{\textbf{Non-probability sample ($S_A$)}}\\ 
\hline 
\hspace{2mm} UW & 65.596 & 65.646 & 0.000 & 10.167 & 1.013 & & 52.643 & 52.693 & 0.000 & 9.483 & 1.052 & & 47.162 & 47.209 & 0.000 & 8.512 & 1.033 & & 54.653 & 54.700 & 0.000 & 9.133 & 1.023\\
\hspace{2mm} FW & 0.189 & 5.124 & 92.460 & 19.090 & 0.951 & & 0.156 & 4.164 & 93.452 & 15.456 & 0.947 & & 0.118 & 3.639 & 93.254 & 13.690 & 0.960 & & 0.295 & 5.043 & 94.345 & 19.700 & 0.998\\
\hline 
\multicolumn{24}{l}{\textbf{Non-robust method}}\\ 
\hline 
\multicolumn{24}{l}{Model specification: QR--True}\\ 
\hline 
\hspace{2mm} IPSW & -4.867 & 14.193 & 91.766 & 51.753 & 0.990 & & -3.670 & 11.036 & 92.560 & 39.534 & 0.969 & & -3.159 & 9.561 & 91.171 & 35.012 & 0.989 & & -4.899 & 14.133 & 91.071 & 52.612 & 1.012\\
\hline 
\multicolumn{24}{l}{Model specification: QR--False}\\ 
\hline 
\hspace{2mm} IPSW & 36.123 & 36.423 & 0.000 & 17.563 & 0.960 & & 25.933 & 26.232 & 0.000 & 14.990 & 0.968 & & 24.127 & 24.396 & 0.000 & 13.830 & 0.975 & & 46.633 & 46.748 & 0.000 & 12.846 & 1.001\\
\hline 
\multicolumn{24}{l}{\textbf{Doubly robust methods}}\\ 
\hline 
\multicolumn{24}{l}{Model specification: QR--True, PM--True}\\ 
\hline 
\hspace{2mm} GPPP & -0.446 & 7.130 & 92.361 & 27.574 & 0.988 & & -0.465 & 5.916 & 92.857 & 22.849 & 0.988 & & -0.371 & 5.540 & 92.659 & 21.555 & 0.994 & & -0.255 & 7.631 & 94.048 & 29.894 & 0.999\\
\hspace{2mm} PSPP & -0.382 & 7.106 & 92.460 & 27.585 & 0.991 & & -0.382 & 5.878 & 92.857 & 22.622 & 0.983 & & -0.489 & 5.573 & 92.758 & 21.585 & 0.991 & & -0.233 & 7.589 & 94.246 & 29.646 & 0.996\\
\hspace{2mm} LWP & -0.571 & 7.228 & 92.361 & 27.723 & 0.981 & & -0.447 & 5.924 & 92.857 & 22.770 & 0.983 & & -0.307 & 5.570 & 93.056 & 21.565 & 0.989 & & -0.289 & 7.659 & 93.452 & 29.760 & 0.991\\
\hspace{2mm} AIPW & -0.542 & 7.258 & 92.262 & 27.894 & 0.983 & & -0.399 & 5.928 & 92.956 & 22.827 & 0.984 & & -0.318 & 5.541 & 93.353 & 21.668 & 0.999 & & -0.287 & 7.691 & 93.155 & 29.800 & 0.989\\
\hline 
\multicolumn{24}{l}{Model specification: QR--True, PM--False}\\ 
\hline 
\hspace{2mm} GPPP & -0.527 & 7.224 & 90.972 & 27.305 & 0.966 & & -0.362 & 5.957 & 92.262 & 22.889 & 0.982 & & -0.371 & 5.538 & 92.460 & 21.580 & 0.996 & & -0.242 & 7.576 & 93.750 & 29.494 & 0.993\\
\hspace{2mm} PSPP & -0.418 & 7.181 & 91.071 & 27.221 & 0.968 & & -0.326 & 5.892 & 92.262 & 22.590 & 0.979 & & -0.462 & 5.526 & 92.262 & 21.489 & 0.995 & & -0.202 & 7.526 & 93.552 & 29.398 & 0.996\\
\hspace{2mm} LWP & -0.439 & 7.204 & 90.873 & 27.226 & 0.965 & & -0.320 & 5.897 & 92.163 & 22.576 & 0.978 & & -0.455 & 5.563 & 92.361 & 21.502 & 0.989 & & -0.234 & 7.527 & 93.651 & 29.419 & 0.997\\
\hspace{2mm} AIPW & -0.174 & 7.144 & 91.468 & 27.063 & 0.966 & & -0.327 & 5.912 & 92.262 & 22.427 & 0.969 & & -0.589 & 5.524 & 91.766 & 21.391 & 0.993 & & -0.145 & 7.516 & 93.552 & 29.321 & 0.995\\
\hline 
\multicolumn{24}{l}{Model specification: QR--False, PM--True}\\ 
\hline 
\hspace{2mm} GPPP & -0.339 & 6.274 & 92.063 & 24.082 & 0.980 & & 0.249 & 5.047 & 93.254 & 19.747 & 0.999 & & -0.101 & 4.573 & 93.056 & 18.084 & 1.009 & & 0.134 & 6.915 & 93.452 & 27.301 & 1.007\\
\hspace{2mm} PSPP & 0.739 & 6.360 & 93.155 & 24.830 & 1.002 & & 0.789 & 5.201 & 94.246 & 20.681 & 1.026 & & 0.650 & 4.777 & 94.048 & 19.098 & 1.029 & & 0.707 & 6.824 & 93.353 & 26.604 & 0.999\\
\hspace{2mm} LWP & -2.228 & 9.094 & 94.147 & 34.246 & 0.990 & & -1.293 & 6.413 & 95.536 & 24.766 & 1.005 & & -1.467 & 6.143 & 94.841 & 23.780 & 1.016 & & -4.775 & 15.831 & 93.452 & 55.293 & 0.934\\
\hspace{2mm} AIPW & -5.685 & 17.931 & 93.750 & 63.563 & 0.953 & & -4.543 & 14.755 & 94.940 & 51.751 & 0.940 & & -3.861 & 12.612 & 94.742 & 44.501 & 0.945 & & -4.997 & 14.525 & 91.171 & 54.283 & 1.015\\
\hline 
\multicolumn{24}{l}{Model specification: QR--False, PM--False}\\ 
\hline 
\hspace{2mm} GPPP & 33.746 & 33.950 & 0.000 & 14.357 & 0.983 & & 24.333 & 24.549 & 0.000 & 12.472 & 0.978 & & 22.495 & 22.681 & 0.000 & 11.546 & 1.015 & & 43.763 & 43.892 & 0.000 & 13.207 & 1.002\\
\hspace{2mm} PSPP & 33.720 & 33.925 & 0.000 & 14.394 & 0.986 & & 24.398 & 24.612 & 0.000 & 12.445 & 0.979 & & 22.530 & 22.715 & 0.000 & 11.541 & 1.016 & & 43.725 & 43.855 & 0.000 & 13.229 & 1.000\\
\hspace{2mm} LWP & 33.761 & 33.972 & 0.000 & 14.581 & 0.983 & & 24.281 & 24.504 & 0.000 & 12.606 & 0.975 & & 22.464 & 22.656 & 0.000 & 11.697 & 1.012 & & 45.559 & 45.658 & 0.000 & 11.969 & 1.017\\
\hspace{2mm} AIPW & 34.909 & 35.127 & 0.000 & 14.878 & 0.971 & & 24.751 & 24.973 & 0.000 & 12.705 & 0.976 & & 23.136 & 23.333 & 0.000 & 11.856 & 1.000 & & 46.416 & 46.524 & 0.000 & 12.466 & 1.004\\
\bottomrule
\end{tabular}}
  \begin{tablenotes}
   \footnotesize
   \item UW: Unweighted; FW: Fully weighted; GPPP: Gaussian Process of Propensity Prediction; PSPP: Penalized Spline of Propensity Prediction; LWP: Linear-in-weight prediction; AIPW: Augmented Inverse Propensity Weighting;
  \end{tablenotes}
 \end{threeparttable}
\end{sidewaystable}

\begin{sidewaystable}[hbt!]
\centering
\caption{Comparing the performance of the bias adjustment methods in the simulation study for the \emph{continuous} outcome with $(n_A, n_R)=(500, 500)$ and $\gamma_1=0.6$}\label{tab:5.6}
\begin{threeparttable}
\scriptsize{\begin{tabular}{l l l l l l l l l l l l l l l l l l l l l l l l l}
\toprule
& \multicolumn{5}{c}{\textbf{$LIN$}} & & \multicolumn{5}{c}{\textbf{$CUB$}} & & \multicolumn{5}{c}{\textbf{$EXP$}} & & \multicolumn{5}{c}{\textbf{$SIN$}}\\\cline{2-6}\cline{8-12}\cline{14-18}\cline{20-24}
\textbf{Measure} & rBias & rMSE & crCI & lCI & rSE & & rBias & rMSE & crCI & lCI & rSE & & rBias & rMSE & crCI & lCI & rSE & & rBias & rMSE & crCI & lCI & rSE \\
\midrule
\multicolumn{24}{l}{\textbf{Probability sample ($S_R$)}}\\ 
\hline 
\hspace{2mm} UW & -32.193 & 32.514 & 0.000 & 13.346 & 0.747 & & -25.780 & 26.036 & 0.000 & 11.118 & 0.780 & & -24.168 & 24.422 & 0.000 & 10.264 & 0.746 & & -34.387 & 34.793 & 0.000 & 15.911 & 0.766\\
\hspace{2mm} FW & -0.478 & 7.302 & 92.560 & 28.264 & 0.989 & & -0.323 & 6.128 & 92.262 & 24.001 & 1.000 & & -0.330 & 5.892 & 92.063 & 22.883 & 0.992 & & -0.277 & 7.854 & 93.452 & 30.285 & 0.984\\
\hline 
\multicolumn{24}{l}{\textbf{Non-probability sample ($S_A$)}}\\ 
\hline 
\hspace{2mm} UW & 66.963 & 67.069 & 0.000 & 14.460 & 0.978 & & 54.004 & 54.111 & 0.000 & 13.565 & 1.018 & & 48.290 & 48.392 & 0.000 & 12.131 & 0.985 & & 54.966 & 55.063 & 0.000 & 12.869 & 1.008\\
\hspace{2mm} FW & 0.267 & 7.234 & 91.964 & 26.459 & 0.933 & & 0.138 & 5.927 & 93.155 & 21.376 & 0.920 & & 0.173 & 5.243 & 92.361 & 19.079 & 0.928 & & 0.312 & 7.228 & 94.841 & 28.042 & 0.99\\
\hline 
\multicolumn{24}{l}{\textbf{Non-robust method}}\\ 
\hline 
\multicolumn{24}{l}{Model specification: QR--True}\\ 
\hline 
\hspace{2mm} IPSW & -5.988 & 16.290 & 92.460 & 57.597 & 0.969 & & -4.537 & 12.766 & 93.155 & 44.298 & 0.947 & & -3.931 & 10.860 & 91.766 & 39.169 & 0.987 & & -6.370 & 15.995 & 90.079 & 59.186 & 1.029\\
\hline 
\multicolumn{24}{l}{Model specification: QR--False}\\ 
\hline 
\hspace{2mm} IPSW & 36.099 & 36.481 & 0.000 & 20.047 & 0.971 & & 25.809 & 26.200 & 0.000 & 17.184 & 0.972 & & 24.049 & 24.419 & 0.000 & 15.911 & 0.958 & & 46.657 & 46.863 & 0.000 & 17.101 & 0.994\\
\hline 
\multicolumn{24}{l}{\textbf{Doubly robust methods}}\\ 
\hline 
\multicolumn{24}{l}{Model specification: QR--True, PM--True}\\ 
\hline 
\hspace{2mm} GPPP & -0.356 & 7.303 & 92.758 & 28.865 & 1.009 & & -0.543 & 6.284 & 92.758 & 25.177 & 1.025 & & -0.334 & 5.741 & 94.643 & 23.626 & 1.051 & & -0.191 & 7.809 & 95.536 & 31.859 & 1.041\\
\hspace{2mm} PSPP & -0.298 & 7.272 & 92.758 & 28.643 & 1.005 & & -0.471 & 6.223 & 92.758 & 24.304 & 0.999 & & -0.521 & 5.763 & 94.147 & 23.141 & 1.028 & & -0.149 & 7.789 & 95.139 & 30.904 & 1.012\\
\hspace{2mm} LWP & -0.530 & 7.521 & 92.857 & 29.081 & 0.988 & & -0.486 & 6.319 & 92.956 & 24.642 & 0.997 & & -0.212 & 5.845 & 94.147 & 23.347 & 1.019 & & -0.218 & 7.917 & 94.841 & 31.093 & 1.002\\
\hspace{2mm} AIPW & -0.481 & 7.478 & 92.361 & 28.902 & 0.988 & & -0.461 & 6.267 & 93.056 & 24.214 & 0.988 & & -0.287 & 5.753 & 94.048 & 23.050 & 1.023 & & -0.200 & 7.935 & 94.345 & 30.681 & 0.986\\
\hline 
\multicolumn{24}{l}{Model specification: QR--True, PM--False}\\ 
\hline 
\hspace{2mm} GPPP & -0.470 & 7.353 & 92.560 & 28.513 & 0.991 & & -0.434 & 6.317 & 93.155 & 24.923 & 1.008 & & -0.352 & 5.793 & 93.254 & 22.718 & 1.002 & & -0.341 & 7.962 & 94.246 & 30.438 & 0.976\\
\hspace{2mm} PSPP & -0.346 & 7.296 & 92.460 & 28.364 & 0.992 & & -0.413 & 6.145 & 92.560 & 23.883 & 0.993 & & -0.490 & 5.751 & 93.056 & 22.460 & 0.999 & & -0.258 & 7.857 & 94.444 & 30.105 & 0.977\\
\hspace{2mm} LWP & -0.374 & 7.329 & 92.560 & 28.405 & 0.989 & & -0.380 & 6.184 & 92.361 & 23.722 & 0.980 & & -0.500 & 5.830 & 92.659 & 22.456 & 0.986 & & -0.303 & 7.824 & 94.246 & 30.134 & 0.983\\
\hspace{2mm} AIPW & -0.127 & 7.252 & 92.460 & 28.115 & 0.989 & & -0.459 & 6.153 & 92.857 & 23.438 & 0.974 & & -0.655 & 5.755 & 92.758 & 22.206 & 0.990 & & -0.210 & 7.837 & 94.147 & 29.890 & 0.973\\
\hline 
\multicolumn{24}{l}{Model specification: QR--False, PM--True}\\ 
\hline 
\hspace{2mm} GPPP & -0.352 & 6.580 & 94.345 & 27.253 & 1.058 & & 0.507 & 5.550 & 94.444 & 22.889 & 1.056 & & 0.021 & 5.145 & 94.940 & 21.731 & 1.077 & & -0.179 & 7.539 & 95.337 & 32.738 & 1.108\\
\hspace{2mm} PSPP & 1.528 & 6.825 & 95.337 & 28.887 & 1.107 & & 1.684 & 6.065 & 94.643 & 25.175 & 1.102 & & 1.533 & 5.678 & 94.345 & 23.655 & 1.103 & & 1.069 & 7.255 & 94.345 & 29.606 & 1.052\\
\hspace{2mm} LWP & -4.549 & 12.067 & 94.940 & 42.565 & 0.971 & & -2.484 & 8.075 & 96.230 & 30.197 & 1.002 & & -2.878 & 8.105 & 94.940 & 30.271 & 1.019 & & -9.622 & 21.835 & 91.567 & 71.398 & 0.929\\
\hspace{2mm} AIPW & -6.795 & 21.022 & 94.841 & 71.617 & 0.918 & & -5.265 & 17.481 & 94.940 & 57.749 & 0.883 & & -4.513 & 14.430 & 94.048 & 49.842 & 0.927 & & -6.441 & 16.546 & 90.179 & 60.223 & 1.008\\
\hline 
\multicolumn{24}{l}{Model specification: QR--False, PM--False}\\ 
\hline 
\hspace{2mm} GPPP & 33.879 & 34.125 & 0.000 & 16.200 & 1.010 & & 24.355 & 24.629 & 0.000 & 14.299 & 0.995 & & 22.560 & 22.814 & 0.000 & 13.346 & 1.003 & & 43.989 & 44.182 & 0.000 & 16.532 & 1.021\\
\hspace{2mm} PSPP & 33.921 & 34.166 & 0.000 & 16.214 & 1.013 & & 24.450 & 24.721 & 0.000 & 14.288 & 0.998 & & 22.653 & 22.905 & 0.000 & 13.332 & 1.003 & & 43.895 & 44.091 & 0.000 & 16.603 & 1.021\\
\hspace{2mm} LWP & 33.794 & 34.049 & 0.000 & 16.405 & 1.006 & & 24.210 & 24.492 & 0.000 & 14.405 & 0.992 & & 22.444 & 22.706 & 0.000 & 13.447 & 0.996 & & 45.676 & 45.853 & 0.000 & 16.073 & 1.018\\
\hspace{2mm} AIPW & 34.925 & 35.195 & 0.000 & 16.849 & 0.987 & & 24.677 & 24.960 & 0.000 & 14.515 & 0.987 & & 23.106 & 23.377 & 0.000 & 13.619 & 0.978 & & 46.466 & 46.662 & 0.000 & 16.718 & 0.998\\
\bottomrule
\end{tabular}}
  \begin{tablenotes}
   \footnotesize
   \item UW: Unweighted; FW: Fully weighted; GPPP: Gaussian Process of Propensity Prediction; PSPP: Penalized Spline of Propensity Prediction; LWP: Linear-in-weight prediction; AIPW: Augmented Inverse Propensity Weighting;
  \end{tablenotes}
 \end{threeparttable}
\end{sidewaystable}


\begin{sidewaystable}[hbt!]
\centering
\caption{Comparing the performance of the bias adjustment methods in the simulation study for the \emph{binary} outcome with $(n_A, n_R)=(500, 1,000)$ and $\gamma_1=0.3$}\label{tab:5.7}
\begin{threeparttable}
\scriptsize{\begin{tabular}{l l l l l l l l l l l l l l l l l l l l l l l l l}
\toprule
& \multicolumn{5}{c}{\textbf{$LIN$}} & & \multicolumn{5}{c}{\textbf{$CUB$}} & & \multicolumn{5}{c}{\textbf{$EXP$}} & & \multicolumn{5}{c}{\textbf{$SIN$}}\\\cline{2-6}\cline{8-12}\cline{14-18}\cline{20-24}
\textbf{Measure} & rBias & rMSE & crCI & lCI & rSE & & rBias & rMSE & crCI & lCI & rSE & & rBias & rMSE & crCI & lCI & rSE & & rBias & rMSE & crCI & lCI & rSE \\
\midrule
\multicolumn{24}{l}{\textbf{Probability sample ($S_R$)}}\\ 
\hline 
\hspace{2mm} UW & -40.760 & 41.085 & 0.000 & 13.610 & 0.673 & & -44.127 & 44.513 & 0.000 & 14.703 & 0.642 & & -42.643 & 43.014 & 0.000 & 14.229 & 0.644 & & -34.513 & 34.804 & 0.000 & 12.322 & 0.700\\
\hspace{2mm} FW & 0.177 & 9.453 & 92.460 & 36.982 & 0.998 & & 0.334 & 11.357 & 93.155 & 44.750 & 1.005 & & 0.058 & 10.753 & 93.353 & 41.452 & 0.983 & & 0.093 & 7.220 & 92.659 & 28.324 & 1\\
\hline 
\multicolumn{24}{l}{\textbf{Non-probability sample ($S_A$)}}\\ 
\hline 
\hspace{2mm} UW & 44.519 & 44.897 & 0.000 & 22.957 & 1.007 & & 45.983 & 46.508 & 0.000 & 26.997 & 0.987 & & 42.958 & 43.413 & 0.000 & 25.222 & 1.026 & & 36.027 & 36.340 & 0.000 & 18.349 & 0.982\\
\hspace{2mm} FW & -0.065 & 5.792 & 95.040 & 23.261 & 1.024 & & -0.186 & 6.463 & 94.940 & 25.507 & 1.007 & & -0.425 & 6.084 & 95.437 & 24.581 & 1.033 & & -0.209 & 5.306 & 94.246 & 20.774 & 0.999\\
\hline 
\multicolumn{24}{l}{\textbf{Non-robust method}}\\ 
\hline 
\multicolumn{24}{l}{Model specification: QR--True}\\ 
\hline 
\hspace{2mm} IPSW & -1.380 & 9.343 & 93.552 & 38.396 & 1.059 & & -1.417 & 9.703 & 93.552 & 39.646 & 1.053 & & -1.530 & 9.478 & 93.155 & 38.062 & 1.038 & & -1.462 & 8.588 & 93.254 & 34.342 & 1.035\\
\hline 
\multicolumn{24}{l}{Model specification: QR--False}\\ 
\hline 
\hspace{2mm} IPSW & 31.422 & 32.111 & 0.298 & 26.210 & 1.010 & & 30.009 & 30.986 & 2.381 & 30.531 & 1.009 & & 28.849 & 29.724 & 1.786 & 28.479 & 1.014 & & 30.020 & 30.469 & 0.000 & 20.056 & 0.981\\
\hline 
\multicolumn{24}{l}{\textbf{Doubly robust methods}}\\ 
\hline 
\multicolumn{24}{l}{Model specification: QR--True, PM--True}\\ 
\hline 
\hspace{2mm} GPPP & 0.305 & 9.095 & 94.444 & 36.708 & 1.030 & & 0.209 & 10.976 & 93.155 & 43.570 & 1.012 & & 0.052 & 10.292 & 93.849 & 40.779 & 1.010 & & 0.206 & 7.339 & 93.155 & 28.380 & 0.986\\
\hspace{2mm} PSPP & 0.308 & 9.099 & 94.345 & 36.695 & 1.029 & & 0.197 & 10.978 & 93.353 & 43.507 & 1.011 & & 0.045 & 10.283 & 93.552 & 40.709 & 1.009 & & 0.200 & 7.325 & 92.956 & 28.328 & 0.986\\
\hspace{2mm} LWP & 0.376 & 9.066 & 94.345 & 36.601 & 1.030 & & 0.249 & 10.988 & 93.452 & 43.597 & 1.012 & & 0.066 & 10.274 & 94.048 & 40.745 & 1.011 & & 0.489 & 7.337 & 92.758 & 28.334 & 0.987\\
\hspace{2mm} AIPW & 0.294 & 9.098 & 94.048 & 36.641 & 1.027 & & 0.170 & 10.962 & 93.651 & 43.504 & 1.012 & & -0.041 & 10.296 & 93.750 & 40.663 & 1.007 & & 0.132 & 7.329 & 92.659 & 28.293 & 0.985\\
\hline 
\multicolumn{24}{l}{Model specification: QR--True, PM--False}\\ 
\hline 
\hspace{2mm} GPPP & 0.498 & 9.123 & 94.048 & 36.881 & 1.032 & & 0.453 & 11.105 & 93.849 & 43.564 & 1.001 & & 0.235 & 10.286 & 94.444 & 40.953 & 1.016 & & 0.614 & 7.277 & 94.544 & 28.324 & 0.996\\
\hspace{2mm} PSPP & 0.367 & 9.127 & 94.147 & 36.845 & 1.030 & & 0.305 & 11.091 & 93.452 & 43.491 & 1.000 & & 0.042 & 10.305 & 94.544 & 40.885 & 1.012 & & 0.309 & 7.277 & 94.246 & 28.312 & 0.993\\
\hspace{2mm} LWP & 0.311 & 9.124 & 94.246 & 36.796 & 1.029 & & 0.283 & 11.084 & 93.552 & 43.373 & 0.998 & & -0.046 & 10.291 & 94.345 & 40.694 & 1.008 & & 0.535 & 7.270 & 94.544 & 28.260 & 0.994\\
\hspace{2mm} AIPW & 0.341 & 9.126 & 94.048 & 36.734 & 1.027 & & 0.344 & 11.075 & 93.452 & 43.352 & 0.999 & & 0.023 & 10.279 & 94.544 & 40.624 & 1.008 & & 0.229 & 7.263 & 94.048 & 28.246 & 0.992\\
\hline 
\multicolumn{24}{l}{Model specification: QR--False, PM--True}\\ 
\hline 
\hspace{2mm} GPPP & 0.210 & 7.589 & 93.849 & 30.850 & 1.037 & & 0.070 & 8.274 & 94.940 & 33.506 & 1.033 & & -0.050 & 8.019 & 93.651 & 31.554 & 1.003 & & 0.242 & 6.494 & 93.948 & 25.292 & 0.994\\
\hspace{2mm} PSPP & 1.632 & 7.675 & 94.246 & 30.754 & 1.046 & & 1.403 & 8.286 & 95.238 & 33.442 & 1.044 & & 1.161 & 8.017 & 94.048 & 31.547 & 1.014 & & 1.202 & 6.556 & 93.353 & 25.177 & 0.996\\
\hspace{2mm} LWP & 0.161 & 7.591 & 93.353 & 30.803 & 1.035 & & -0.019 & 8.269 & 94.544 & 33.496 & 1.033 & & -0.114 & 8.025 & 93.948 & 31.474 & 1.000 & & 0.119 & 6.409 & 93.849 & 25.012 & 0.995\\
\hspace{2mm} AIPW & -1.744 & 11.622 & 94.544 & 47.373 & 1.051 & & -1.869 & 12.454 & 95.238 & 50.703 & 1.050 & & -1.878 & 11.830 & 94.643 & 47.349 & 1.034 & & -1.389 & 9.396 & 93.155 & 37.347 & 1.025\\
\hline 
\multicolumn{24}{l}{Model specification: QR--False, PM--False}\\ 
\hline 
\hspace{2mm} GPPP & 30.673 & 31.269 & 0.099 & 24.600 & 1.032 & & 29.385 & 30.261 & 1.587 & 28.489 & 1.005 & & 28.263 & 29.047 & 1.091 & 26.805 & 1.019 & & 29.357 & 29.783 & 0.000 & 19.439 & 0.988\\
\hspace{2mm} PSPP & 30.670 & 31.264 & 0.099 & 24.584 & 1.034 & & 29.369 & 30.243 & 1.587 & 28.467 & 1.005 & & 28.260 & 29.044 & 1.091 & 26.808 & 1.020 & & 29.338 & 29.765 & 0.000 & 19.417 & 0.985\\
\hspace{2mm} LWP & 30.671 & 31.262 & 0.099 & 24.389 & 1.029 & & 29.349 & 30.220 & 1.488 & 28.355 & 1.003 & & 28.241 & 29.023 & 1.091 & 26.674 & 1.017 & & 29.585 & 30.003 & 0.000 & 19.114 & 0.977\\
\hspace{2mm} AIPW & 30.827 & 31.416 & 0.099 & 24.312 & 1.024 & & 29.456 & 30.326 & 1.290 & 28.331 & 1.001 & & 28.355 & 29.135 & 0.992 & 26.651 & 1.015 & & 29.875 & 30.299 & 0.000 & 19.381 & 0.978\\
\bottomrule
\end{tabular}}
  \begin{tablenotes}
   \footnotesize
   \item UW: Unweighted; FW: Fully weighted; GPPP: Gaussian Process of Propensity Prediction; PSPP: Penalized Spline of Propensity Prediction; LWP: Linear-in-weight prediction; AIPW: Augmented Inverse Propensity Weighting;
  \end{tablenotes}
 \end{threeparttable}
\end{sidewaystable}

\begin{sidewaystable}[hbt!]
\centering
\caption{Comparing the performance of the bias adjustment methods in the simulation study for the \emph{binary} outcome with $(n_A, n_R)=(1,000, 500)$ and $\gamma_1=0.3$}\label{tab:5.8}
\begin{threeparttable}
\scriptsize{\begin{tabular}{l l l l l l l l l l l l l l l l l l l l l l l l l}
\toprule
& \multicolumn{5}{c}{\textbf{$LIN$}} & & \multicolumn{5}{c}{\textbf{$CUB$}} & & \multicolumn{5}{c}{\textbf{$EXP$}} & & \multicolumn{5}{c}{\textbf{$SIN$}}\\\cline{2-6}\cline{8-12}\cline{14-18}\cline{20-24}
\textbf{Measure} & rBias & rMSE & crCI & lCI & rSE & & rBias & rMSE & crCI & lCI & rSE & & rBias & rMSE & crCI & lCI & rSE & & rBias & rMSE & crCI & lCI & rSE \\
\midrule
\multicolumn{24}{l}{\textbf{Probability sample ($S_R$)}}\\ 
\hline 
\hspace{2mm} UW & -41.047 & 41.487 & 0.000 & 19.213 & 0.813 & & -44.574 & 45.028 & 0.000 & 20.730 & 0.828 & & -42.945 & 43.406 & 0.000 & 20.084 & 0.812 & & -34.898 & 35.315 & 0.000 & 17.395 & 0.820\\
\hspace{2mm} FW & -0.300 & 10.798 & 93.849 & 42.547 & 1.005 & & -0.588 & 12.476 & 93.353 & 50.758 & 1.039 & & -0.555 & 11.617 & 93.651 & 47.476 & 1.043 & & -0.358 & 8.568 & 93.651 & 33.456 & 0.996\\
\hline 
\multicolumn{24}{l}{\textbf{Non-probability sample ($S_A$)}}\\ 
\hline 
\hspace{2mm} UW & 44.057 & 44.254 & 0.000 & 16.239 & 0.993 & & 45.304 & 45.566 & 0.000 & 19.086 & 0.998 & & 42.865 & 43.094 & 0.000 & 17.834 & 1.026 & & 35.889 & 36.043 & 0.000 & 12.980 & 0.992\\
\hspace{2mm} FW & 0.170 & 4.160 & 95.139 & 16.418 & 1.007 & & -0.039 & 4.564 & 95.139 & 18.010 & 1.006 & & 0.158 & 4.328 & 95.635 & 17.384 & 1.025 & & 0.142 & 3.739 & 95.635 & 14.658 & 1.000\\
\hline 
\multicolumn{24}{l}{\textbf{Non-robust method}}\\ 
\hline 
\multicolumn{24}{l}{Model specification: QR--True}\\ 
\hline 
\hspace{2mm} IPSW & -2.157 & 10.471 & 92.857 & 42.950 & 1.069 & & -2.257 & 10.919 & 92.956 & 43.494 & 1.038 & & -1.980 & 10.273 & 92.857 & 41.675 & 1.054 & & -2.058 & 9.476 & 94.246 & 38.427 & 1.059\\
\hline 
\multicolumn{24}{l}{Model specification: QR--False}\\ 
\hline 
\hspace{2mm} IPSW & 30.837 & 31.398 & 0.099 & 23.746 & 1.025 & & 29.353 & 30.150 & 0.694 & 27.671 & 1.024 & & 28.686 & 29.364 & 0.496 & 25.303 & 1.029 & & 29.823 & 30.082 & 0.000 & 16.020 & 1.038\\
\hline 
\multicolumn{24}{l}{\textbf{Doubly robust methods}}\\ 
\hline 
\multicolumn{24}{l}{Model specification: QR--True, PM--True}\\ 
\hline 
\hspace{2mm} GPPP & -0.329 & 9.828 & 93.452 & 39.581 & 1.027 & & -0.491 & 11.391 & 92.361 & 45.621 & 1.022 & & -0.279 & 10.520 & 93.948 & 42.782 & 1.037 & & -0.347 & 8.042 & 93.552 & 31.670 & 1.005\\
\hspace{2mm} PSPP & -0.331 & 9.825 & 93.552 & 39.571 & 1.028 & & -0.497 & 11.397 & 92.460 & 45.574 & 1.021 & & -0.275 & 10.517 & 93.849 & 42.718 & 1.036 & & -0.353 & 8.030 & 93.452 & 31.655 & 1.006\\
\hspace{2mm} LWP & -0.318 & 9.819 & 93.452 & 39.545 & 1.027 & & -0.468 & 11.408 & 92.560 & 45.582 & 1.020 & & -0.278 & 10.516 & 93.849 & 42.698 & 1.036 & & -0.249 & 8.029 & 93.552 & 31.664 & 1.006\\
\hspace{2mm} AIPW & -0.344 & 9.827 & 93.254 & 39.557 & 1.027 & & -0.500 & 11.402 & 92.560 & 45.548 & 1.020 & & -0.313 & 10.522 & 93.651 & 42.663 & 1.034 & & -0.380 & 8.020 & 93.651 & 31.635 & 1.007\\
\hline 
\multicolumn{24}{l}{Model specification: QR--True, PM--False}\\ 
\hline 
\hspace{2mm} GPPP & -0.243 & 9.885 & 93.651 & 39.576 & 1.021 & & -0.446 & 11.315 & 92.758 & 45.663 & 1.030 & & -0.181 & 10.569 & 93.948 & 42.585 & 1.028 & & -0.248 & 8.029 & 93.452 & 31.783 & 1.010\\
\hspace{2mm} PSPP & -0.293 & 9.889 & 93.552 & 39.552 & 1.020 & & -0.514 & 11.319 & 92.758 & 45.628 & 1.029 & & -0.272 & 10.575 & 93.948 & 42.539 & 1.026 & & -0.368 & 8.036 & 93.552 & 31.802 & 1.010\\
\hspace{2mm} LWP & -0.306 & 9.892 & 93.651 & 39.546 & 1.020 & & -0.523 & 11.309 & 92.659 & 45.595 & 1.029 & & -0.305 & 10.573 & 93.849 & 42.463 & 1.024 & & -0.290 & 8.028 & 93.651 & 31.771 & 1.010\\
\hspace{2mm} AIPW & -0.288 & 9.902 & 93.452 & 39.494 & 1.017 & & -0.476 & 11.304 & 92.659 & 45.534 & 1.028 & & -0.289 & 10.563 & 93.750 & 42.418 & 1.024 & & -0.422 & 8.042 & 93.651 & 31.768 & 1.009\\
\hline 
\multicolumn{24}{l}{Model specification: QR--False, PM--True}\\ 
\hline 
\hspace{2mm} GPPP & -0.027 & 8.255 & 93.651 & 33.254 & 1.027 & & -0.268 & 8.722 & 93.552 & 34.618 & 1.012 & & -0.050 & 8.168 & 93.948 & 32.870 & 1.026 & & -0.187 & 7.194 & 94.048 & 28.743 & 1.019\\
\hspace{2mm} PSPP & 0.664 & 8.282 & 93.552 & 33.140 & 1.024 & & 0.324 & 8.716 & 93.353 & 34.499 & 1.010 & & 0.514 & 8.163 & 94.345 & 32.827 & 1.027 & & 0.297 & 7.201 & 93.948 & 28.654 & 1.016\\
\hspace{2mm} LWP & -0.052 & 8.262 & 93.750 & 33.260 & 1.026 & & -0.296 & 8.754 & 93.552 & 34.672 & 1.010 & & -0.084 & 8.174 & 93.948 & 32.824 & 1.024 & & -0.211 & 7.029 & 94.841 & 28.203 & 1.024\\
\hspace{2mm} AIPW & -2.397 & 13.400 & 94.048 & 53.981 & 1.044 & & -2.748 & 14.221 & 93.948 & 57.030 & 1.042 & & -2.459 & 13.296 & 94.742 & 53.741 & 1.049 & & -2.081 & 10.389 & 94.345 & 42.436 & 1.063\\
\hline 
\multicolumn{24}{l}{Model specification: QR--False, PM--False}\\ 
\hline 
\hspace{2mm} GPPP & 30.818 & 31.348 & 0.099 & 22.751 & 1.010 & & 29.492 & 30.201 & 0.496 & 26.142 & 1.025 & & 28.797 & 29.405 & 0.298 & 24.074 & 1.032 & & 29.590 & 29.893 & 0.000 & 16.838 & 1.012\\
\hspace{2mm} PSPP & 30.815 & 31.347 & 0.099 & 22.752 & 1.008 & & 29.486 & 30.196 & 0.496 & 26.129 & 1.023 & & 28.792 & 29.401 & 0.298 & 24.082 & 1.032 & & 29.580 & 29.879 & 0.000 & 16.685 & 1.009\\
\hspace{2mm} LWP & 30.840 & 31.366 & 0.099 & 22.408 & 0.999 & & 29.484 & 30.195 & 0.397 & 26.043 & 1.020 & & 28.791 & 29.397 & 0.198 & 23.909 & 1.027 & & 29.853 & 30.103 & 0.000 & 15.380 & 1.014\\
\hspace{2mm} AIPW & 30.885 & 31.393 & 0.099 & 22.123 & 1.003 & & 29.522 & 30.224 & 0.298 & 25.858 & 1.018 & & 28.836 & 29.433 & 0.198 & 23.739 & 1.026 & & 29.903 & 30.143 & 0.000 & 15.305 & 1.027\\
\bottomrule
\end{tabular}}
  \begin{tablenotes}
   \footnotesize
   \item UW: Unweighted; FW: Fully weighted; GPPP: Gaussian Process of Propensity Prediction; PSPP: Penalized Spline of Propensity Prediction; LWP: Linear-in-weight prediction; AIPW: Augmented Inverse Propensity Weighting;
  \end{tablenotes}
 \end{threeparttable}
\end{sidewaystable}

\begin{sidewaystable}[hbt!]
\centering
\caption{Comparing the performance of the bias adjustment methods in the simulation study for the \emph{binary} outcome with $(n_A, n_R)=(500, 500)$ and $\gamma_1=0.3$}\label{tab:5.9}
\begin{threeparttable}
\scriptsize{\begin{tabular}{l l l l l l l l l l l l l l l l l l l l l l l l l}
\toprule
& \multicolumn{5}{c}{\textbf{$LIN$}} & & \multicolumn{5}{c}{\textbf{$CUB$}} & & \multicolumn{5}{c}{\textbf{$EXP$}} & & \multicolumn{5}{c}{\textbf{$SIN$}}\\\cline{2-6}\cline{8-12}\cline{14-18}\cline{20-24}
\textbf{Measure} & rBias & rMSE & crCI & lCI & rSE & & rBias & rMSE & crCI & lCI & rSE & & rBias & rMSE & crCI & lCI & rSE & & rBias & rMSE & crCI & lCI & rSE \\
\midrule
\multicolumn{24}{l}{\textbf{Probability sample ($S_R$)}}\\ 
\hline 
\hspace{2mm} UW & -41.047 & 41.487 & 0.000 & 19.213 & 0.813 & & -44.574 & 45.028 & 0.000 & 20.730 & 0.828 & & -42.945 & 43.406 & 0.000 & 20.084 & 0.812 & & -34.898 & 35.315 & 0.000 & 17.395 & 0.820\\
\hspace{2mm} FW & -0.335 & 10.901 & 93.552 & 42.588 & 0.997 & & -0.575 & 12.433 & 92.560 & 50.551 & 1.038 & & -0.570 & 11.625 & 93.254 & 47.208 & 1.037 & & -0.326 & 8.522 & 94.544 & 33.469 & 1.002\\
\hline 
\multicolumn{24}{l}{\textbf{Non-probability sample ($S_A$)}}\\ 
\hline 
\hspace{2mm} UW & 44.575 & 44.964 & 0.000 & 22.956 & 0.992 & & 46.016 & 46.554 & 0.000 & 26.997 & 0.976 & & 43.285 & 43.785 & 0.000 & 25.219 & 0.974 & & 36.188 & 36.468 & 0.000 & 18.343 & 1.037\\
\hspace{2mm} FW & 0.188 & 5.912 & 95.238 & 23.269 & 1.004 & & 0.094 & 6.541 & 94.742 & 25.533 & 0.995 & & 0.038 & 6.316 & 95.734 & 24.616 & 0.994 & & 0.156 & 5.027 & 95.337 & 20.785 & 1.055\\
\hline 
\multicolumn{24}{l}{\textbf{Non-robust method}}\\ 
\hline 
\multicolumn{24}{l}{Model specification: QR--True}\\ 
\hline 
\hspace{2mm} IPSW & -2.243 & 11.249 & 92.659 & 45.293 & 1.048 & & -2.251 & 11.567 & 92.758 & 46.382 & 1.042 & & -2.284 & 11.271 & 92.262 & 44.587 & 1.030 & & -2.216 & 10.028 & 93.155 & 40.735 & 1.062\\
\hline 
\multicolumn{24}{l}{Model specification: QR--False}\\ 
\hline 
\hspace{2mm} IPSW & 31.049 & 31.894 & 1.290 & 28.822 & 1.008 & & 29.561 & 30.735 & 6.151 & 33.381 & 1.011 & & 28.698 & 29.785 & 5.060 & 31.006 & 0.992 & & 30.014 & 30.449 & 0.000 & 20.837 & 1.036\\
\hline 
\multicolumn{24}{l}{\textbf{Doubly robust methods}}\\ 
\hline 
\multicolumn{24}{l}{Model specification: QR--True, PM--True}\\ 
\hline 
\hspace{2mm} GPPP & -0.304 & 10.252 & 93.948 & 40.870 & 1.017 & & -0.278 & 11.735 & 94.246 & 47.397 & 1.030 & & -0.374 & 10.829 & 95.040 & 44.445 & 1.047 & & -0.276 & 8.244 & 93.254 & 32.312 & 1.000\\
\hspace{2mm} PSPP & -0.312 & 10.254 & 93.849 & 40.855 & 1.016 & & -0.309 & 11.731 & 94.246 & 47.293 & 1.028 & & -0.401 & 10.822 & 94.940 & 44.352 & 1.046 & & -0.297 & 8.219 & 93.452 & 32.270 & 1.002\\
\hspace{2mm} LWP & -0.241 & 10.233 & 94.048 & 40.783 & 1.016 & & -0.239 & 11.747 & 94.147 & 47.290 & 1.027 & & -0.373 & 10.819 & 94.643 & 44.304 & 1.045 & & -0.021 & 8.216 & 93.750 & 32.302 & 1.003\\
\hspace{2mm} AIPW & -0.334 & 10.246 & 94.147 & 40.796 & 1.016 & & -0.340 & 11.717 & 94.147 & 47.183 & 1.027 & & -0.487 & 10.837 & 94.444 & 44.240 & 1.042 & & -0.349 & 8.213 & 93.353 & 32.233 & 1.002\\
\hline 
\multicolumn{24}{l}{Model specification: QR--True, PM--False}\\ 
\hline 
\hspace{2mm} GPPP & -0.022 & 10.274 & 93.552 & 40.822 & 1.013 & & -0.232 & 11.714 & 93.353 & 47.716 & 1.039 & & -0.065 & 10.868 & 93.651 & 44.175 & 1.036 & & -0.030 & 8.185 & 93.651 & 32.685 & 1.018\\
\hspace{2mm} PSPP & -0.161 & 10.277 & 93.452 & 40.787 & 1.012 & & -0.368 & 11.703 & 93.353 & 47.627 & 1.038 & & -0.276 & 10.853 & 93.849 & 44.023 & 1.035 & & -0.291 & 8.197 & 93.651 & 32.673 & 1.017\\
\hspace{2mm} LWP & -0.189 & 10.269 & 93.552 & 40.775 & 1.013 & & -0.403 & 11.701 & 93.254 & 47.504 & 1.036 & & -0.367 & 10.855 & 93.849 & 43.904 & 1.032 & & -0.108 & 8.174 & 93.948 & 32.588 & 1.017\\
\hspace{2mm} AIPW & -0.180 & 10.278 & 93.155 & 40.695 & 1.010 & & -0.327 & 11.655 & 93.056 & 47.403 & 1.037 & & -0.312 & 10.815 & 94.048 & 43.809 & 1.033 & & -0.372 & 8.203 & 93.452 & 32.545 & 1.013\\
\hline 
\multicolumn{24}{l}{Model specification: QR--False, PM--True}\\ 
\hline 
\hspace{2mm} GPPP & -0.074 & 8.802 & 93.353 & 34.932 & 1.012 & & -0.108 & 9.361 & 94.147 & 37.706 & 1.027 & & -0.099 & 8.807 & 94.444 & 35.413 & 1.025 & & -0.271 & 7.442 & 93.452 & 29.490 & 1.011\\
\hspace{2mm} PSPP & 1.298 & 8.884 & 93.750 & 34.903 & 1.013 & & 1.242 & 9.401 & 94.742 & 37.718 & 1.032 & & 1.110 & 8.819 & 94.742 & 35.405 & 1.032 & & 0.675 & 7.426 & 93.948 & 29.415 & 1.014\\
\hspace{2mm} LWP & -0.113 & 8.812 & 93.056 & 34.873 & 1.009 & & -0.168 & 9.391 & 93.750 & 37.675 & 1.023 & & -0.169 & 8.795 & 94.742 & 35.309 & 1.024 & & -0.325 & 7.269 & 93.651 & 29.004 & 1.018\\
\hspace{2mm} AIPW & -2.559 & 14.081 & 95.139 & 57.536 & 1.060 & & -2.702 & 15.082 & 94.643 & 61.474 & 1.056 & & -2.527 & 14.254 & 93.849 & 57.570 & 1.046 & & -2.289 & 11.031 & 94.048 & 45.025 & 1.064\\
\hline 
\multicolumn{24}{l}{Model specification: QR--False, PM--False}\\ 
\hline 
\hspace{2mm} GPPP & 31.065 & 31.834 & 0.595 & 27.121 & 0.994 & & 29.748 & 30.772 & 2.679 & 31.444 & 1.019 & & 28.823 & 29.783 & 2.778 & 29.269 & 0.995 & & 29.821 & 30.265 & 0.000 & 20.999 & 1.037\\
\hspace{2mm} PSPP & 31.076 & 31.846 & 0.595 & 27.108 & 0.993 & & 29.738 & 30.764 & 2.877 & 31.423 & 1.017 & & 28.818 & 29.779 & 2.579 & 29.258 & 0.994 & & 29.783 & 30.231 & 0.000 & 20.989 & 1.032\\
\hspace{2mm} LWP & 31.060 & 31.825 & 0.595 & 26.790 & 0.985 & & 29.720 & 30.751 & 2.877 & 31.231 & 1.009 & & 28.806 & 29.764 & 2.480 & 29.036 & 0.988 & & 30.003 & 30.404 & 0.000 & 19.992 & 1.035\\
\hspace{2mm} AIPW & 31.111 & 31.865 & 0.595 & 26.638 & 0.986 & & 29.743 & 30.769 & 2.877 & 31.144 & 1.008 & & 28.847 & 29.804 & 2.877 & 28.960 & 0.985 & & 30.078 & 30.482 & 0.000 & 20.037 & 1.033\\
\bottomrule
\end{tabular}}
  \begin{tablenotes}
   \footnotesize
   \item UW: Unweighted; FW: Fully weighted; GPPP: Gaussian Process of Propensity Prediction; PSPP: Penalized Spline of Propensity Prediction; LWP: Linear-in-weight prediction; AIPW: Augmented Inverse Propensity Weighting;
  \end{tablenotes}
 \end{threeparttable}
\end{sidewaystable}

\begin{sidewaystable}[hbt!]
\centering
\caption{Comparing the performance of the bias adjustment methods in the simulation study for the \emph{binary} outcome with $(n_A, n_R)=(500, 1,000)$ and $\gamma_1=0.6$}\label{tab:5.10}
\begin{threeparttable}
\scriptsize{\begin{tabular}{l l l l l l l l l l l l l l l l l l l l l l l l l}
\toprule
& \multicolumn{5}{c}{\textbf{$LIN$}} & & \multicolumn{5}{c}{\textbf{$CUB$}} & & \multicolumn{5}{c}{\textbf{$EXP$}} & & \multicolumn{5}{c}{\textbf{$SIN$}}\\\cline{2-6}\cline{8-12}\cline{14-18}\cline{20-24}
\textbf{Measure} & rBias & rMSE & crCI & lCI & rSE & & rBias & rMSE & crCI & lCI & rSE & & rBias & rMSE & crCI & lCI & rSE & & rBias & rMSE & crCI & lCI & rSE \\
\midrule
\multicolumn{24}{l}{\textbf{Probability sample ($S_R$)}}\\ 
\hline 
\hspace{2mm} UW & -40.760 & 41.085 & 0.000 & 13.610 & 0.673 & & -44.127 & 44.513 & 0.000 & 14.703 & 0.642 & & -42.643 & 43.014 & 0.000 & 14.229 & 0.644 & & -34.513 & 34.804 & 0.000 & 12.322 & 0.700\\
\hspace{2mm} FW & 0.256 & 9.389 & 92.758 & 37.069 & 1.007 & & 0.371 & 11.344 & 93.254 & 45.111 & 1.014 & & 0.101 & 10.820 & 92.758 & 41.488 & 0.978 & & 0.095 & 7.111 & 93.849 & 28.415 & 1.019\\
\hline 
\multicolumn{24}{l}{\textbf{Non-probability sample ($S_A$)}}\\ 
\hline 
\hspace{2mm} UW & 86.387 & 86.546 & 0.000 & 20.970 & 1.019 & & 93.225 & 93.459 & 0.000 & 26.157 & 1.011 & & 86.223 & 86.433 & 0.000 & 24.102 & 1.021 & & 63.287 & 63.420 & 0.000 & 16.349 & 1.015\\
\hspace{2mm} FW & 0.131 & 7.764 & 95.238 & 30.716 & 1.009 & & 0.046 & 8.475 & 94.841 & 32.731 & 0.985 & & -0.118 & 8.148 & 94.147 & 32.023 & 1.002 & & -0.025 & 7.263 & 95.734 & 28.730 & 1.009\\
\hline 
\multicolumn{24}{l}{\textbf{Non-robust method}}\\ 
\hline 
\multicolumn{24}{l}{Model specification: QR--True}\\ 
\hline 
\hspace{2mm} IPSW & -3.227 & 13.982 & 93.353 & 56.871 & 1.066 & & -3.152 & 14.463 & 92.460 & 57.679 & 1.042 & & -3.082 & 13.955 & 92.758 & 55.562 & 1.041 & & -3.285 & 12.805 & 92.956 & 52.146 & 1.074\\
\hline 
\multicolumn{24}{l}{Model specification: QR--False}\\ 
\hline 
\hspace{2mm} IPSW & 53.562 & 54.137 & 0.000 & 31.653 & 1.025 & & 50.993 & 51.855 & 0.000 & 37.317 & 1.010 & & 49.050 & 49.812 & 0.000 & 34.351 & 1.009 & & 51.063 & 51.358 & 0.000 & 21.820 & 1.012\\
\hline 
\multicolumn{24}{l}{\textbf{Doubly robust methods}}\\ 
\hline 
\multicolumn{24}{l}{Model specification: QR--True, PM--True}\\ 
\hline 
\hspace{2mm} GPPP & 0.337 & 9.267 & 94.742 & 37.662 & 1.037 & & 0.357 & 11.473 & 93.254 & 44.914 & 0.999 & & 0.391 & 10.659 & 94.444 & 42.397 & 1.015 & & 0.313 & 7.382 & 93.056 & 28.939 & 1.000\\
\hspace{2mm} PSPP & 0.307 & 9.265 & 94.742 & 37.616 & 1.036 & & 0.312 & 11.436 & 93.254 & 44.827 & 1.000 & & 0.383 & 10.639 & 94.742 & 42.158 & 1.011 & & 0.277 & 7.361 & 93.056 & 28.724 & 0.996\\
\hspace{2mm} LWP & 0.887 & 9.303 & 95.139 & 37.769 & 1.040 & & 0.679 & 11.588 & 93.452 & 45.211 & 0.997 & & 0.563 & 10.697 & 94.544 & 42.513 & 1.015 & & 2.122 & 7.827 & 93.552 & 29.736 & 1.006\\
\hspace{2mm} AIPW & 0.242 & 9.280 & 94.940 & 37.601 & 1.033 & & 0.147 & 11.439 & 93.254 & 44.799 & 0.999 & & -0.006 & 10.682 & 94.444 & 42.023 & 1.003 & & 0.133 & 7.387 & 92.956 & 28.793 & 0.994\\
\hline 
\multicolumn{24}{l}{Model specification: QR--True, PM--False}\\ 
\hline 
\hspace{2mm} GPPP & 1.021 & 9.354 & 94.643 & 38.386 & 1.053 & & 0.790 & 11.546 & 92.857 & 45.111 & 0.999 & & 0.739 & 10.591 & 94.444 & 42.842 & 1.034 & & 1.411 & 7.633 & 93.948 & 29.422 & 1.000\\
\hspace{2mm} PSPP & 0.469 & 9.290 & 94.147 & 38.002 & 1.044 & & 0.322 & 11.365 & 93.056 & 44.503 & 0.999 & & 0.285 & 10.541 & 94.147 & 42.234 & 1.022 & & 0.298 & 7.440 & 93.452 & 29.029 & 0.996\\
\hspace{2mm} LWP & 0.282 & 9.303 & 94.048 & 37.791 & 1.036 & & 0.296 & 11.384 & 92.857 & 44.395 & 0.995 & & 0.206 & 10.517 & 94.048 & 41.911 & 1.016 & & 0.650 & 7.450 & 94.048 & 29.049 & 0.998\\
\hspace{2mm} AIPW & 0.335 & 9.278 & 93.948 & 37.612 & 1.034 & & 0.472 & 11.339 & 93.155 & 44.319 & 0.997 & & 0.396 & 10.522 & 93.750 & 41.799 & 1.014 & & 0.072 & 7.418 & 93.651 & 28.780 & 0.989\\
\hline 
\multicolumn{24}{l}{Model specification: QR--False, PM--True}\\ 
\hline 
\hspace{2mm} GPPP & 0.279 & 7.814 & 95.635 & 32.201 & 1.051 & & 0.515 & 9.122 & 94.048 & 35.787 & 1.002 & & 0.337 & 8.482 & 94.742 & 34.458 & 1.037 & & 0.259 & 6.655 & 93.552 & 26.511 & 1.017\\
\hspace{2mm} PSPP & 4.350 & 9.046 & 93.948 & 33.320 & 1.071 & & 4.859 & 10.448 & 92.857 & 37.065 & 1.022 & & 4.419 & 9.750 & 94.048 & 35.700 & 1.047 & & 2.881 & 7.177 & 93.155 & 26.425 & 1.025\\
\hspace{2mm} LWP & 0.031 & 7.792 & 95.238 & 31.935 & 1.045 & & 0.010 & 9.015 & 93.849 & 35.327 & 0.999 & & -0.070 & 8.411 & 94.544 & 33.866 & 1.027 & & 0.116 & 6.618 & 93.353 & 26.025 & 1.003\\
\hspace{2mm} AIPW & -3.794 & 17.578 & 94.940 & 70.650 & 1.050 & & -4.091 & 19.231 & 94.246 & 75.598 & 1.026 & & -3.969 & 17.898 & 95.536 & 70.722 & 1.033 & & -3.231 & 13.525 & 94.345 & 54.283 & 1.054\\
\hline 
\multicolumn{24}{l}{Model specification: QR--False, PM--False}\\ 
\hline 
\hspace{2mm} GPPP & 46.258 & 46.727 & 0.000 & 26.958 & 1.041 & & 43.406 & 44.150 & 0.000 & 31.503 & 0.995 & & 42.182 & 42.813 & 0.000 & 29.504 & 1.027 & & 46.251 & 46.541 & 0.000 & 20.569 & 1.011\\
\hspace{2mm} PSPP & 46.229 & 46.695 & 0.000 & 26.939 & 1.043 & & 43.369 & 44.112 & 0.000 & 31.478 & 0.995 & & 42.150 & 42.779 & 0.000 & 29.486 & 1.028 & & 46.153 & 46.445 & 0.000 & 20.470 & 1.004\\
\hspace{2mm} LWP & 46.320 & 46.787 & 0.000 & 26.826 & 1.038 & & 43.358 & 44.107 & 0.000 & 31.488 & 0.992 & & 42.145 & 42.777 & 0.000 & 29.429 & 1.024 & & 48.246 & 48.528 & 0.000 & 20.352 & 0.994\\
\hspace{2mm} AIPW & 46.840 & 47.297 & 0.000 & 26.698 & 1.038 & & 43.745 & 44.481 & 0.000 & 31.341 & 0.992 & & 42.734 & 43.354 & 0.000 & 29.297 & 1.022 & & 50.152 & 50.439 & 0.000 & 21.149 & 1.003\\
\bottomrule
\end{tabular}}
  \begin{tablenotes}
   \footnotesize
   \item UW: Unweighted; FW: Fully weighted; GPPP: Gaussian Process of Propensity Prediction; PSPP: Penalized Spline of Propensity Prediction; LWP: Linear-in-weight prediction; AIPW: Augmented Inverse Propensity Weighting;
  \end{tablenotes}
 \end{threeparttable}
\end{sidewaystable}

\begin{sidewaystable}[hbt!]
\centering
\caption{Comparing the performance of the bias adjustment methods in the simulation study for the \emph{binary} outcome with $(n_A, n_R)=(1,000, 500)$ and $\gamma_1=0.6$}\label{tab:5.11}
\begin{threeparttable}
\scriptsize{\begin{tabular}{l l l l l l l l l l l l l l l l l l l l l l l l l}
\toprule
& \multicolumn{5}{c}{\textbf{$LIN$}} & & \multicolumn{5}{c}{\textbf{$CUB$}} & & \multicolumn{5}{c}{\textbf{$EXP$}} & & \multicolumn{5}{c}{\textbf{$SIN$}}\\\cline{2-6}\cline{8-12}\cline{14-18}\cline{20-24}
\textbf{Measure} & rBias & rMSE & crCI & lCI & rSE & & rBias & rMSE & crCI & lCI & rSE & & rBias & rMSE & crCI & lCI & rSE & & rBias & rMSE & crCI & lCI & rSE \\
\midrule
\multicolumn{24}{l}{\textbf{Probability sample ($S_R$)}}\\ 
\hline 
\hspace{2mm} UW & -41.047 & 41.487 & 0.000 & 19.213 & 0.813 & & -44.574 & 45.028 & 0.000 & 20.730 & 0.828 & & -42.945 & 43.406 & 0.000 & 20.084 & 0.812 & & -34.898 & 35.315 & 0.000 & 17.395 & 0.820\\
\hspace{2mm} FW & -0.326 & 10.847 & 93.552 & 42.776 & 1.006 & & -0.675 & 12.615 & 92.758 & 51.013 & 1.033 & & -0.509 & 11.780 & 93.353 & 47.385 & 1.027 & & -0.403 & 8.536 & 93.849 & 33.518 & 1.002\\
\hline 
\multicolumn{24}{l}{\textbf{Non-probability sample ($S_A$)}}\\ 
\hline 
\hspace{2mm} UW & 84.864 & 84.948 & 0.000 & 14.914 & 1.009 & & 91.029 & 91.148 & 0.000 & 18.565 & 1.016 & & 84.651 & 84.761 & 0.000 & 17.102 & 1.007 & & 62.735 & 62.802 & 0.000 & 11.602 & 1.017\\
\hspace{2mm} FW & 0.203 & 5.406 & 95.536 & 21.575 & 1.018 & & 0.015 & 5.762 & 95.933 & 22.988 & 1.017 & & 0.090 & 5.638 & 95.437 & 22.508 & 1.018 & & 0.231 & 5.039 & 95.337 & 20.205 & 1.023\\
\hline 
\multicolumn{24}{l}{\textbf{Non-robust method}}\\ 
\hline 
\multicolumn{24}{l}{Model specification: QR--True}\\ 
\hline 
\hspace{2mm} IPSW & -3.783 & 15.498 & 90.179 & 62.509 & 1.061 & & -3.878 & 15.620 & 89.782 & 62.778 & 1.058 & & -3.599 & 15.118 & 90.675 & 60.239 & 1.046 & & -3.782 & 14.133 & 91.964 & 57.700 & 1.080\\
\hline 
\multicolumn{24}{l}{Model specification: QR--False}\\ 
\hline 
\hspace{2mm} IPSW & 52.198 & 52.687 & 0.000 & 28.997 & 1.032 & & 49.275 & 50.008 & 0.000 & 34.156 & 1.021 & & 47.956 & 48.604 & 0.000 & 31.095 & 1.002 & & 50.472 & 50.655 & 0.000 & 17.591 & 1.042\\
\hline 
\multicolumn{24}{l}{\textbf{Doubly robust methods}}\\ 
\hline 
\multicolumn{24}{l}{Model specification: QR--True, PM--True}\\ 
\hline 
\hspace{2mm} GPPP & -0.261 & 9.930 & 94.643 & 40.048 & 1.029 & & -0.448 & 11.556 & 93.552 & 46.462 & 1.026 & & -0.204 & 10.641 & 94.444 & 43.177 & 1.035 & & -0.335 & 8.120 & 93.056 & 32.132 & 1.010\\
\hspace{2mm} PSPP & -0.276 & 9.928 & 94.643 & 40.018 & 1.028 & & -0.482 & 11.562 & 93.452 & 46.369 & 1.024 & & -0.172 & 10.636 & 94.643 & 43.046 & 1.032 & & -0.314 & 8.105 & 93.452 & 32.015 & 1.008\\
\hspace{2mm} LWP & -0.010 & 9.904 & 94.940 & 39.981 & 1.029 & & -0.272 & 11.604 & 93.552 & 46.460 & 1.021 & & -0.072 & 10.646 & 94.643 & 43.106 & 1.032 & & 0.617 & 8.144 & 94.345 & 32.321 & 1.015\\
\hspace{2mm} AIPW & -0.313 & 9.942 & 94.544 & 40.021 & 1.027 & & -0.535 & 11.589 & 93.452 & 46.411 & 1.022 & & -0.346 & 10.660 & 94.444 & 42.987 & 1.029 & & -0.396 & 8.128 & 93.353 & 32.166 & 1.010\\
\hline 
\multicolumn{24}{l}{Model specification: QR--True, PM--False}\\ 
\hline 
\hspace{2mm} GPPP & 0.112 & 9.911 & 94.544 & 40.299 & 1.037 & & -0.195 & 11.524 & 94.048 & 46.627 & 1.032 & & -0.101 & 10.628 & 93.750 & 43.280 & 1.038 & & 0.391 & 8.149 & 93.651 & 32.036 & 1.004\\
\hspace{2mm} PSPP & -0.165 & 9.926 & 94.444 & 40.142 & 1.031 & & -0.429 & 11.490 & 93.552 & 46.458 & 1.032 & & -0.299 & 10.611 & 93.849 & 43.080 & 1.036 & & -0.240 & 8.113 & 94.147 & 31.929 & 1.004\\
\hspace{2mm} LWP & -0.258 & 9.942 & 94.345 & 40.075 & 1.028 & & -0.475 & 11.489 & 93.849 & 46.395 & 1.031 & & -0.297 & 10.567 & 93.948 & 42.907 & 1.036 & & -0.094 & 8.127 & 93.849 & 31.970 & 1.003\\
\hspace{2mm} AIPW & -0.210 & 9.956 & 93.552 & 39.995 & 1.025 & & -0.298 & 11.469 & 93.849 & 46.334 & 1.030 & & -0.230 & 10.601 & 94.147 & 42.867 & 1.031 & & -0.412 & 8.136 & 93.651 & 31.864 & 1.000\\
\hline 
\multicolumn{24}{l}{Model specification: QR--False, PM--True}\\ 
\hline 
\hspace{2mm} GPPP & 0.021 & 8.403 & 93.849 & 33.606 & 1.020 & & 0.028 & 8.886 & 94.246 & 35.807 & 1.027 & & 0.084 & 8.297 & 94.643 & 33.793 & 1.039 & & -0.331 & 7.326 & 93.452 & 29.197 & 1.017\\
\hspace{2mm} PSPP & 2.302 & 8.643 & 94.841 & 33.926 & 1.038 & & 2.340 & 9.163 & 93.948 & 36.205 & 1.042 & & 2.233 & 8.580 & 94.643 & 34.188 & 1.052 & & 1.294 & 7.406 & 93.452 & 29.067 & 1.016\\
\hspace{2mm} LWP & -0.093 & 8.411 & 93.651 & 33.554 & 1.017 & & -0.196 & 8.926 & 93.948 & 35.751 & 1.021 & & -0.119 & 8.295 & 94.345 & 33.557 & 1.032 & & -0.269 & 7.155 & 92.659 & 28.604 & 1.020\\
\hspace{2mm} AIPW & -4.750 & 18.998 & 94.742 & 77.083 & 1.069 & & -5.189 & 20.238 & 95.139 & 81.947 & 1.068 & & -4.784 & 18.933 & 94.643 & 77.221 & 1.075 & & -3.902 & 14.639 & 92.560 & 60.317 & 1.090\\
\hline 
\multicolumn{24}{l}{Model specification: QR--False, PM--False}\\ 
\hline 
\hspace{2mm} GPPP & 46.307 & 46.727 & 0.000 & 24.810 & 1.012 & & 43.480 & 44.046 & 0.000 & 28.364 & 1.027 & & 42.571 & 43.072 & 0.000 & 26.363 & 1.027 & & 46.321 & 46.541 & 0.000 & 18.121 & 1.023\\
\hspace{2mm} PSPP & 46.282 & 46.703 & 0.000 & 24.788 & 1.009 & & 43.452 & 44.019 & 0.000 & 28.360 & 1.026 & & 42.547 & 43.049 & 0.000 & 26.356 & 1.026 & & 46.336 & 46.552 & 0.000 & 17.919 & 1.018\\
\hspace{2mm} LWP & 46.391 & 46.808 & 0.000 & 24.587 & 1.005 & & 43.422 & 43.994 & 0.000 & 28.379 & 1.023 & & 42.579 & 43.081 & 0.000 & 26.285 & 1.022 & & 48.301 & 48.464 & 0.000 & 16.010 & 1.029\\
\hspace{2mm} AIPW & 46.885 & 47.282 & 0.000 & 24.089 & 1.005 & & 43.817 & 44.373 & 0.000 & 28.079 & 1.023 & & 43.130 & 43.613 & 0.000 & 25.882 & 1.020 & & 49.826 & 49.990 & 0.000 & 16.533 & 1.041\\
\bottomrule
\end{tabular}}
  \begin{tablenotes}
   \footnotesize
   \item UW: Unweighted; FW: Fully weighted; GPPP: Gaussian Process of Propensity Prediction; PSPP: Penalized Spline of Propensity Prediction; LWP: Linear-in-weight prediction; AIPW: Augmented Inverse Propensity Weighting;
  \end{tablenotes}
 \end{threeparttable}
\end{sidewaystable}

\begin{sidewaystable}[hbt!]
\centering
\caption{Comparing the performance of the bias adjustment methods in the simulation study for the \emph{binary} outcome with $(n_A, n_R)=(500, 500)$ and $\gamma_1=0.6$}\label{tab:5.12}
\begin{threeparttable}
\scriptsize{\begin{tabular}{l l l l l l l l l l l l l l l l l l l l l l l l l}
\toprule
& \multicolumn{5}{c}{\textbf{$LIN$}} & & \multicolumn{5}{c}{\textbf{$CUB$}} & & \multicolumn{5}{c}{\textbf{$EXP$}} & & \multicolumn{5}{c}{\textbf{$SIN$}}\\\cline{2-6}\cline{8-12}\cline{14-18}\cline{20-24}
\textbf{Measure} & rBias & rMSE & crCI & lCI & rSE & & rBias & rMSE & crCI & lCI & rSE & & rBias & rMSE & crCI & lCI & rSE & & rBias & rMSE & crCI & lCI & rSE \\
\midrule
\multicolumn{24}{l}{\textbf{Probability sample ($S_R$)}}\\ 
\hline 
\hspace{2mm} UW & -41.047 & 41.487 & 0.000 & 19.213 & 0.813 & & -44.574 & 45.028 & 0.000 & 20.730 & 0.828 & & -42.945 & 43.406 & 0.000 & 20.084 & 0.812 & & -34.898 & 35.315 & 0.000 & 17.395 & 0.820\\
\hspace{2mm} FW & -0.178 & 10.901 & 93.651 & 42.539 & 0.995 & & -0.620 & 12.474 & 92.560 & 50.979 & 1.043 & & -0.508 & 11.640 & 93.750 & 47.428 & 1.040 & & -0.302 & 8.481 & 93.651 & 33.327 & 1.003\\
\hline 
\multicolumn{24}{l}{\textbf{Non-probability sample ($S_A$)}}\\ 
\hline 
\hspace{2mm} UW & 86.167 & 86.338 & 0.000 & 20.986 & 0.985 & & 92.869 & 93.119 & 0.000 & 26.171 & 0.979 & & 86.198 & 86.428 & 0.000 & 24.101 & 0.977 & & 63.186 & 63.318 & 0.000 & 16.359 & 1.019\\
\hspace{2mm} FW & 0.257 & 7.843 & 94.643 & 30.712 & 0.999 & & 0.106 & 8.183 & 96.032 & 32.740 & 1.020 & & 0.146 & 8.318 & 93.948 & 32.044 & 0.982 & & 0.248 & 7.226 & 95.139 & 28.741 & 1.015\\
\hline 
\multicolumn{24}{l}{\textbf{Non-robust method}}\\ 
\hline 
\multicolumn{24}{l}{Model specification: QR--True}\\ 
\hline 
\hspace{2mm} IPSW & -5.218 & 16.959 & 90.675 & 68.487 & 1.082 & & -5.192 & 17.348 & 90.873 & 69.336 & 1.068 & & -5.013 & 16.767 & 90.079 & 66.988 & 1.068 & & -5.259 & 15.947 & 91.270 & 63.328 & 1.073\\
\hline 
\multicolumn{24}{l}{Model specification: QR--False}\\ 
\hline 
\hspace{2mm} IPSW & 52.213 & 52.937 & 0.000 & 34.832 & 1.018 & & 49.369 & 50.447 & 0.397 & 40.870 & 1.005 & & 47.969 & 48.914 & 0.099 & 37.609 & 1.002 & & 50.477 & 50.805 & 0.000 & 23.045 & 1.019\\
\hline 
\multicolumn{24}{l}{\textbf{Doubly robust methods}}\\ 
\hline 
\multicolumn{24}{l}{Model specification: QR--True, PM--True}\\ 
\hline 
\hspace{2mm} GPPP & -0.041 & 10.399 & 92.857 & 41.628 & 1.021 & & -0.195 & 12.013 & 93.651 & 48.898 & 1.038 & & 0.067 & 11.263 & 94.048 & 45.943 & 1.040 & & -0.094 & 8.322 & 94.048 & 33.008 & 1.011\\
\hspace{2mm} PSPP & -0.071 & 10.392 & 92.857 & 41.589 & 1.020 & & -0.253 & 12.007 & 92.956 & 48.684 & 1.034 & & 0.042 & 11.238 & 94.643 & 45.765 & 1.038 & & -0.151 & 8.278 & 93.849 & 32.812 & 1.011\\
\hspace{2mm} LWP & 0.585 & 10.393 & 92.758 & 41.847 & 1.028 & & 0.224 & 12.129 & 93.849 & 49.125 & 1.033 & & 0.310 & 11.319 & 94.048 & 46.165 & 1.040 & & 1.629 & 8.595 & 93.750 & 33.723 & 1.019\\
\hspace{2mm} AIPW & -0.141 & 10.366 & 92.659 & 41.535 & 1.022 & & -0.401 & 12.071 & 93.254 & 48.666 & 1.029 & & -0.318 & 11.284 & 94.345 & 45.620 & 1.031 & & -0.274 & 8.394 & 93.452 & 33.054 & 1.005\\
\hline 
\multicolumn{24}{l}{Model specification: QR--True, PM--False}\\ 
\hline 
\hspace{2mm} GPPP & 0.595 & 10.442 & 94.544 & 42.427 & 1.038 & & 0.203 & 12.135 & 93.452 & 49.055 & 1.031 & & 0.345 & 11.208 & 93.948 & 46.329 & 1.054 & & 0.882 & 8.415 & 93.353 & 33.316 & 1.015\\
\hspace{2mm} PSPP & 0.019 & 10.409 & 93.948 & 42.033 & 1.030 & & -0.266 & 12.046 & 93.651 & 48.487 & 1.027 & & -0.131 & 11.131 & 93.750 & 45.665 & 1.046 & & -0.187 & 8.351 & 92.956 & 32.960 & 1.007\\
\hspace{2mm} LWP & -0.122 & 10.423 & 93.750 & 41.834 & 1.023 & & -0.303 & 12.042 & 93.353 & 48.357 & 1.024 & & -0.183 & 11.103 & 93.849 & 45.407 & 1.043 & & 0.162 & 8.377 & 93.155 & 33.080 & 1.007\\
\hspace{2mm} AIPW & -0.140 & 10.431 & 93.353 & 41.598 & 1.017 & & -0.075 & 12.003 & 93.254 & 48.215 & 1.024 & & -0.019 & 11.115 & 93.948 & 45.235 & 1.038 & & -0.407 & 8.372 & 92.560 & 32.762 & 0.999\\
\hline 
\multicolumn{24}{l}{Model specification: QR--False, PM--True}\\ 
\hline 
\hspace{2mm} GPPP & 0.199 & 8.950 & 94.643 & 36.139 & 1.030 & & 0.267 & 9.903 & 94.643 & 39.802 & 1.025 & & 0.346 & 9.341 & 94.246 & 37.453 & 1.023 & & -0.179 & 7.651 & 93.750 & 30.244 & 1.008\\
\hspace{2mm} PSPP & 4.124 & 9.791 & 93.948 & 37.276 & 1.070 & & 4.632 & 11.070 & 94.048 & 40.979 & 1.039 & & 4.346 & 10.435 & 93.452 & 38.781 & 1.042 & & 2.419 & 7.961 & 93.849 & 30.197 & 1.015\\
\hspace{2mm} LWP & -0.042 & 8.924 & 94.147 & 35.868 & 1.025 & & -0.202 & 9.907 & 94.147 & 39.346 & 1.013 & & -0.022 & 9.262 & 93.948 & 36.989 & 1.018 & & -0.209 & 7.591 & 93.651 & 29.763 & 1.000\\
\hspace{2mm} AIPW & -6.490 & 21.428 & 94.643 & 87.422 & 1.092 & & -6.913 & 23.067 & 95.933 & 93.312 & 1.081 & & -6.310 & 21.508 & 95.536 & 87.184 & 1.081 & & -5.406 & 16.466 & 92.262 & 66.288 & 1.087\\
\hline 
\multicolumn{24}{l}{Model specification: QR--False, PM--False}\\ 
\hline 
\hspace{2mm} GPPP & 46.582 & 47.202 & 0.000 & 29.649 & 0.991 & & 43.813 & 44.673 & 0.000 & 34.430 & 1.006 & & 42.785 & 43.578 & 0.099 & 32.147 & 0.990 & & 46.636 & 46.972 & 0.000 & 22.280 & 1.013\\
\hspace{2mm} PSPP & 46.553 & 47.174 & 0.000 & 29.601 & 0.989 & & 43.779 & 44.640 & 0.000 & 34.406 & 1.006 & & 42.753 & 43.547 & 0.099 & 32.117 & 0.989 & & 46.536 & 46.875 & 0.000 & 22.174 & 1.005\\
\hspace{2mm} LWP & 46.613 & 47.232 & 0.000 & 29.416 & 0.984 & & 43.758 & 44.630 & 0.000 & 34.367 & 0.998 & & 42.752 & 43.549 & 0.099 & 32.022 & 0.985 & & 48.424 & 48.721 & 0.000 & 21.085 & 1.002\\
\hspace{2mm} AIPW & 47.053 & 47.652 & 0.000 & 29.122 & 0.985 & & 44.101 & 44.957 & 0.000 & 34.142 & 0.997 & & 43.241 & 44.019 & 0.099 & 31.741 & 0.983 & & 49.864 & 50.171 & 0.000 & 21.994 & 1.012\\
\bottomrule
\end{tabular}}
  \begin{tablenotes}
   \footnotesize
   \item UW: Unweighted; FW: Fully weighted; GPPP: Gaussian Process of Propensity Prediction; PSPP: Penalized Spline of Propensity Prediction; LWP: Linear-in-weight prediction; AIPW: Augmented Inverse Propensity Weighting;
  \end{tablenotes}
 \end{threeparttable}
\end{sidewaystable}

\newpage
\clearpage

\begin{table}[hbt!]
\subsection{Supplemental results on CIREN/CDS data}\label{S:5.6.2}
\centering\caption{Percentage of severe injury in HEAD and associated 95\% CIs by different covariates across DR adjustment methods}\label{tab:5.13}
\begin{threeparttable}
\scriptsize{\begin{tabular}{l l l l l l}
\toprule
\textbf{Covariate} & \textbf{n} & \textbf{Unweighted (95\% CI)} & \textbf{GPPP (95\% CI)} & \textbf{LWP (95\% CI)} & \textbf{AIPW (95\% CI)}\\
\midrule
\textbf{Total} & 6,271 & 25.036(23.964,26.108) & 9.552(7.33,11.774) & 10.36(4.559,16.161) & 9.658(6.398,12.918)\\
\hline
\multicolumn{3}{l}{\textbf{Gender}} & & & \\ 
\hspace{2mm} Male & 3,230 & 28.359(26.805,29.914) & 11.464(7.982,14.947) & 12.526(4.69,20.361) & 10.81(6.382,15.239)\\
\hspace{2mm} Female & 3,041 & 21.506(20.046,22.966) & 8.245(5.697,10.794) & 8.882(3.84,13.924) & 8.918(4.303,13.533)\\
\hline
\multicolumn{3}{l}{\textbf{Age group}} & & & \\ 
\hspace{2mm} 16-19 & 659 & 32.625(29.046,36.205) & 9.498(4.117,14.878) & 10.874(0.606,21.143) & 8.557(0,18.879)\\
\hspace{2mm} 20-39 & 2,732 & 26.354(24.702,28.006) & 10.292(7.258,13.326) & 11.027(5.256,16.799) & 10.137(4.987,15.286)\\
\hspace{2mm} 40-64 & 1,946 & 20.863(19.058,22.669) & 8.527(5.688,11.365) & 9.442(2.439,16.445) & 9.222(3.92,14.525)\\
\hspace{2mm} 65+ & 934 & 24.518(21.759,27.277) & 9.999(6.371,13.626) & 10.43(5.369,15.491) & 10.024(1.537,18.511)\\
\hline
\multicolumn{3}{l}{\textbf{Occupant role}} & & & \\ 
\hspace{2mm} Driver & 1,547 & 32.515(30.18,34.849) & 9.681(4.677,14.685) & 10.958(0.535,21.382) & 9.924(0,20.727)\\
\hspace{2mm} Passenger & 4,724 & 22.587(21.394,23.779) & 9.594(7.445,11.743) & 10.296(5.322,15.27) & 9.614(6.267,12.961)\\
\hline
\multicolumn{3}{l}{\textbf{Seating Row}} & & & \\ 
\hspace{2mm} Front & 5,853 & 24.056(22.961,25.151) & 9.4(7.326,11.474) & 10.125(4.791,15.458) & 9.41(6.121,12.699)\\
\hspace{2mm} Rear & 418 & 38.756(34.086,43.426) & 12.632(0.87,24.394) & 15.661(0,36.448) & 14.995(3.168,26.822)\\
\hline
\multicolumn{3}{l}{\textbf{Restraint use}} & & & \\ 
\hspace{2mm} Restraint & 2,275 & 32.967(31.035,34.899) & 15.351(9.796,20.906) & 17.135(4.552,29.718) & 14.327(7.613,21.042)\\
\hspace{2mm} Unrestraint & 3,996 & 20.521(19.268,21.773) & 7.01(4.926,9.094) & 7.346(4.113,10.579) & 7.34(3.457,11.222)\\
\hline
\multicolumn{3}{l}{\textbf{Air bag}} & & & \\ 
\hspace{2mm} Deployed & 2,194 & 30.447(28.521,32.372) & 10.956(5.005,16.907) & 13.067(0.993,25.14) & 13.113(6.087,20.138)\\
\hspace{2mm} Not deployed & 4,077 & 22.124(20.85,23.398) & 8.995(6.466,11.524) & 9.057(6.161,11.953) & 7.804(3.654,11.955)\\
\hline
\multicolumn{3}{l}{\textbf{Injury scale (AIS)}} & & & \\ 
\hspace{2mm} 3 & 1,977 & 18.412(16.703,20.12) & 13.609(9.765,17.453) & 13.61(8.5,18.719) & 12.903(8.281,17.525)\\
\hspace{2mm} 4 & 761 & 55.979(52.452,59.506) & 31.675(24.511,38.839) & 31.549(23.938,39.161) & 33.75(21.739,45.761)\\
\hspace{2mm} 5+ & 781 & 76.056(73.063,79.049) & 59.405(41.668,77.143) & 59.683(37.787,81.578) & 59.896(39.338,80.453)\\
\hline
\multicolumn{3}{l}{\textbf{Days hospitalized}} & & & \\ 
\hspace{2mm} 0 & 2,259 & 26.737(24.912,28.563) & 2.633(0,11.319) & 3.842(0,16.924) & 4.406(0,18.412)\\
\hspace{2mm} 1-3 & 1,757 & 17.758(15.971,19.544) & 8.564(4.978,12.15) & 8.619(4.918,12.32) & 9.408(5.25,13.566)\\
\hspace{2mm} 4-7 & 1,153 & 20.902(18.555,23.249) & 10.174(6.976,13.371) & 9.981(6.845,13.117) & 10.826(6.88,14.771)\\
\hspace{2mm} 8-14 & 640 & 28.438(24.943,31.932) & 16.411(11.916,20.907) & 16.313(11.886,20.74) & 17.117(11.688,22.547)\\
\hspace{2mm} 15+ & 462 & 50(45.441,54.559) & 27.032(21.157,32.907) & 26.964(21.18,32.747) & 23.857(17.808,29.905)\\
\hline
\multicolumn{3}{l}{\textbf{Delta-v (Km/h)}} & & & \\ 
\hspace{2mm} Minor & 404 & 14.109(10.714,17.503) & 9.347(0.209,18.485) & 11.878(0,26.253) & 17.364(1.382,33.346)\\
\hspace{2mm} Moderate & 3,304 & 19.613(18.259,20.967) & 7.748(5.154,10.343) & 8.597(2.33,14.864) & 6.326(2.517,10.135)\\
\hspace{2mm} Severe & 2,563 & 33.75(31.919,35.58) & 15.234(11.167,19.301) & 15.395(10.571,20.219) & 15.357(8.964,21.749)\\
\hline
\multicolumn{3}{l}{\textbf{Damage distribution}} & & & \\ 
\hspace{2mm} Wide & 4,657 & 23.148(21.937,24.359) & 8.304(6.187,10.421) & 9.104(3.819,14.389) & 7.276(3.906,10.647)\\
\hspace{2mm} Narrow & 320 & 27.812(22.903,32.722) & 10.447(1.981,18.912) & 11.33(0,23.056) & 11.031(4.628,17.434)\\
\hspace{2mm} Corner & 584 & 20.205(16.949,23.462) & 12.282(7.027,17.537) & 12.667(6.537,18.798) & 14.807(3.911,25.702)\\
\hspace{2mm} Other & 710 & 40.141(36.535,43.746) & 15.842(6.849,24.835) & 17.071(3.194,30.949) & 21.066(6.792,35.341)\\
\hline
\multicolumn{3}{l}{\textbf{Rollover}} & & & \\ 
\hspace{2mm} Yes & 4,942 & 21.833(20.681,22.985) & 8.566(6.467,10.665) & 9.295(4.417,14.173) & 7.526(4.333,10.719)\\
\hspace{2mm} No & 1,329 & 36.945(34.35,39.54) & 12.579(6.805,18.354) & 13.867(2.222,25.512) & 17.555(8.204,26.906)\\
\hline
\multicolumn{3}{l}{\textbf{Deformation location}} & & & \\ 
\hspace{2mm} Front & 3,671 & 16.78(15.571,17.989) & 7.353(4.777,9.93) & 7.905(3.188,12.621) & 6.853(2.596,11.11)\\
\hspace{2mm} Right & 1,123 & 32.591(29.85,35.333) & 12.15(8.516,15.783) & 13.916(3.853,23.979) & 10.703(5.523,15.883)\\
\hspace{2mm} Left & 857 & 39.44(36.168,42.712) & 13.239(1.592,24.887) & 14.049(1.539,26.559) & 16.421(2.379,30.463)\\
\hspace{2mm} Rear & 620 & 40.323(36.461,44.184) & 15.445(7.088,23.801) & 16.007(4.144,27.869) & 17.129(2.342,31.916)\\
\hline
\multicolumn{3}{l}{\textbf{Model year}} & & & \\ 
\hspace{2mm} $\leq$2003 & 2,529 & 28.035(26.284,29.785) & 13.405(8.529,18.282) & 14.961(2.984,26.938) & 17.958(10.59,25.325)\\
\hspace{2mm} 2004-2007 & 2,377 & 24.443(22.715,26.17) & 10.117(7.061,13.174) & 10.558(5.971,15.146) & 7.776(4.489,11.063)\\
\hspace{2mm} 2008-2011 & 1,019 & 20.51(18.031,22.989) & 3.33(0,8.253) & 3.656(0,9.11) & 1.094(0,7.874)\\
\hspace{2mm} 2012+ & 346 & 20.52(16.265,24.776) & 8.831(3.093,14.569) & 9.234(3.369,15.1) & 3.904(0,11.404)\\
\hline
\multicolumn{3}{l}{\textbf{Vehicle make}} & & & \\ 
\hspace{2mm} American & 3,635 & 25.777(24.355,27.199) & 10.091(6.99,13.192) & 11.175(3.082,19.268) & 9.535(4.856,14.213)\\
\hspace{2mm} Japanese & 1,927 & 24.754(22.827,26.68) & 9.997(6.843,13.15) & 10.479(5.934,15.025) & 12.601(7.211,17.991)\\
\hspace{2mm} Korean & 387 & 21.189(17.117,25.26) & 5.661(0,12.117) & 6.576(0,16.798) & 5.023(0,12.299)\\
\hspace{2mm} Other & 322 & 22.981(18.386,27.577) & 5.801(0,13.635) & 5.954(0,13.893) & 4.832(0,11.801)\\
\hline
\multicolumn{3}{l}{\textbf{Vehicle type}} & & & \\ 
\hspace{2mm} Passenger car & 4,063 & 25.966(24.618,27.314) & 9.167(6.918,11.415) & 9.928(4.629,15.228) & 7.384(3.613,11.154)\\
\hspace{2mm} Light truck & 1,213 & 22.341(19.997,24.685) & 9.113(5.987,12.239) & 9.52(3.705,15.335) & 13.447(6.271,20.623)\\
\hspace{2mm} Van & 290 & 27.241(22.117,32.365) & 4.829(0,10.283) & 6.556(0,17.997) & 7.793(0,18.304)\\
\hspace{2mm} SUV & 705 & 23.404(20.279,26.53) & 14.224(9.007,19.441) & 15.617(5.833,25.401) & 16.67(6.355,26.985)\\
\bottomrule
\end{tabular}}
    \begin{tablenotes}
      \small
      \item
    \end{tablenotes}
  \end{threeparttable}
\end{table}

\begin{table}[hbt!]
\centering\caption{Percentage of severe injury in ABDOMEN and associated 95\% CIs by different covariates across DR adjustment methods}\label{tab:5.14}
\begin{threeparttable}
\scriptsize{\begin{tabular}{l l l l l l}
\toprule
\textbf{Covariate} & \textbf{n} & \textbf{Unweighted  (95\% CI)} & \textbf{GPPP  (95\% CI)} & \textbf{LWP  (95\% CI)} & \textbf{AIPW  (95\% CI)}\\
\midrule 
\textbf{Total} & 6,271 & 9.919(9.179,10.658) & 6.677(4.408,8.946) & 7.621(1.93,13.313) & 6.479(2.354,10.604)\\
\hline 
\multicolumn{3}{l}{\textbf{Gender}} & & & \\ 
\hspace{2mm} Male & 3,230 & 10.464(9.409,11.52) & 7.694(4.134,11.253) & 8.837(1.872,15.802) & 3.552(0,11.317)\\
\hspace{2mm} Female & 3,041 & 9.339(8.305,10.373) & 5.931(3.485,8.376) & 6.724(1.411,12.036) & 8.852(3.96,13.744)\\
\hline 
\multicolumn{3}{l}{\textbf{Age group}} & & & \\ 
\hspace{2mm} 16-19 & 659 & 11.836(9.37,14.302) & 6.592(2.583,10.601) & 7.886(0,16.685) & 0(0,22.12)\\
\hspace{2mm} 20-39 & 2,732 & 10.505(9.355,11.655) & 6.8(4.066,9.535) & 7.782(1.873,13.691) & 7.539(1.553,13.524)\\
\hspace{2mm} 40-64 & 1,946 & 8.941(7.674,10.209) & 7.109(4.012,10.205) & 8.157(1.388,14.926) & 7.149(3.55,10.749)\\
\hspace{2mm} 65+ & 934 & 8.887(7.062,10.711) & 5.75(2.911,8.589) & 6.179(2.056,10.301) & 8.945(0.561,17.329)\\
\hline 
\multicolumn{3}{l}{\textbf{Occupant role}} & & & \\ 
\hspace{2mm} Driver & 1,547 & 12.864(11.195,14.532) & 8.85(4.445,13.255) & 10.308(1.336,19.28) & 13.363(2.809,23.917)\\
\hspace{2mm} Passenger & 4,724 & 8.954(8.14,9.768) & 6.1(3.73,8.47) & 6.921(1.726,12.116) & 4.778(0,9.568)\\
\hline 
\multicolumn{3}{l}{\textbf{Seating Row}} & & & \\ 
\hspace{2mm} Front & 5,853 & 9.465(8.715,10.215) & 6.414(4.287,8.541) & 7.28(2.052,12.507) & 6.264(2.088,10.44)\\
\hspace{2mm} Rear & 418 & 16.268(12.73,19.806) & 13.822(2.075,25.57) & 16.764(0,38.476) & 12.916(3.318,22.514)\\
\hline 
\multicolumn{3}{l}{\textbf{Restraint use}} & & & \\ 
\hspace{2mm} Restraint & 2,275 & 11.648(10.33,12.967) & 8.836(4.185,13.488) & 10.761(0,22.227) & 13.03(5.326,20.733)\\
\hspace{2mm} Unrestraint & 3,996 & 8.934(8.05,9.818) & 5.629(3.583,7.675) & 6.103(2.64,9.565) & 2.976(0,8.312)\\
\hline 
\multicolumn{3}{l}{\textbf{Air bag}} & & & \\ 
\hspace{2mm} Deployed & 2,194 & 10.255(8.986,11.525) & 5.091(0,10.404) & 6.952(0,19.7) & 0.928(0,8.901)\\
\hspace{2mm} Not deployed & 4,077 & 9.738(8.828,10.648) & 7.718(5.486,9.95) & 7.907(5.302,10.512) & 10.032(5.212,14.852)\\
\hline 
\multicolumn{3}{l}{\textbf{Injury scale (AIS)}} & & & \\ 
\hspace{2mm} 3 & 1,977 & 7.081(5.951,8.212) & 9.233(5.979,12.487) & 9.28(5.232,13.327) & 10.554(5.247,15.861)\\
\hspace{2mm} 4 & 761 & 21.682(18.754,24.61) & 27.838(18.134,37.542) & 28.076(19.007,37.146) & 24.953(13.553,36.352)\\
\hspace{2mm} 5+ & 781 & 33.291(29.986,36.596) & 35.512(15.655,55.37) & 37.705(13.063,62.348) & 31.665(5.388,57.942)\\
\hline 
\multicolumn{3}{l}{\textbf{Days hospitalized}} & & & \\ 
\hspace{2mm} 0 & 2,259 & 11.554(10.236,12.872) & 13.306(4.864,21.747) & 16.519(2.786,30.252) & 6.87(0,26.651)\\
\hspace{2mm} 1-3 & 1,757 & 4.098(3.171,5.025) & 3.898(0.778,7.017) & 3.158(0.132,6.183) & 3.129(0.835,5.424)\\
\hspace{2mm} 4-7 & 1,153 & 9.193(7.526,10.861) & 7.787(4.405,11.169) & 7.487(4.375,10.598) & 9.753(5.341,14.166)\\
\hspace{2mm} 8-14 & 640 & 15.156(12.378,17.934) & 12.72(8.472,16.968) & 13.127(8.583,17.671) & 13.053(8.499,17.606)\\
\hspace{2mm} 15+ & 462 & 18.615(15.066,22.164) & 17.194(11.338,23.05) & 17.839(11.819,23.858) & 16.739(9.864,23.615)\\
\hline 
\multicolumn{3}{l}{\textbf{Delta-v (Km/h)}} & & & \\ 
\hspace{2mm} Minor & 404 & 4.208(2.25,6.166) & 6.388(1.47,11.306) & 8.723(0,21.201) & 3.851(0,10.325)\\
\hspace{2mm} Moderate & 3,304 & 5.72(4.928,6.512) & 4.55(2.512,6.588) & 5.446(0,11.202) & 4.565(1.801,7.329)\\
\hspace{2mm} Severe & 2,563 & 16.231(14.803,17.659) & 11.707(8.402,15.011) & 12.236(7.964,16.508) & 11.988(2.973,21.003)\\
\hline 
\multicolumn{3}{l}{\textbf{Damage distribution}} & & & \\ 
\hspace{2mm} Wide & 4,657 & 10.715(9.827,11.603) & 6.978(4.739,9.217) & 7.867(2.418,13.317) & 6.987(1.455,12.519)\\
\hspace{2mm} Narrow & 320 & 10.312(6.98,13.645) & 6.759(0.837,12.681) & 8.316(0,19.15) & 7.17(0,17.036)\\
\hspace{2mm} Corner & 584 & 8.219(5.992,10.447) & 9.113(3.785,14.441) & 9.514(3.29,15.737) & 9.046(1.952,16.139)\\
\hspace{2mm} Other & 710 & 5.915(4.18,7.651) & 2.464(0,6.822) & 4.08(0,13.134) & 0.848(0,3.768)\\
\hline 
\multicolumn{3}{l}{\textbf{Rollover}} & & & \\ 
\hspace{2mm} Yes & 4,942 & 9.996(9.16,10.832) & 7.036(4.712,9.36) & 7.864(2.674,13.054) & 9.018(5.334,12.703)\\
\hspace{2mm} No & 1,329 & 9.631(8.045,11.217) & 5.026(0.376,9.677) & 6.532(0,15.904) & 0(0,8.629)\\
\hline 
\multicolumn{3}{l}{\textbf{Deformation location}} & & & \\ 
\hspace{2mm} Front & 3,671 & 8.172(7.286,9.058) & 7.344(4.903,9.785) & 8.009(3.232,12.787) & 9.683(4.673,14.693)\\
\hspace{2mm} Right & 1,123 & 14.604(12.538,16.669) & 9.325(5.04,13.609) & 10.978(1.525,20.431) & 1.682(0,14.697)\\
\hspace{2mm} Left & 857 & 14.002(11.679,16.326) & 4.597(0,9.713) & 5.728(0,14.156) & 5.434(0,12)\\
\hspace{2mm} Rear & 620 & 6.129(4.241,8.017) & 2.253(0,6.802) & 3.513(0,11.383) & 1.418(0,5.151)\\
\hline 
\multicolumn{3}{l}{\textbf{Model year}} & & & \\ 
\hspace{2mm} $\leq$2003 & 2,529 & 10.874(9.661,12.087) & 5.716(1.128,10.303) & 6.679(0,16.721) & 1.923(0,10.724)\\
\hspace{2mm} 2004-2007 & 2,377 & 9.34(8.17,10.509) & 8.557(5.391,11.724) & 8.984(3.825,14.143) & 12.216(6.554,17.878)\\
\hspace{2mm} 2008-2011 & 1,019 & 10.01(8.167,11.853) & 5.989(1.648,10.33) & 7.077(2.546,11.607) & 6.495(0,13.18)\\
\hspace{2mm} 2012+ & 346 & 6.647(4.023,9.272) & 2.479(0,7.511) & 3.11(0,8.306) & 3.667(0,12.272)\\
\hline 
\multicolumn{3}{l}{\textbf{Vehicle make}} & & & \\ 
\hspace{2mm} American & 3,635 & 9.684(8.722,10.645) & 6.702(3.843,9.56) & 7.917(0.565,15.268) & 7.181(0.778,13.584)\\
\hspace{2mm} Japanese & 1,927 & 10.171(8.822,11.521) & 7.5(4.529,10.471) & 8.118(3.518,12.719) & 6.951(2.257,11.645)\\
\hspace{2mm} Korean & 387 & 9.561(6.631,12.49) & 5.316(2.03,8.602) & 6.271(0,13.168) & 0.991(0,7.926)\\
\hspace{2mm} Other & 322 & 11.491(8.007,14.974) & 3.667(0.61,6.723) & 4.116(0.173,8.06) & 1.609(0,4.707)\\
\hline 
\multicolumn{3}{l}{\textbf{Vehicle type}} & & & \\ 
\hspace{2mm} Passenger car & 4,063 & 11.149(10.182,12.117) & 6.76(4.635,8.885) & 7.677(2.315,13.039) & 9.024(4.753,13.294)\\
\hspace{2mm} Light truck & 1,213 & 7.997(6.47,9.523) & 7.318(3.225,11.411) & 8.158(1.289,15.028) & 0(0,13.862)\\
\hspace{2mm} Van & 290 & 8.276(5.105,11.447) & 2.859(0,7.441) & 4.148(0,13.173) & 5.129(0,12.484)\\
\hspace{2mm} SUV & 705 & 6.809(4.949,8.668) & 6.509(2.655,10.362) & 7.666(0,15.435) & 4.766(1.067,8.466)\\
\bottomrule
\end{tabular}}
    \begin{tablenotes}
      \small
      \item
    \end{tablenotes}
  \end{threeparttable}
\end{table}

\begin{table}[hbt!]
\centering\caption{Percentage of severe injury in THORAX and associated 95\% CIs by different covariates across DR adjustment methods}\label{tab:5:15}
\begin{threeparttable}
\scriptsize{\begin{tabular}{l l l l l l}
\toprule
\textbf{Covariate} & \textbf{n} & \textbf{Unweighted  (95\% CI)} & \textbf{GPPP  (95\% CI)} & \textbf{LWP  (95\% CI)} & \textbf{AIPW  (95\% CI)}\\
\midrule 
\textbf{Total} & 6,271 & 33.934(32.762,35.106) & 21.087(14.925,27.248) & 29.62(17.969,41.272) & 21.39(18.238,24.541)\\
\hline 
\multicolumn{3}{l}{\textbf{Gender}} & & & \\ 
\hspace{2mm} Male & 3,230 & 37.337(35.669,39.006) & 23.605(17.945,29.264) & 31.925(20.364,43.486) & 21.954(18.66,25.249)\\
\hspace{2mm} Female & 3,041 & 30.319(28.685,31.953) & 19.353(12.307,26.399) & 28.049(15.549,40.548) & 21.074(16.628,25.52)\\
\hline 
\multicolumn{3}{l}{\textbf{Age group}} & & & \\ 
\hspace{2mm} 16-19 & 659 & 37.329(33.636,41.022) & 17.62(6.253,28.988) & 34.815(9.195,60.436) & 21.41(13.6,29.219)\\
\hspace{2mm} 20-39 & 2,732 & 31.442(29.701,33.183) & 15.872(8.609,23.136) & 22.863(11.708,34.017) & 15.318(9.955,20.681)\\
\hspace{2mm} 40-64 & 1,946 & 32.066(29.992,34.139) & 22.877(16.442,29.312) & 31.709(18.647,44.771) & 21.864(16.993,26.734)\\
\hspace{2mm} 65+ & 934 & 42.719(39.547,45.892) & 35.715(29.256,42.175) & 41.792(31.461,52.122) & 38.108(31.117,45.098)\\
\hline 
\multicolumn{3}{l}{\textbf{Occupant role}} & & & \\ 
\hspace{2mm} Driver & 1,547 & 41.047(38.596,43.498) & 25.589(15.135,36.044) & 39.822(19.233,60.411) & 22.188(13.747,30.629)\\
\hspace{2mm} Passenger & 4,724 & 31.605(30.279,32.93) & 20.007(13.585,26.429) & 27.096(16.702,37.489) & 21.249(17.329,25.17)\\
\hline 
\multicolumn{3}{l}{\textbf{Seating Row}} & & & \\ 
\hspace{2mm} Front & 5,853 & 33.18(31.973,34.386) & 20.399(14.202,26.596) & 28.514(17.063,39.964) & 20.94(17.786,24.095)\\
\hspace{2mm} Rear & 418 & 44.498(39.733,49.262) & 40.118(25.011,55.225) & 59.85(33.897,85.803) & 34.6(20.286,48.914)\\
\hline 
\multicolumn{3}{l}{\textbf{Restraint use}} & & & \\ 
\hspace{2mm} Restraint & 2,275 & 39.253(37.246,41.259) & 25.078(15.779,34.377) & 40.585(21.1,60.069) & 28.13(21.873,34.387)\\
\hspace{2mm} Unrestraint & 3,996 & 30.906(29.473,32.339) & 18.765(12.572,24.958) & 23.818(14.563,33.074) & 17.835(14.413,21.256)\\
\hline 
\multicolumn{3}{l}{\textbf{Air bag}} & & & \\ 
\hspace{2mm} Deployed & 2,194 & 36.372(34.359,38.385) & 22.703(12.088,33.318) & 38.464(17.814,59.114) & 23.947(17.058,30.836)\\
\hspace{2mm} Not deployed & 4,077 & 32.622(31.183,34.061) & 20.517(14.113,26.92) & 24.339(14.383,34.294) & 20.319(15.73,24.908)\\
\hline 
\multicolumn{3}{l}{\textbf{Injury scale (AIS)}} & & & \\ 
\hspace{2mm} 3 & 1,977 & 42.742(40.561,44.922) & 44.551(39.909,49.194) & 45.565(41.221,49.909) & 44.497(38.612,50.382)\\
\hspace{2mm} 4 & 761 & 72.405(69.229,75.581) & 71.412(63.902,78.923) & 71.434(63.938,78.931) & 74.925(64.879,84.971)\\
\hspace{2mm} 5+ & 781 & 70.679(67.486,73.871) & 82.191(74.038,90.344) & 83.318(73.489,93.148) & 79.427(68.421,90.433)\\
\hline 
\multicolumn{3}{l}{\textbf{Days hospitalized}} & & & \\ 
\hspace{2mm} 0 & 2,259 & 30.102(28.21,31.993) & 16.397(3.115,29.679) & 33.637(10.187,57.087) & 20.958(15.035,26.88)\\
\hspace{2mm} 1-3 & 1,757 & 24.986(22.961,27.01) & 16.58(10.727,22.433) & 16.916(10.974,22.857) & 15.234(10.013,20.456)\\
\hspace{2mm} 4-7 & 1,153 & 35.386(32.626,38.146) & 30.08(23.956,36.203) & 30.835(24.826,36.843) & 34.479(27.631,41.326)\\
\hspace{2mm} 8-14 & 640 & 49.375(45.502,53.248) & 41.973(34.31,49.636) & 41.683(34.216,49.15) & 39.04(31.172,46.909)\\
\hspace{2mm} 15+ & 462 & 61.688(57.255,66.121) & 53.435(46.093,60.777) & 52.866(45.985,59.748) & 48.834(40.281,57.387)\\
\hline 
\multicolumn{3}{l}{\textbf{Delta-v (Km/h)}} & & & \\ 
\hspace{2mm} Minor & 404 & 19.059(15.229,22.889) & 11.995(1.156,22.834) & 30.193(8.357,52.029) & 8.541(0,20.813)\\
\hspace{2mm} Moderate & 3,304 & 25.938(24.444,27.433) & 17.681(10.649,24.712) & 27.582(13.704,41.46) & 16.779(13.611,19.947)\\
\hspace{2mm} Severe & 2,563 & 46.586(44.655,48.517) & 32.809(27.293,38.324) & 34.293(28.012,40.575) & 35.216(29.025,41.407)\\
\hline 
\multicolumn{3}{l}{\textbf{Damage distribution}} & & & \\ 
\hspace{2mm} Wide & 4,657 & 33.906(32.546,35.266) & 21.063(14.911,27.214) & 28.624(18.042,39.206) & 23.114(19.528,26.7)\\
\hspace{2mm} Narrow & 320 & 36.562(31.286,41.839) & 27.796(19.029,36.564) & 37.404(23.902,50.906) & 22.562(10.103,35.022)\\
\hspace{2mm} Corner & 584 & 26.884(23.288,30.479) & 17.923(10.176,25.67) & 24.838(10.202,39.473) & 15.681(5.905,25.458)\\
\hspace{2mm} Other & 710 & 38.732(35.149,42.316) & 22.155(10.491,33.82) & 38.694(15.227,62.161) & 16.249(4.619,27.879)\\
\hline 
\multicolumn{3}{l}{\textbf{Rollover}} & & & \\ 
\hspace{2mm} Yes & 4,942 & 32.436(31.131,33.741) & 20.109(13.803,26.415) & 27.518(16.931,38.105) & 20.769(16.931,24.608)\\
\hspace{2mm} No & 1,329 & 39.503(36.875,42.132) & 25.651(17.024,34.278) & 39.077(20.706,57.449) & 24.238(15.953,32.522)\\
\hline 
\multicolumn{3}{l}{\textbf{Deformation location}} & & & \\ 
\hspace{2mm} Front & 3,671 & 26.832(25.399,28.265) & 19.037(12.539,25.536) & 25.505(14.472,36.537) & 19.842(14.653,25.03)\\
\hspace{2mm} Right & 1,123 & 45.77(42.856,48.684) & 30.62(20.887,40.352) & 42.05(25.575,58.525) & 30.289(22.563,38.016)\\
\hspace{2mm} Left & 857 & 44.924(41.594,48.254) & 20.7(2.847,38.554) & 30.117(7.802,52.431) & 20.237(4.837,35.637)\\
\hspace{2mm} Rear & 620 & 39.355(35.509,43.2) & 21.911(8.05,35.771) & 37.303(12.58,62.027) & 20.181(7.315,33.046)\\
\hline 
\multicolumn{3}{l}{\textbf{Model year}} & & & \\ 
\hspace{2mm} $\leq$2003 & 2,529 & 35.389(33.526,37.253) & 24.05(16.137,31.962) & 37.778(20.449,55.108) & 22.382(17.842,26.922)\\
\hspace{2mm} 2004-2007 & 2,377 & 34.034(32.13,35.939) & 22.589(15.129,30.049) & 29.434(17.015,41.853) & 24.151(17.576,30.727)\\
\hspace{2mm} 2008-2011 & 1,019 & 31.403(28.554,34.253) & 15.693(6.189,25.198) & 20.036(6.962,33.11) & 14.735(5.492,23.979)\\
\hspace{2mm} 2012+ & 346 & 30.058(25.227,34.889) & 18.322(7.572,29.073) & 20.336(7.789,32.883) & 23.776(12.118,35.433)\\
\hline 
\multicolumn{3}{l}{\textbf{Vehicle make}} & & & \\ 
\hspace{2mm} American & 3,635 & 33.92(32.381,35.459) & 25.321(18.17,32.471) & 37.287(20.562,54.012) & 24.194(19.862,28.526)\\
\hspace{2mm} Japanese & 1,927 & 34.51(32.387,36.632) & 17.178(8.904,25.451) & 21.274(9.996,32.551) & 17.272(10.022,24.522)\\
\hspace{2mm} Korean & 387 & 32.558(27.89,37.227) & 21.338(10.317,32.358) & 30.873(11.583,50.162) & 25.404(14.335,36.473)\\
\hspace{2mm} Other & 322 & 32.298(27.191,37.406) & 11.894(0.98,22.807) & 13.358(0.14,26.576) & 14.575(5.943,23.206)\\
\hline 
\multicolumn{3}{l}{\textbf{Vehicle type}} & & & \\ 
\hspace{2mm} Passenger car & 4,063 & 35.565(34.093,37.037) & 20.78(14.442,27.117) & 28.72(17.684,39.755) & 21.834(18.109,25.559)\\
\hspace{2mm} Light truck & 1,213 & 31.904(29.281,34.527) & 20.445(10.611,30.28) & 28.5(12.712,44.289) & 20.931(14.712,27.149)\\
\hspace{2mm} Van & 290 & 26.897(21.793,32) & 25.071(10.311,39.832) & 39.315(14.396,64.234) & 19.598(6.469,32.727)\\
\hspace{2mm} SUV & 705 & 30.922(27.51,34.334) & 21.916(12.779,31.053) & 32.376(15.318,49.435) & 20.32(12.643,27.996)\\
\bottomrule
\end{tabular}}
    \begin{tablenotes}
      \small
      \item
    \end{tablenotes}
  \end{threeparttable}
\end{table}

\begin{table}[hbt!]
\centering\caption{Percentage of severe injury in SPINE and associated 95\% CIs by different covariates across DR adjustment methods}\label{tab:5.16}
\begin{threeparttable}
\scriptsize{\begin{tabular}{l l l l l l}
\toprule
\textbf{Covariate} & \textbf{n} & \textbf{Unweighted  (95\% CI)} & \textbf{GPPP  (95\% CI)} & \textbf{LWP  (95\% CI)} & \textbf{AIPW  (95\% CI)}\\
\midrule 
\textbf{Total} & 6,271 & 3.716(3.247,4.184) & 6.735(4.696,8.774) & 6.978(2.777,11.18) & 6.205(3.15,9.261)\\
\hline 
\multicolumn{3}{l}{\textbf{Gender}} & & & \\ 
\hspace{2mm} Male & 3,230 & 4.18(3.489,4.87) & 6.819(4.049,9.588) & 7.154(2.499,11.809) & 8.946(5.07,12.822)\\
\hspace{2mm} Female & 3,041 & 3.223(2.595,3.85) & 6.719(3.861,9.577) & 6.891(2.261,11.522) & 4.21(0,9.072)\\
\hline 
\multicolumn{3}{l}{\textbf{Age group}} & & & \\ 
\hspace{2mm} 16-19 & 659 & 3.49(2.089,4.891) & 1.823(0,4.007) & 2.358(0,9.212) & 0(0,3.847)\\
\hspace{2mm} 20-39 & 2,732 & 3.807(3.089,4.524) & 5.099(2.982,7.217) & 5.406(1.363,9.448) & 3.964(0.399,7.529)\\
\hspace{2mm} 40-64 & 1,946 & 3.443(2.633,4.253) & 7.68(5.417,9.943) & 7.948(3.292,12.603) & 9.501(5.524,13.479)\\
\hspace{2mm} 65+ & 934 & 4.176(2.893,5.458) & 12.795(7.935,17.655) & 12.616(7.457,17.775) & 11.194(1.223,21.164)\\
\hline 
\multicolumn{3}{l}{\textbf{Occupant role}} & & & \\ 
\hspace{2mm} Driver & 1,547 & 5.624(4.476,6.772) & 7.059(2.952,11.166) & 7.549(0,16.523) & 3.584(0,10.498)\\
\hspace{2mm} Passenger & 4,724 & 3.091(2.597,3.584) & 6.683(4.361,9.005) & 6.868(3.323,10.412) & 6.93(3.137,10.722)\\
\hline 
\multicolumn{3}{l}{\textbf{Seating Row}} & & & \\ 
\hspace{2mm} Front & 5,853 & 3.434(2.968,3.901) & 6.498(4.559,8.438) & 6.73(2.905,10.554) & 5.999(2.883,9.116)\\
\hspace{2mm} Rear & 418 & 7.656(5.107,10.204) & 12.752(4.811,20.692) & 13.399(0,29.785) & 9.724(1.107,18.341)\\
\hline 
\multicolumn{3}{l}{\textbf{Restraint use}} & & & \\ 
\hspace{2mm} Restraint & 2,275 & 4.308(3.473,5.142) & 5.92(1.798,10.042) & 6.805(0,16.123) & 6.475(0,14.4)\\
\hspace{2mm} Unrestraint & 3,996 & 3.378(2.818,3.939) & 6.766(4.721,8.812) & 6.797(4.24,9.353) & 6.268(3.289,9.247)\\
\hline 
\multicolumn{3}{l}{\textbf{Air bag}} & & & \\ 
\hspace{2mm} Deployed & 2,194 & 3.874(3.067,4.682) & 8.504(3.145,13.864) & 10.364(0.927,19.801) & 6.919(0,14.085)\\
\hspace{2mm} Not deployed & 4,077 & 3.63(3.056,4.204) & 5.876(4.41,7.343) & 5.826(4.01,7.641) & 5.954(3.153,8.756)\\
\hline 
\multicolumn{3}{l}{\textbf{Injury scale (AIS)}} & & & \\ 
\hspace{2mm} 3 & 1,977 & 5.513(4.507,6.52) & 17.854(14.154,21.553) & 17.101(12.953,21.249) & 18.924(12.225,25.623)\\
\hspace{2mm} 4 & 761 & 4.731(3.222,6.239) & 16.246(9.105,23.387) & 15.862(8.16,23.564) & 11.393(0.586,22.2)\\
\hspace{2mm} 5+ & 781 & 8.707(6.729,10.684) & 22.462(10.034,34.889) & 24.098(8.787,39.409) & 22.264(8.191,36.336)\\
\hline 
\multicolumn{3}{l}{\textbf{Days hospitalized}} & & & \\ 
\hspace{2mm} 0 & 2,259 & 2.745(2.071,3.418) & 3.815(0.804,6.826) & 4.92(0,12.6) & 0(0,6.667)\\
\hspace{2mm} 1-3 & 1,757 & 2.732(1.97,3.494) & 4.476(2.538,6.415) & 4.054(2.23,5.877) & 5.118(1.926,8.31)\\
\hspace{2mm} 4-7 & 1,153 & 4.51(3.312,5.708) & 13.038(9.061,17.015) & 12.105(8.546,15.665) & 13.629(8.565,18.692)\\
\hspace{2mm} 8-14 & 640 & 5.625(3.84,7.41) & 14.223(10.23,18.215) & 13.837(10.044,17.63) & 13.278(8.82,17.737)\\
\hspace{2mm} 15+ & 462 & 7.576(5.163,9.989) & 25.109(19.802,30.416) & 24.763(19.61,29.915) & 25.392(20.03,30.754)\\
\hline 
\multicolumn{3}{l}{\textbf{Delta-v (Km/h)}} & & & \\ 
\hspace{2mm} Minor & 404 & 2.228(0.789,3.667) & 11.229(1.136,21.322) & 12.934(0,26.013) & 8.643(0,30.853)\\
\hspace{2mm} Moderate & 3,304 & 2.542(2.006,3.079) & 6.269(3.382,9.156) & 6.597(1.454,11.741) & 6.322(3.002,9.642)\\
\hspace{2mm} Severe & 2,563 & 5.462(4.583,6.342) & 8.53(5.654,11.406) & 8.535(5.507,11.562) & 7.466(3.136,11.796)\\
\hline 
\multicolumn{3}{l}{\textbf{Damage distribution}} & & & \\ 
\hspace{2mm} Wide & 4,657 & 3.393(2.873,3.913) & 5.28(3.515,7.045) & 5.532(1.939,9.125) & 3.803(0.55,7.057)\\
\hspace{2mm} Narrow & 320 & 5(2.612,7.388) & 2.417(0,5.39) & 2.967(0,8.287) & 2.673(0,5.984)\\
\hspace{2mm} Corner & 584 & 1.884(0.781,2.986) & 7.107(2.759,11.454) & 7.097(2.231,11.963) & 15.586(0.739,30.432)\\
\hspace{2mm} Other & 710 & 6.761(4.914,8.607) & 18.668(10.843,26.494) & 19.005(6.446,31.564) & 17.857(7.203,28.512)\\
\hline 
\multicolumn{3}{l}{\textbf{Rollover}} & & & \\ 
\hspace{2mm} Yes & 4,942 & 3.157(2.669,3.644) & 5.129(3.248,7.01) & 5.354(1.836,8.872) & 4.668(1.216,8.119)\\
\hspace{2mm} No & 1,329 & 5.794(4.538,7.05) & 13.905(9.303,18.508) & 14.245(5.965,22.524) & 13.444(6.794,20.094)\\
\hline 
\multicolumn{3}{l}{\textbf{Deformation location}} & & & \\ 
\hspace{2mm} Front & 3,671 & 3.242(2.669,3.815) & 5.676(3.614,7.738) & 5.811(2.379,9.244) & 3.488(0,7.784)\\
\hspace{2mm} Right & 1,123 & 3.562(2.478,4.646) & 3.673(0,7.795) & 4.021(0,10.545) & 7.553(1.56,13.546)\\
\hspace{2mm} Left & 857 & 3.851(2.562,5.139) & 7.337(0.385,14.289) & 7.933(0.269,15.598) & 7.972(0,17.665)\\
\hspace{2mm} Rear & 620 & 6.613(4.657,8.569) & 22.12(12.372,31.869) & 22.254(9.529,34.98) & 20.896(8.037,33.755)\\
\hline 
\multicolumn{3}{l}{\textbf{Model year}} & & & \\ 
\hspace{2mm} $\leq$2003 & 2,529 & 3.598(2.872,4.324) & 7.792(5.022,10.561) & 8.376(0.769,15.983) & 4.647(0,10.046)\\
\hspace{2mm} 2004-2007 & 2,377 & 3.366(2.641,4.091) & 8.083(5.475,10.691) & 8.184(4.72,11.647) & 10.824(6.32,15.328)\\
\hspace{2mm} 2008-2011 & 1,019 & 4.514(3.239,5.789) & 3.349(1.139,5.558) & 3.365(1.098,5.633) & 4.816(0.267,9.366)\\
\hspace{2mm} 2012+ & 346 & 4.624(2.411,6.837) & 5.411(1.644,9.178) & 5.207(1.219,9.195) & 0(0,15.701)\\
\hline 
\multicolumn{3}{l}{\textbf{Vehicle make}} & & & \\ 
\hspace{2mm} American & 3,635 & 3.301(2.72,3.882) & 7.388(4.517,10.258) & 7.965(1.09,14.839) & 8.747(4.478,13.016)\\
\hspace{2mm} Japanese & 1,927 & 4.307(3.401,5.214) & 6.099(2.908,9.29) & 6.032(2.707,9.357) & 1.562(0,7.543)\\
\hspace{2mm} Korean & 387 & 3.359(1.564,5.154) & 8.952(2.792,15.112) & 9.172(2.504,15.841) & 8.469(0.439,16.499)\\
\hspace{2mm} Other & 322 & 5.28(2.837,7.722) & 3.023(0,8.831) & 2.777(0,8.478) & 3.854(0,9.39)\\
\hline 
\multicolumn{3}{l}{\textbf{Vehicle type}} & & & \\ 
\hspace{2mm} Passenger car & 4,063 & 3.569(2.998,4.139) & 6.075(3.814,8.337) & 6.245(2.521,9.969) & 4.639(0.478,8.8)\\
\hspace{2mm} Light truck & 1,213 & 4.122(3.003,5.241) & 8.549(3.935,13.164) & 8.846(2.517,15.175) & 9.311(2.577,16.044)\\
\hspace{2mm} Van & 290 & 1.379(0.037,2.722) & 3.755(0,11.694) & 4.766(0,15.574) & 12.787(0,27.049)\\
\hspace{2mm} SUV & 705 & 4.823(3.241,6.404) & 8.681(1.348,16.013) & 9.09(0.315,17.865) & 8.533(0,19.472)\\
\bottomrule
\end{tabular}}
    \begin{tablenotes}
      \small
      \item
    \end{tablenotes}
  \end{threeparttable}
\end{table}

\end{document}